\documentclass[a4paper,12pt]{article}
\pdfoutput=1
\usepackage{jheppub}
\usepackage[utf8]{inputenc}
\usepackage{url}
\usepackage{mathrsfs}  
\usepackage{pdflscape}
\usepackage{refcount}
\usepackage{booktabs}
\usepackage{todonotes}
\usepackage{enumitem}
\usepackage{adjustbox}
\usepackage{afterpage}
\usepackage{mathdots}
\usepackage{ytableau, youngtab} 

\usepackage{amsfonts}
\usepackage{bbm}

\usepackage{pdfpages}
\usepackage{caption}
\usepackage{subcaption}

\usepackage{tikz}
\usetikzlibrary{positioning}
\usetikzlibrary{arrows}
\usetikzlibrary{decorations.pathreplacing}
\usetikzlibrary{shapes.geometric}
\usetikzlibrary{calc}
\usetikzlibrary{decorations.pathmorphing}
\usetikzlibrary{shapes.misc}

\usetikzlibrary{positioning,trees,decorations.pathmorphing,decorations.markings,decorations.pathreplacing,calc,shapes,patterns,arrows,chains,arrows.meta,fit,fadings,decorations.markings,graphs,graphs.standard,quotes,plotmarks,calligraphy,decorations.pathreplacing}
\usetikzlibrary{intersections, calc}
\usetikzlibrary{shapes.geometric}
\usetikzlibrary{matrix, arrows.meta}
\usetikzlibrary{shapes.geometric}
\usetikzlibrary{decorations.pathmorphing}
\usetikzlibrary{shapes.misc}
\usetikzlibrary{arrows.meta}
\usetikzlibrary{angles, quotes}

\definecolor{goodgreen}{RGB}{55,169,49}

\definecolor{darkyellow}{RGB}{230,170,10}

\definecolor{brightyellow}{RGB}{255,240,190}

\tikzset{flavour/.style={draw=none,minimum size=0.3mm,fill=white, regular polygon,regular polygon sides=4,draw}}
\tikzset{gaugeBig/.style={inner sep=1mm,draw=none,fill=white,minimum size=2mm,circle, draw}}
\tikzset{bd/.style={circle, draw=black, inner sep=0pt, fill=black, minimum size=2mm}}
\tikzset{wd/.style={circle, draw=black, inner sep=0pt, fill=white, minimum size=2mm}}
\tikzset{Dynkin/.style={circle, draw=black, inner sep=0pt, fill=white, minimum size=2mm}}
\tikzstyle{ligne}=[draw, very thick] 
\tikzstyle{gridline}=[draw, gray] 
\tikzset{gauge/.style={circle, draw,inner sep=2.5pt}}
\tikzset{gaugeo/.style={circle, draw,inner sep=2.5pt,fill=orange}}
\tikzset{gaugec/.style={circle, draw,inner sep=2.5pt,fill=cyan}}
\tikzset{gauger/.style={circle, draw,inner sep=2.5pt,fill=red}}
\tikzset{gaugeb/.style={circle, draw,inner sep=2.5pt,fill=blue}}
\tikzset{gaugeg/.style={circle, draw,inner sep=2.5pt,fill=green}}
\tikzset{gaugem/.style={circle, draw,inner sep=2.5pt,fill=magenta}}
\tikzset{hasse/.style={circle, fill,inner sep=2pt}}
\tikzset{shrinky/.style={circle, fill,inner sep=1pt}}
\tikzset{sized/.style={circle, draw, inner sep=1.5pt}}
\tikzset{seven/.style={circle, draw,inner sep=3pt}}

\tikzset{dotto/.style={circle, orange, draw,inner sep=1.5pt,fill=orange}}
\tikzset{dottp/.style={circle, purple, draw,inner sep=1.5pt,fill=purple}}
\tikzset{dottc/.style={circle, cyan, draw,inner sep=1.5pt,fill=cyan}}
\tikzset{dottr/.style={circle, red, draw,inner sep=1.5pt,fill=red}}
\tikzset{dottb/.style={circle, blue, draw,inner sep=1.5pt,fill=blue}}
\tikzset{dottg/.style={circle, green, draw,inner sep=1.5pt,fill=green}}
\tikzset{dottm/.style={circle, magenta, draw,inner sep=1.5pt,fill=magenta}}

\tikzset{redgauge/.style={draw=none,minimum size=0.4cm,fill=red,circle, draw}}
\tikzset{gauge3/.style={draw=none,minimum size=0.4cm,fill=white,circle, draw}}
\tikzset{bluegauge/.style={draw=none,minimum size=0.4cm,fill=blue,circle, draw}}
\tikzset{redflavor/.style={draw=none,minimum size=0.6cm,fill=red, regular polygon,regular polygon sides=4,draw}}
\tikzset{blueflavor/.style={draw=none,minimum size=0.6cm,fill=blue, regular polygon,regular polygon sides=4,draw}}
\tikzset{flavour2/.style={draw=none,minimum size=0.6cm,fill=white, regular polygon,regular polygon sides=4,draw}}
\tikzset{rede/.style={line width=0.5mm,red}}
\tikzset{bluee/.style={line width=0.5mm,blue}}
\tikzset{pinkline/.style={line width=0.5mm,purple}}
\pgfdeclarelayer{edgelayer}
\pgfdeclarelayer{nodelayer}

\DeclareMathOperator{\U}{U}
\DeclareMathOperator{\SU}{SU}
\DeclareMathOperator{\SO}{SO}

\DeclareMathOperator{\USp}{USp}

\usepackage{epsfig}
\usepackage{amsmath}
\usepackage{amssymb}
\usepackage{amsfonts}
\usepackage{amsxtra}
\usepackage{amsthm}
\usepackage{mathrsfs}
\usepackage{makeidx}
\usepackage{graphics}
\usepackage{dsfont}
\usepackage{mathtools}
\usepackage{graphicx}
\usepackage{placeins}
\usepackage{bm}
\usepackage[capitalise]{cleveref}
\usepackage{empheq}
\usepackage{colortbl}
\usepackage{xcolor}
\usepackage{titlesec}
\usepackage{longtable}
\usepackage{hhline}
\usepackage{float}
\usepackage{color}
\usepackage{tikz}
\usepackage{xfrac}
\usepackage{footnote}
\usepackage{rotating}
\usepackage{lscape}
\usepackage{makecell}
\usepackage{environ}
\usepackage{tabularx}
\usepackage{subfiles}
\usepackage{ytableau}
\usepackage{tikz-3dplot}
\usepackage{slashed}
\usepackage{pifont}
\usepackage{multirow}
\usepackage{mdframed}
\usepackage{bbm}
\usepackage[normalem]{ulem}
\usepackage{arydshln}
\usepackage{enumitem} 
\usepackage{fancybox}

\usetikzlibrary{positioning,trees,decorations.pathmorphing,decorations.markings,decorations.pathreplacing,calc,shapes,patterns,arrows,chains,arrows.meta,fit,fadings,decorations.markings,graphs,graphs.standard,quotes,plotmarks,calligraphy,decorations.pathreplacing}

\usetikzlibrary{positioning}
\usetikzlibrary{chains}
\usetikzlibrary{arrows, arrows.meta ,fit,decorations.pathreplacing}
\tikzstyle{every picture}+=[remember picture]
\tikzstyle{na} = [baseline]
\tikzstyle{ligne}=[draw, thick]

\usetikzlibrary{arrows, decorations.markings, calc, fadings, decorations.pathreplacing, patterns, decorations.pathmorphing, positioning}
\tikzset{>={Latex[width=1.5mm,length=1.5mm]}}
\tikzset{bd/.style={circle, draw=black, inner sep=0pt, fill=black, minimum size=1.2mm}}
\tikzset{bld/.style={circle, draw=blue, inner sep=0pt, fill=blue, minimum size=1.2mm}}
\tikzset{wd/.style={circle, draw=black, inner sep=0pt, fill=white, minimum size=1.2mm}}
\tikzset{rd/.style={circle, draw=red, inner sep=0pt, fill=red, minimum size=.9mm}}
\tikzset{wrd/.style={circle, draw=red, inner sep=0pt, fill=white, minimum size=.9mm}}
\usetikzlibrary{graphs,graphs.standard,quotes}

\def\node#1#2{\overset{#1}{\underset{#2}{{\color{gray} \bullet}}}}

\def\node#1#2{\overset{#1}{\underset{#2}{\circ}}}

\tikzstyle{every picture}+=[remember picture]
\tikzstyle{na} = [baseline=-.5ex]

\newcommand{\eg}{e.g. }

\newcommand{\ie}{i.e. }

\numberwithin{equation}{section}
\newcommand{\bes}[1]{\begin{equation} \begin{split} #1\end{split} \end{equation}}

\newcommand{\be}{\begin{equation}} \newcommand{\ee}{\end{equation}}
\newcommand{\bea}{\begin{equation} \begin{aligned}} \newcommand{\eea}{\end{aligned} \end{equation}}

\def\tilde{\widetilde}

\def\hat{\widehat}

\def\rt2{\sqrt{2}}

\def\det{\mathop{\rm det}}

\def\CA{{\cal A}}

\def\CC{{\cal C}}

\def\CI{{\cal I}}

\def\CM{{\cal M}}
\def\CN{{\cal N}}

\def\CT{{\cal T}}

\def\CZ{{\cal Z}}

\def\1{{\ds 1}}

\newcommand{\cA}{\mathcal{A}}

\newcommand{\cC}{\mathcal{C}}

\def\SO{\mathrm{SO}}

\def\O{\mathrm{O}}
\def\SU{\mathrm{SU}}

\def\Spin{\mathrm{Spin}}
\def\Pin{\mathrm{Pin}}
\def\Usp{\mathrm{\USp}}

\def\su{\mathfrak{su}}

\def\so{\mathfrak{so}}

\def\usp{\mathfrak{usp}}

\def\repa{\raise4pt\hbox{$\square$}\mkern-14mu\raise-4pt\hbox{$\square$}}
\def\repab{\overline{\raise4pt\hbox{$\square$}\mkern-14mu\raise-4pt\hbox{$\square$}\mkern-1mu}}

\def\smileface{\ensuremath{\hbox{\large$\bigcirc$}\mkern-15mu\raise-1pt\hbox{\scriptsize$\smallsmile$}%
\mkern-10mu\raise4pt\hbox{..}\mkern4mu}}
\def\frownface{\ensuremath{\hbox{\large$\bigcirc$}\mkern-15mu\raise-1pt\hbox{\scriptsize$\smallfrown$}%
\mkern-10mu\raise4pt\hbox{..}\mkern4mu}}

\newcommand{\ba}{\begin{array}}
\newcommand{\ea}{\end{array}}
\newcommand{\bi}{\begin{itemize}}
\newcommand{\ei}{\end{itemize}}
\def\vec#1{\bm{#1}}
\def\bea#1\eea{\allowdisplaybreaks \begin{align}#1\end{align}}
 \newcommand{\ben}{\begin{enumerate}}
\newcommand{\een}{\end{enumerate}}
\newcommand{\bean}{\begin{eqnarray*}}
\newcommand{\eean}{\end{eqnarray*}}
\newcommand{\eref}[1]{(\ref{#1})}

\newcommand{\PE}{\mathop{\rm PE}}

\newcommand{\BC}{\mathbb{C}}
\newcommand{\BR}{\mathbb{R}}

\newcommand{\BZ}{\mathbb{Z}}

\newcommand{\BH}{\mathbb{H}}

\newcommand{\diag}{\mathrm{diag}}

\newcommand{\Sym}{\mathrm{Sym}}

\definecolor{light-gray}{gray}{0.5}
\definecolor{new-green}{rgb}{0,0.7,0.3}
\definecolor{cerulean}{rgb}{0.0, 0.48, 0.65}
\definecolor{claret}{rgb}{0.50, 0.09, 0.20}
\definecolor{darkred}{rgb}{0.7, 0.11, 0.11}
\definecolor{scarlet}{rgb}{1.0, 0.13, 0.0}
\definecolor{orange-red}{rgb}{1.0, 0.27, 0.0}
\definecolor{blue-green}{rgb}{0.0, 0.5, 0.65}
\definecolor{green-red}{rgb}{0.5, 0.65, 0.0}

\newcommand{\purple}{\color{purple}}

\newcommand{\blue}{\color{blue}}

\newcommand{\red}{\color{red}}
\newcommand{\green}{\color{new-green}}
\newcommand{\violet}{\color{violet}}

\newcommand{\cerulean}{\color{cerulean}}
\newcommand{\claret}{\color{claret}}

\newcommand{\bluegreen}{\color{blue-green}}
\newcommand{\greenred}{\color{green-red}}

\def\aup#1 {\overset{#1}{\uparrow} \, \overset{\tilde{#1}}{\downarrow}}

\tikzset{snake it/.style={decorate, decoration={snake, amplitude=.4mm, segment length=2mm,
                       post length=0mm,pre length=0mm}}}

\DeclareMathAlphabet{\mymathds}{U}{BOONDOX-ds}{m}{n}
\tikzstyle{double_border} = [draw, double, double distance=1pt]

\theoremstyle{plain}

\title{Interplay of Generalised Symmetries and Moduli Spaces in 3d $\mathcal{N}=5$ SCFTs}
\author[a,b]{Sebastiano Garavaglia,}
\author[a,b]{William Harding,}
\author[c]{Deshuo Liu,}
\author[b,d]{\\ and Noppadol Mekareeya}
\affiliation[a]{Dipartimento di Fisica, Universit\`a di Milano-Bicocca, Piazza della Scienza 3, I-20126 Milano, Italy}
\affiliation[b]{INFN, sezione di Milano-Bicocca,
Piazza della Scienza 3, I-20126 Milano, Italy}
\affiliation[c]{Theoretical Physics Group, The Blackett Laboratory, Imperial College London,
Prince Consort Road London, SW7 2AZ, UK}
\affiliation[d]{Department of Physics, Faculty of Science, Chulalongkorn University, Phayathai Road, Pathumwan, Bangkok 10330, Thailand}
\emailAdd{s.garavaglia18@campus.unimib.it}
\emailAdd{w.harding@campus.unimib.it}
\emailAdd{deshuo.liu21@imperial.ac.uk}
\emailAdd{n.mekareeya@gmail.com}

\abstract{
The moduli space and generalised global symmetries of 3d $\mathcal{N} = 5$ superconformal field theories are investigated, with a focus on the orthosymplectic ABJ theories and their discrete gauging variants. We extend the known classification of $\mathcal{N}=5$ moduli spaces as orbifolds $\mathbb{H}^{2N}/\Gamma$, where $\Gamma$ is a quaternionic reflection group, to theories incorporating $\mathrm{Spin}$, $\mathrm{O}^-$, and $\mathrm{Pin}$-type gauge groups. In these cases, we find that the moduli space is governed not by $\Gamma$ itself, but by a $\mathbb{Z}_2$ central extension thereof, for which we explicitly describe the generators. We provide a systematic method to construct the group $\Gamma'$ governing the moduli space of a theory $\mathcal{T}'$ obtained by gauging a $\mathbb{Z}_2$ zero-form symmetry of an original theory $\mathcal{T}$. This is achieved by identifying the specific generator that must be added to $\Gamma$. We compute the Hilbert series for these moduli spaces and verify them against the corresponding limits of the superconformal index, finding perfect agreement. We also discuss how 't Hooft anomalies for the zero-form symmetries manifest in the superconformal index and the moduli space. Furthermore, we revisit the symmetry category of the $\mathfrak{so}(2N)_{2k} \times \mathfrak{usp}(2N)_{-k}$ theories. Building on previous work that identified the symmetry category for all parities of $N$ and $k$, we provide the explicit symmetry webs for the opposite parity $D_8$ case. We find that the details of these webs differ from the previously studied $D_8$ webs corresponding to the both even parity case. Finally, we analyse theories with unequal ranks, those containing the $\mathfrak{so}(2N+1)$ gauge algebra, and the two SCFT variants based on the $F(4)$ superalgebra.
}

\begin{document}
\maketitle

\section{Introduction}
Three-dimensional superconformal field theories (SCFTs) with $\CN \geq 5$ supersymmetry possess rich moduli space structures and generalised global symmetries. It was pointed out in \cite{Tachikawa:2019dvq} that, upon appropriate gauging of finite symmetry groups, the moduli spaces of these 3d SCFTs are given by orbifolds of reflection groups. For $\CN=8$, the moduli space is $\BR^{8N}/\Gamma$, with $\Gamma$ a real reflection group. For $\CN=6$, it is $\BC^{4N}/\Gamma$, with $\Gamma$ a complex reflection group. Similarly, for the $\CN=5$ case, the moduli space can be identified with $\BH^{2N}/\Gamma$, with $\Gamma$ a quaternionic reflection group \cite{Deb:2024zay}. For $\CN=8$, the list of real reflection groups provides a classification scheme for $\CN=8$ SCFTs, in the sense that any two $\CN=8$ SCFTs with the same moduli space must either be the same or be related to each other by gauging some finite group \cite{Tachikawa:2019dvq} (see also \cite{Bashkirov:2011pt, Gang:2011xp, Agmon:2017lga}). However, as supersymmetry decreases to $\CN=6$ or $\CN=5$, this classification becomes weaker: distinct SCFTs, not related by discrete gauging, may share the same moduli space. Nevertheless, the correspondence between $\Gamma$ and the SCFTs suggests that novel theories may remain to be discovered, see \eg~ \cite{Kaidi:2022uux}.

Let us focus on the Lagrangian subclass of these SCFTs. Their gauge algebra and matter content are classified by Lie superalgebras \cite{KAC19778, Gaiotto:2008sd, Hosomichi:2008jb, Schnabl:2008wj}, where the former corresponds to the bosonic generators and the latter to the fermionic generators. Of course, many variants can exist that are related by gauging finite discrete symmetries for a given superalgebra. For $\CN \geq 6$ SCFTs, two superalgebras are relevant: $\SU(M|N)$ and $\mathrm{PSU}(N|N)$. A notable theory in the $\SU(M|N)$ class is the $\U(M)_k \times \U(N)_{-k}$ theory with two bifundamental hypermultiplets, known as the ABJM theory \cite{Aharony:2008ug} when $M=N$, and the unitary ABJ theory \cite{Aharony:2008gk} when $M\neq N$. These theories are realised on the worldvolume of M2-branes probing the $\BC^4/\BZ_k$ singularity. Related variants also exist that can be obtained by gauging a one-form symmetry, \eg~ $[\U(N)_k \times \U(N)_{-k}]/\BZ_k$. The $\mathrm{PSU}(N|N)$ class corresponds to the $[\SU(N)_k \times \SU(N)_{-k}]/\BZ_N$ theory, which is dual to $[\U(N)_k \times \U(N)_{-k}]/\BZ_k$ \cite{Tachikawa:2019dvq, Bergman:2020ifi} (see also \cite{Lambert:2010ji}); thus, this class is effectively equivalent to $\SU(N|N)$. The $N=2$ case corresponds to the BLG theories \cite{Bagger:2006sk, Gustavsson:2007vu, VanRaamsdonk:2008ft}.

For $\CN \geq 5$ SCFTs, there are four known superalgebra classes: $\mathrm{OSp}(M|2N)$, $D(2|1; \alpha)$, $F(4)$, and $G(3)$. The $\mathrm{OSp}(M|2N)$ class includes the theories with gauge groups $\O(2M)^+_{2k} \times \USp(2N)_{-k}$ and $\O(2M+1)^+_{2k} \times \USp(2N)_{-k}$ with two bifundamental half-hypermultiplets, which are realised on M2-branes probing the $\BC^4/\hat{D}_k$ singularity. These are the {\it orthosymplectic ABJ theories} \cite{Aharony:2008gk}. These theories, along with variants obtained by discrete gauging, are the main protagonists of this paper. We will also examine the $F(4)$ superalgebra class, which contains the $\Spin(7)_{-3k} \times \SU(2)_{2k}$ and $[\Spin(7)_{-3k} \times \SU(2)_{2k}]/\BZ_2$ gauge theories with two half-hypermultiplets in the $(\mathbf{8}, \mathbf{2})$ representation. The $D(2|1;\alpha)$ class, containing the $\SU(2)_{k_1} \times \SU(2)_{k_2} \times \SU(2)_{k_3}$ gauge group with the trifundamental hypermultiplet and $\sum_{i=1}^3 k_i^{-1}=0$, was analysed in \cite{Assel:2022row, Comi:2023lfm, Deb:2024zay}, and the $G(3)$ class, containing the $\SU(2)_{3k} \times (G_2)_{-4k}$ gauge theory with two half-hypermultiplets in $(\mathbf{2}, \mathbf{7})$ representation, was studied in \cite{Deb:2024zay}.

The orthosymplectic ABJ theories and their variants also possess rich generalised global symmetry structures. Some aspects of these, including the presence of two-groups and non-invertible symmetries, were studied in \cite{Beratto:2021xmn, Mekareeya:2022spm} using the superconformal index (see also \cite{Bhardwaj:2022dyt, Bhardwaj:2023zix}). The authors of \cite{Bergman:2024its} subsequently pointed out that the underlying finite non-Abelian global symmetry of the theories with gauge algebra $\so(2N)_{2k} \times \usp(2N)_{-k}$ is either the quaternion group $Q_8$ (when $k$ and $N$ are both odd) or the dihedral group $D_8$ of order eight (for other parities). Sequentially gauging subgroups of these non-Abelian finite symmetries produces an intricate symmetry web and associated symmetry categories. The $D_8$ case, in particular, was previously discussed in \cite{Bhardwaj:2022maz, Bartsch:2022ytj} (see also \cite{Tachikawa:2017gyf, Bhardwaj:2017xup, Buican:2020who, Bhardwaj:2022yxj, Bhardwaj:2022lsg, Bartsch:2022mpm, Bhardwaj:2022kot, Bartsch:2023pzl, Bartsch:2023wvv, Bhardwaj:2023ayw, Schafer-Nameki:2023jdn, Buican:2023bzl, Choi:2024rjm, Bullimore:2024khm}). Moreover, \cite{Deb:2024zay} investigated the moduli space of the orthosymplectic ABJ theories for several forms of the gauge group, including those with $\SO$ and $\mathrm{O}^+$ types, along with possible $\mathbb{Z}_2$ quotients. That study found the moduli space to be of the form $\mathbb{H}^{2N}/\Gamma$, where $\Gamma$ is a quaternionic reflection group.

This paper extends existing results in the literature in many directions. First, we investigate the moduli space of the orthosymplectic ABJ theories with $\Spin$, $\O^-$ and $\Pin$-type gauge groups. We find that, in these cases, the corresponding $\Gamma$ may not be a quaternionic reflection group itself, but a $\BZ_2$ central extension thereof. We describe the generators of such extensions in detail. Second, when gauging a $\BZ^{[0]}_2$ zero-form symmetry of a theory $\CT$ (whose moduli space is $\BH^{2N}/\Gamma$) which leads to another theory $\CT'$ (whose moduli space is $\BH^{2N}/\Gamma'$), we provide a systematic way to construct $\Gamma'$ from $\Gamma$. Specifically, one simply needs to add an appropriate generator to $\Gamma$ to obtain $\Gamma'$, and we provide an explicit expression for this generator. We compute the Hilbert series for the Higgs or Coulomb branch (viewed as an $\CN=4$ moduli space) of these theories, which is isomorphic to $\BH^N/\Gamma$, and verify it against the corresponding limit of the superconformal index, finding agreement in all cases. We also identify situations where the $\BZ^{[0]}_2$ zero-form symmetry in question is not gaugable, and discuss how to detect the corresponding anomaly from the perspective of the superconformal index and the moduli space. Third, we revisit the symmetry category of the orthosymplectic ABJ theories with the gauge algebra $\so(2N)_{2k} \times \usp(2N)_{-k}$ for every parity of $N$ and $k$. The explicit $D_8$ and $Q_8$ symmetry webs for the case where $N$ and $k$ have the same parity were provided in \cite{Bergman:2024its}. That work also established that the symmetry category is $D_8$ for $N$ and $k$ with opposite parities. However, the details of the symmetry webs, which we provide explicitly in this work, differ from the case where both $N$ and $k$ are even. We also analyse unequal-rank orthosymplectic ABJ theories. In contrast to the equal-rank case, some symmetries do not act faithfully on the moduli space. We also briefly discuss theories containing the $\so(2N+1)$ gauge algebra. Finally, we discuss the two variants of the SCFTs based on the $F(4)$ superalgebra.

The paper is organised as follows. In Section \ref{sec_fti}, we discuss 't Hooft anomalies in general orthosymplectic ABJ theories. Section \ref{sec:ABJMequalranks} is devoted to the analysis of the orthosymplectic ABJ theories with equal ranks. We discuss the quaternionic reflection groups, their extensions, and their generators in Section \ref{quatrefgroupmodspace}. Theories with unequal ranks are discussed in Section \ref{sec:unequalranks}. We discuss theories with the $\so(2N+1)$ gauge algebra in Section \ref{Sec:Soodd}. Section \ref{sec:F4} discusses the SCFTs based on the $F(4)$ superalgebra. We collect the formulae for computing the superconformal index of the theories in this paper in Appendix \ref{app:index}.

\section{'t Hooft anomalies of the orthosymplectic ABJ theories}
\label{sec_fti}
We begin by considering the 3d $\mathcal{N}=3$ $\SO(2L)_{2k_1} \times \USp(2M)_{k_2}$ theory, which includes two bifundamental half-hypermultiplets:
\begin{equation} \label{SO2LUSp2M}
\vcenter{\hbox{\begin{tikzpicture}[font=\footnotesize]
        \node[gauge,label={below,xshift=-0.3cm}:{$\SO(2L)_{2k_1}$}] (so) at (0,0) {};
        \node[gauge,label={below,xshift=0.6cm}:{$\USp(2M)_{k_2}$}] (usp) at (5,0) {};
        \draw (so) to [bend right=20] node[midway, above] {} (usp);
        \draw (so) to [bend left=20] node[midway, above] {} (usp);
\end{tikzpicture}}}
\end{equation}
where we take $k_1$ and $k_2$ to be integers. This theory has $\mathbb{Z}^{[0]}_{2, \mathcal{M}}$ (magnetic) and $\mathbb{Z}^{[0]}_{2, \mathcal{C}}$ (charge conjugation) zero-form symmetries, both associated with the $\SO(2L)$ gauge node. Furthermore, the $\mathbb{Z}_2 \times \mathbb{Z}_2$ centre symmetry of the $\SO(2L) \times \USp(2M)$ gauge group is screened by the bifundamental fields, reducing it to the diagonal $\mathbb{Z}_2$ subgroup. This subgroup is identified as a $\mathbb{Z}^{[1]}_2$ one-form symmetry \cite{Tachikawa:2019dvq,Beratto:2021xmn} (see also \cite{Bergman:2020ifi}).

These discrete $p$-form symmetries (for $p=0,1$) can be coupled to $(p+1)$-form background gauge fields. Specifically, we denote the one-form background gauge fields for $\mathbb{Z}^{[0]}_{2, \mathcal{M}}$ and $\mathbb{Z}^{[0]}_{2, \mathcal{C}}$ as $\mathcal{A}^{\mathcal{M}}_1$ and $\mathcal{A}^{\mathcal{C}}_1$, respectively. The $\mathbb{Z}^{[1]}_2$ one-form symmetry couples to a two-form background gauge field, denoted as $\mathcal{A}^{B}_2$. The mixed 't Hooft anomaly for these discrete global symmetries in the theory \eqref{SO2LUSp2M} is described by the action:
\bes{\label{anomSO2LUSp2M}
i \pi \int_{M_4} \mathcal{A}^B_2 \cup \Big[L \mathcal{A}^{\mathcal{M}}_1 \cup \mathcal{A}^{\mathcal{M}}_1 &+ k_1 \mathcal{A}^{\mathcal{C}}_1 \cup \mathcal{A}^{\mathcal{C}}_1 + \mathcal{A}^{\mathcal{M}}_1 \cup \mathcal{A}^{\mathcal{C}}_1 + \left(k_1 L + k_2 M\right) \mathcal{A}^B_2 \Big]~,}
where $M_4$ is the 4d bulk whose boundary is the 3d spacetime of the theory. The terms of the form $\mathcal{A}^B_2 \cup \mathcal{A}^{\mathcal{M},\mathcal{C}}_1 \cup \mathcal{A}^{\mathcal{M},\mathcal{C}}_1$ can also be written as $\mathcal{A}^B_2 \cup e(\mathcal{A}^{\mathcal{M},\mathcal{C}}_1)$, where $e(\mathcal{A}^{\mathcal{M},\mathcal{C}}_1)$ is the non-trivial element defined by the extension class $e \in H^2(\mathbb{Z}_2, \mathbb{Z}_2) = \mathbb{Z}_2$, corresponding to the short exact sequence $0 \rightarrow \mathbb{Z}_2 \rightarrow \mathbb{Z}_4 \rightarrow \mathbb{Z}_2 \rightarrow 0$.\footnote{In other words, consider an anomaly theory of the form $\int_{M_4} \mathcal{A}_2 \cup \mathcal{A}_1 \cup \mathcal{A}_1$, also known as a $(2+1)$d Type III anomaly (see \cite{Choi:2024rjm}), where $\mathcal{A}_1$ and $\mathcal{A}_2$ are background fields for a $\mathbb{Z}_2$ zero-form and one-form symmetry, respectively. When the $\mathbb{Z}_2$ one-form symmetry is gauged, the zero-form symmetry is enhanced to $\mathbb{Z}_4$. This $\mathbb{Z}_4$ is an extension of $\mathbb{Z}_2$ by $\mathbb{Z}_2$, as described by the short exact sequence in the main text.}

The anomaly in \eqref{anomSO2LUSp2M} can be derived from \cite{Cordova:2017vab} and matches the result in \cite[(3.5)]{Bergman:2024its} for the special case where $L=M$ and $k_1 = -k_2$.\footnote{In that reference, the anomaly theory is expressed in terms of $\delta \tilde{\mathcal{A}}^{\mathcal{M},\mathcal{C}}_1 \sim e(\mathcal{A}^{\mathcal{M},\mathcal{C}}_1)$, where $\tilde{\mathcal{A}}^{\mathcal{M},\mathcal{C}}_1$ is the one-cochain responsible for the lift of $\mathcal{A}^{\mathcal{M},\mathcal{C}}_1$ to $\mathbb{Z}_4$.} For general values of $L, M, k_1,$ and $k_2$, the last terms of \eqref{anomSO2LUSp2M}, \ie the ones containing the round bracket, correspond to self-anomalies of the one-form symmetry. Specifically, the term $k_1 L \mathcal{A}^B_2 \cup \mathcal{A}^{B}_2$ is the self-anomaly from the $\SO(2L)_{2k_1}$ gauge factor \cite[(2.17)]{Cordova:2017vab}, while $k_2 M \mathcal{A}^B_2 \cup \mathcal{A}^{B}_2$ is the self-anomaly from the $\USp(2M)_{k_2}$ gauge factor.\footnote{As explained in \cite[Section 2.3]{Benini:2017dus}, the anomaly for the $\USp(2M)_{k}$ gauge theory with $N_f$ scalars in the vector representation is given by $i \pi \int_{M_4} w_2 \cup w_2$ when $k M$ is odd and $N_f$ is even, where $w_2$ is the second Stiefel-Whitney class, which is an obstruction to lifting the $\USp(2M)/\BZ_2$ bundles to the $\USp(2M)$ bundles. In our case, we identify $k = k_2$, $w_2 = \CA^B_2$, and the $\USp(2M)$ gauge node is attached to an even number of vector flavours $N_f = 2 L$, resulting into a non-vanishing self-anomaly for the one-form symmetry when $k_2 M$ is odd.} 

Consequently, the $\mathbb{Z}^{[1]}_2$ one-form symmetry can be gauged if the total self-anomaly vanishes, which occurs when the following condition is met (see \cite[Footnote 4]{Comi:2023lfm}):
\begin{equation} \label{Z21formcond}
\frac{k_1}{2} L + \frac{k_2}{2} M \in \mathbb{Z}~.
\end{equation}
This condition arises from a slight modification of the argument in \cite[(3.27)]{Tachikawa:2019dvq} (see also \cite{Deb:2024zay}), which we briefly review here. For the $\mathbb{Z}_2$ quotient to be non-anomalous in the $\SO(2L)_{2k_1} \times \USp(2M)_{k_2}$ gauge theory, the variation of the Chern-Simons action under a gauge transformation must be trivial. This variation can produce a phase factor $\exp\left[2 \pi i \left(2k_1 l_{\SO(2L)/\mathbb{Z}_2} + k_2 l_{\USp(2M)/\mathbb{Z}_2}\right)\right]$, where $l_G$ is the instanton number for the bundle of gauge group $G$. While $l_G$ is an integer for a simply connected gauge group $G$, it can be fractional for non-simply connected groups \cite{Witten:2000nv} (see also \cite{Aharony:2013hda}). Specifically, $l_{\SO(2L)/\mathbb{Z}_2}$ can be half-integer (for even $L$) or a multiple of $1/4$ (for odd $L$) \cite{Witten:2000nv}. Similarly, $l_{\USp(2M)/\mathbb{Z}_2}$ is an integer for even $M$, but can be half-integer for odd $M$.

When condition \eqref{Z21formcond} holds, gauging the $\mathbb{Z}^{[1]}_2$ one-form symmetry yields the $\left[\SO(2L)_{2k_1} \times \USp(2M)_{k_2}\right]/\mathbb{Z}_2$ theory. This resulting theory features a dual $\mathbb{Z}^{[0]}_{2,B}$ zero-form symmetry,\footnote{Recall that gauging a $p$-form symmetry in $d$ dimensions gives rise to a dual $(d-p-2)$-form symmetry.} which, combined with the existing magnetic and charge conjugation symmetries from the $\SO(2L)$ node, appears to form a $\mathbb{Z}^{[0]}_{2,B} \times \mathbb{Z}^{[0]}_{2, \mathcal{M}} \times \mathbb{Z}^{[0]}_{2,\mathcal{C}}$ Abelian group of zero-form symmetries. However, the non-trivial relationships dictated by the anomaly \eqref{anomSO2LUSp2M} cause these three $\mathbb{Z}_2$ symmetries to combine and enhance into a finite non-Abelian group, such as the dihedral group $D_8$ or the quaternion group $Q_8$, as discussed in \cite{Bergman:2024its}.\footnote{In order for this enhancement to take place, an essential role is played by the Type III anomaly term $\int_{M_4} \cA^B_2 \cup \cA^{\CM}_1 \cup \cA^{\cC}_1$. Let us suppose that condition \eref{Z21formcond} is satisfied, and consider the following modification of the anomaly theory \eref{anomSO2LUSp2M}: $i \pi \int_{M_4} \cA^B_2 \cup \left(L \cA^{\CM}_1 \cup \cA^{\CM}_1 + k_1 \cA^{\cC}_1 \cup \cA^{\cC}_1 + t \cA^{\CM}_1 \cup \cA^{\cC}_1\right)$, where $t = \{0, 1\}$. As explained around \cite[(2.4)]{Bergman:2024its}, the various possible extensions of an order four element of $\BZ_2 \times \BZ_2$ by an order two element are classified by $H^2(\BZ_2 \times \BZ_2, \BZ_2)$, with the following outcome: 1) The trivial extension $\BZ_2 \times \BZ_2 \times \BZ_2$ corresponds to $L$, $k_1$ and $t$ all equal to zero. 2) The elements $(L, k_1, t) ) = \{(1,0,0), (0,1,0), (1,1,0)\}$ give rise to the $\BZ_4 \times \BZ_2$ extension. 3) The $D_8$ group is associated with $(L, k_1, t) ) = \{(0,0,1), (1,0,1), (0,1,1)\}$. 4) Finally, the $Q_8$ group arises from the element in which $L$, $k_1$ and $t$ all equal to one. In particular, observe that, as long as $t$ is different from zero, or, in other words, the term $\int_{M_4} \cA^B_2 \cup \cA^{\CM}_1 \cup \cA^{\cC}_1$ is non-vanishing, then $\BZ_2 \times \BZ_2 \times \BZ_2$ is always extended to a non-Abelian finite group, either $D_8$ or $Q_8$.}

\begin{figure}
\centering
\scalebox{0.75}{
\begin{tikzpicture} 
			\node[draw] (ord8) at (0,-4.5) {\begin{tabular}{c}
                Whole group \\ $\langle r, s \rangle$ \end{tabular} }; 
            \node[draw] (ord4ch) at (0,-1.5) {\begin{tabular}{c}
                Order four, characteristic \\ $\langle r \rangle$ \end{tabular} };
            \node[draw] (ord4normr) at (6.5,-1.5) {\begin{tabular}{c}
                Order four, normal \\ $\langle r^2, r s \rangle$ \end{tabular} };
            \node[draw] (ord4norml) at (-6.5,-1.5) {\begin{tabular}{c}
                Order four, normal \\ $\langle r^2, s \rangle$ \end{tabular} };
            \node[draw] (ord2ch) at (0,1.5) {\begin{tabular}{c}
                Order two, characteristic \\ centre $\langle r^2 \rangle$ \end{tabular} };
            \node[draw] (ord2subr) at (6.5,1.5) {\begin{tabular}{c}
                Order two, two-subnormal \\ $\langle r s \rangle \overset{s \cdot s^{-1}}{\longleftrightarrow} \langle r^3 s \rangle$ \end{tabular} };
            \node[draw] (ord2subl) at (-6.5,1.25) {\begin{tabular}{c}
                Order two, two-subnormal \\ $\langle s \rangle \overset{r s \cdot s r^3}{\longleftrightarrow} \langle r^2 s \rangle$ \end{tabular} };
            \node[draw] (trivial) at (0,4.5) {\begin{tabular}{c}
                Trivial subgroup \\ $\langle 1 \rangle$ \end{tabular} }; 
            \draw[->] (trivial) to [bend right=10] (ord2subl);
            \draw[->] (trivial) to [bend left=10] (ord2subr);
            \draw[->] (trivial) to (ord2ch);
            \draw[->] (ord2subl) to (ord4norml);
            \draw[->] (ord2subr) to (ord4normr);
            \draw[->] (ord2ch) to [bend right=10] (ord4norml);
            \draw[->] (ord2ch) to [bend left=10] (ord4normr);
            \draw[->] (ord2ch) to (ord4ch);
            \draw[->] (ord4norml) to [bend right=10] (ord8);
            \draw[->] (ord4normr) to [bend left=10] (ord8);
            \draw[->] (ord4ch) to (ord8);
\end{tikzpicture}
}
    \caption[D8lattice]{Lattice of subgroups of $D_8$, where each box represents a distinct conjugacy class of subgroups. Observe that the order two two-subnormal subgroups enjoy an inner automorphism generated by conjugations $x = s x s^{-1} = s x s$, $y = (r s) y (s r^3) = (r s) y (r s)$, where $x \in \langle rs \rangle$ and $y \in \langle s \rangle$.} \label{fig:D8lattice}
\end{figure}
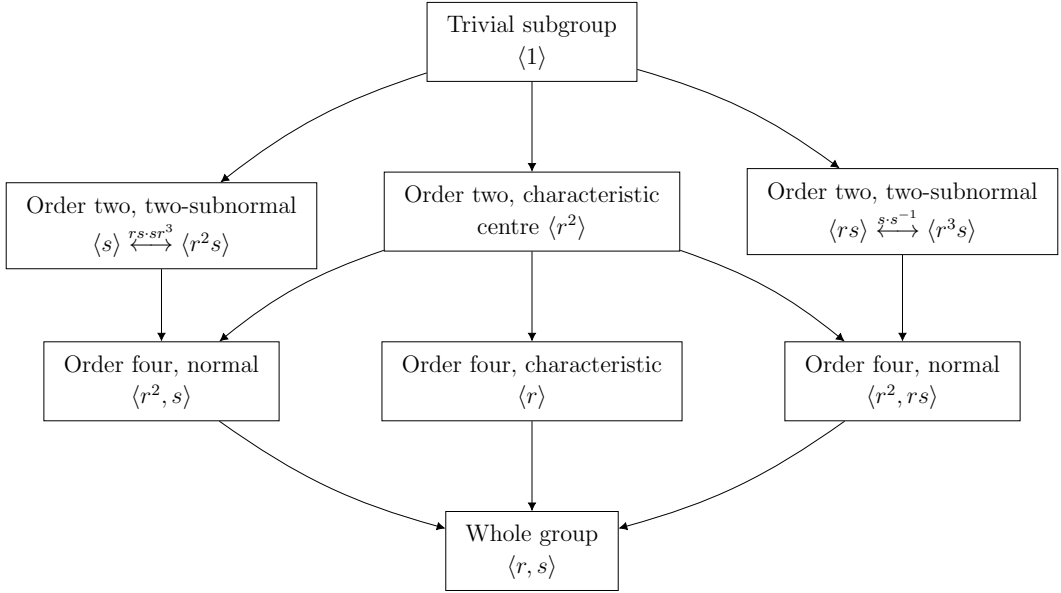

The dihedral group $D_8$ of order eight represents the symmetries of a square and can be generated by a rotation $r$ and a reflection $s$, with the presentation
\begin{equation}
D_8 = \langle r, s| r^4 = 1, \, s^2 = 1, \, s r s^{-1} = r^{-1}\rangle ~.
\end{equation}
There are five conjugacy classes, explicitly $\{1\}$, $\{r^2\}$, $\{r, r^3\}$, $\{s, r^2 s\}$ and $\{r s, r^3 s\}$, hence there are five (four one-dimensional and one two-dimensional) irreducible representations. There are ten subgroups of $D_8$, including the whole group itself, divided into eight conjugacy classes, which can be organised in the lattice of subgroups depicted in Figure \ref{fig:D8lattice}.

\begin{figure}
\centering
\scalebox{0.75}{
\begin{tikzpicture} 
			\node[draw] (ord8) at (0,-4.5) {\begin{tabular}{c}
                Whole group \\ $\{1, -1, i, -i, j, -j, k, -k\} \equiv \langle x, y \rangle$ \end{tabular} }; 
            \node[draw] (ord4norm) at (0,-1.5) {\begin{tabular}{c}
                Order four, normal \\ $\{1, j, -1, -j \} \equiv \langle x \rangle$ \end{tabular} };
            \node[draw] (ord4norml) at (-6.5,-1.5) {\begin{tabular}{c}
                Order four, normal \\ $\{1, i, -1, -i\} \equiv \langle y \rangle$ \end{tabular} };
            \node[draw] (ord4normr) at (6.5,-1.5) {\begin{tabular}{c}
                Order four, normal \\ $\{1, k, -1, -k\} \equiv \langle x y \rangle$ \end{tabular} };
            \node[draw] (ord2ch) at (0,1.5) {\begin{tabular}{c}
                Order two, characteristic \\ centre $\{1, -1\} \equiv \langle x^2 \rangle$ \end{tabular} };
            \node[draw] (trivial) at (0,4.5) {\begin{tabular}{c}
                Trivial subgroup \\ $\langle 1 \rangle$ \end{tabular} }; 
            \draw[->] (trivial) to (ord2ch);
            \draw[->] (ord2ch) to [bend right=10] (ord4norml);
            \draw[->] (ord2ch) to [bend left=10] (ord4normr);
            \draw[->] (ord2ch) to (ord4norm);
            \draw[->] (ord4norml) to [bend right=10] (ord8);
            \draw[->] (ord4normr) to [bend left=10] (ord8);
            \draw[->] (ord4norm) to (ord8);
\end{tikzpicture}
}
    \caption[Q8lattice]{Lattice of subgroups of $Q_8$, where each box represents a distinct subgroup.} \label{fig:Q8lattice}
\end{figure}
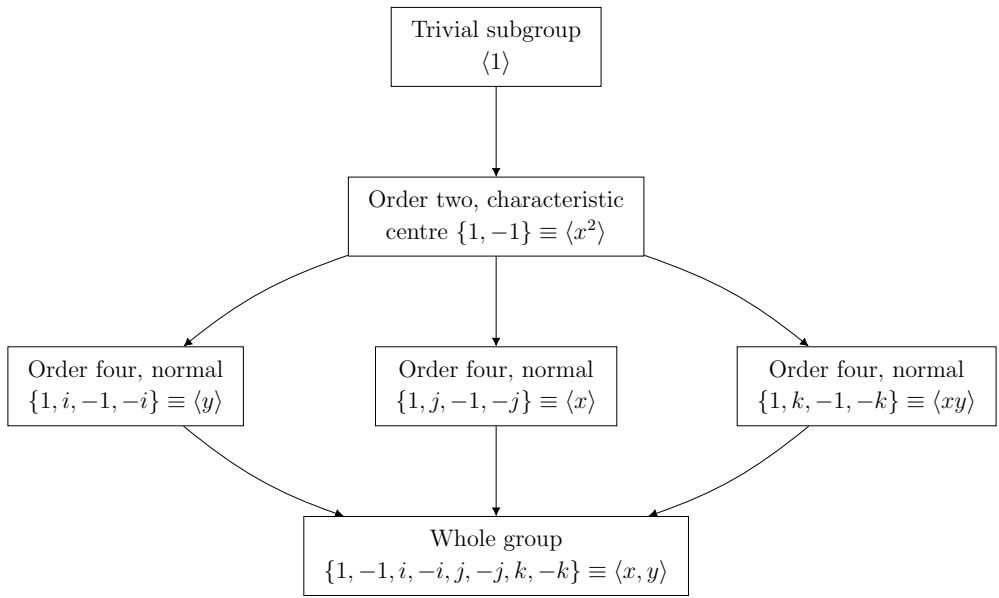

Let us also introduce the quaternion group $Q_8$ of order eight, with elements $\{1, -1, i, -i, j, -j, k, -k\}$ satisfying $i^2 = j^2 = k^2 = -1$, $i j = - j i =k$, $j k = - k j = i$ and $k i = -i k = j$. This also admits the presentation
\bes{
\langle x, y| x^4 = y^4 = 1, \, x^2 = y^2, \, y x y^{-1} = y^{-1}\rangle ~,
}
where $x$ and $y$ can be identified with any pair of distinct elements from $\{\pm i, \pm j, \pm k\}$. Also in this case, there are five conjugacy classes, \ie $\{1\}$, $\{-1\}$, $\{i, -i\}$, $\{j, -j\}$, $\{k, -k\}$, resulting into four one-dimensional and a single two-dimensional irreducible representations. The quaternion group $Q_8$ possesses five proper subgroups, which, together the whole group itself, give rise to the lattice of subgroups reported in Figure \ref{fig:Q8lattice}.

As we are now going to discuss extensively, the various global forms of the gauge group of theories with $\so(2L)_{2 k_1} \times \usp(2M)_{k_2}$ gauge algebra, where the Chern-Simons levels $k_1$ and $k_2$ satisfy condition \eref{Z21formcond}, perfectly fit into the lattice of subgroups of either $D_8$ or $Q_8$, depending on the parity of $L$ and $k_1$, where the analogy with Figures \ref{fig:D8lattice} and \ref{fig:Q8lattice} is as follows: each box coincides with a particular global variant of the theory, and each black arrow corresponds to the gauging of a $\BZ_2$ zero-form symmetry constructed from $\mathbb{Z}^{[0]}_{2,B} \times \mathbb{Z}^{[0]}_{2, \mathcal{M}} \times \mathbb{Z}^{[0]}_{2,\mathcal{C}}$.

\section{Orthosymplectic ABJ theories with equal ranks}\label{sec:ABJMequalranks}
We now specialise to the case of equal ranks, setting $L=M=N$, and choose opposite Chern-Simons levels, $k_1 = k$ and $k_2 = -k$. This parameter choice defines the orthosymplectic ABJ theories with equal ranks \cite{Hosomichi:2008jb,Aharony:2008gk}. These theories are 3d SCFTs with $\mathcal{N}=5$ supersymmetry for $k \geq 2$ and $\mathcal{N}=6$ supersymmetry for $k=1$.\footnote{In the special case $N = 1$ and $k = 1$, the $\SO(2)_{2} \times \USp(2)_{-1}$ ABJ theory actually possesses $\CN = 8$ supersymmetry. Indeed, the theory in question is dual to the $\left[\U(1)_4 \times \U(1)_{-4}\right]/\BZ_2$ variant of the ABJM theory \cite{Beratto:2021xmn}, which is in turn dual to the $\U(1)_2 \times \U(1)_{-2}$ ABJM theory, with $\CN=8$ supersymmetry \cite{Aharony:2008ug}.} The case of unequal ranks will be discussed in the next section. Under these constraints, the 't Hooft anomaly in \eqref{anomSO2LUSp2M} simplifies to
\begin{equation} \label{anomABJequalranks}
i \pi \int_{M_4} \mathcal{A}^B_2 \cup \Big[N \mathcal{A}^{\mathcal{M}}_1 \cup \mathcal{A}^{\mathcal{M}}_1 + k \mathcal{A}^{\mathcal{C}}_1 \cup \mathcal{A}^{\mathcal{C}}_1 + \mathcal{A}^{\mathcal{M}}_1 \cup \mathcal{A}^{\mathcal{C}}_1  \Big]~.
\end{equation}

Recent studies have explored the rich symmetry structures of these theories. The presence of two-groups and non-invertible symmetries in various orthosymplectic ABJ theories was highlighted in \cite{Mekareeya:2022spm}, while the $D_8$ and $Q_8$ categorical symmetries were analysed in \cite{Bergman:2024its}. Furthermore, the work of \cite{Deb:2024zay} investigated the moduli space of ABJ theories with gauge algebra $\mathfrak{so}(2N)_{2k} \times \mathfrak{usp}(2N)_{-k}$ for several global forms of the gauge group, including $\SO(2N)$ and $\mathrm{O}(2N)^+$ for the orthogonal gauge algebra, along with possible $\mathbb{Z}_2$ quotients. That study found the moduli space to be of the form $\mathbb{H}^{2N}/\Gamma$, where $\Gamma = G_N(H, K)$ is a quaternionic reflection group.

This raises a natural question: what is the structure of the moduli space for other choices of the orthogonal gauge group, including $\Spin(2N)$, $\mathrm{O}(2N)^-$, and $\Pin(2N)$? In this section, we provide a complete answer. We demonstrate that, for these variants, the group $\Gamma$ is not a quaternionic reflection group itself, but rather a $\mathbb{Z}_2$ extension thereof. Our findings for all parity combinations of $N$ and $k$ are summarised below.
\bes{ \label{tab:summaryNk}
\scalebox{0.8}{
\renewcommand{\arraystretch}{1.4} 
\begin{tabular}{c|c|c|c}
\hline
Parity of $N$ & Parity of $k$ & Figure & Anomalous variant \\
\hline
Even & Even & \ref{fig:D8Nevenkeven} & 
\makecell{$[\mathrm{O}(2N)^-_{2k} \times \USp(2N)_{-k}]/\mathbb{Z}_2$ \\ $[\Pin(2N)_{2k} \times \USp(2N)_{-k}]/\BZ_2$} \\
\hline
Even & Odd & \ref{fig:D8Nevenkodd} & \makecell{$[\mathrm{O}(2N)^+_{2k} \times \USp(2N)_{-k}]/\mathbb{Z}_2$ \\ $[\Pin(2N)_{2k} \times \USp(2N)_{-k}]/\BZ_2$} \\
\hline
Odd & Even & \ref{fig:D8Noddkeven} & \makecell{$[\Spin(2N)_{2k} \times \USp(2N)_{-k}]/\mathbb{Z}_2$ \\ $[\Pin(2N)_{2k} \times \USp(2N)_{-k}]/\BZ_2$ }\\
\hline
Odd & Odd & \ref{fig:Q8Noddkodd} & \makecell{$[\mathrm{O}(2N)^\pm_{2k} \times \USp(2N)_{-k}]/\mathbb{Z}_2$ \\ $[\Spin(2N)_{2k} \times \USp(2N)_{-k}]/\mathbb{Z}_2$ \\ $[\Pin(2N)_{2k} \times \USp(2N)_{-k}]/\BZ_2$}  \\
\hline
\end{tabular}}}

In Table \eqref{tab:summaryNk}, we also identify the anomalous variants. These are obtained by attempting to gauge discrete symmetries of the $\SO(2N)_{2k} \times \USp(2N)_{-k}$ theory in a way that is forbidden by the 't Hooft anomalies \eqref{anomABJequalranks}.\footnote{For variants with an $\mathrm{O}(2N)^-$ gauge group, we consider the background field $\mathcal{A}_1^{\mathcal{MC}}$ corresponding to the diagonal subgroup of the magnetic and charge conjugation symmetries. The relevant 't Hooft anomaly is found by setting $\mathcal{A}_1^{\mathcal{M}} = \mathcal{A}_1^{\mathcal{C}} = \mathcal{A}_1^{\mathcal{MC}}$ in \eqref{anomABJequalranks}, which results in $i \pi (N+k+1) \int_{M_4} \mathcal{A}_2^B \cup \mathcal{A}_1^{\mathcal{MC}} \cup \mathcal{A}_1^{\mathcal{MC}}$. If $N$ and $k$ have the same parity, then $N+k+1$ is odd, and this anomaly is non-trivial. It therefore \textit{forbids} the simultaneous gauging of the $\mathbb{Z}^{[1]}_{2, B}$ and the diagonal combination of the $\mathbb{Z}^{[0]}_{2, \mathcal{M}}$ and $\mathbb{Z}^{[0]}_{2, \mathcal{C}}$ symmetries in the $\SO(2N)_{2k} \times \USp(2N)_{-k}$ theory required to obtain the $[\mathrm{O}(2N)^{-}_{2k} \times \USp(2N)_{-k}]/\mathbb{Z}_2$ theory. Conversely, if $N$ and $k$ have different parities, the anomaly vanishes, and this gauging is permitted. \label{foot:A1MC}} These anomalous variants, therefore, do not correspond to consistent quantum field theories. A key goal is to understand how these 't Hooft anomalies and the resulting inconsistencies are reflected in the superconformal index and in the structure of the quotient by $\Gamma$ (the quaternionic reflection group or its $\mathbb{Z}_2$ extension). The details of the index are provided in Appendix \ref{app:index}, and those of the group $\Gamma$ can be found in Section \ref{quatrefgroupmodspace}.

\subsection*{Index, symmetries and anomalies}

Let us first consider the case of $(N,k) =$ (even, even) or (even, odd).  We observe that the coefficient of the terms $a^{\pm kN} x^{\frac{1}{2}kN}$ of the index of the $[\SO(2N)_{2k} \times \USp(2N)_{-k}]/\BZ_2$ theory, where $a$ is the fugacity associated with the ``axial symmetry'', under which each of the bifundamental $\SO(2N) \times \USp(2N)$ half-hypermultiplets carries charges $1$ and $-1$,\footnote{The global symmetry associated with fugacity $a$ is actually $\SO(3)$ in the $\CN=2$ formalism. Let us denote this by $\SO(3)_a$. This, however, does not commute with the manifest $\CN=3$ $\SO(3)$ $R$-symmetry. Thus, the $\SO(2)$ $R$-symmetry of the $\CN=2$ formalism combines with $\SO(3)_a$ to form the full $\SO(5)$ $R$-symmetry of the $\CN=5$ SCFT. In fact, the index for the $\CN=5$ SCFT always takes the form $1+x+[ \ldots - (a^2+1+a^{-2})] x^2 +\ldots$; see, \eg~, \eqref{indexexamples}. The negative terms in the round bracket at order $x^2$ are indeed the character of the adjoint representation of $\SO(3)$. In terms of the $\CN=3$ formalism, the coefficients $1$ of $x$ and $-1$ of $x^2$ are the contribution of the $\U(1)$ $\CN=3$ flavour current, and the terms $-a^{\pm 2}$ at order $x^2$ are the contributions of the $\CN=3$ extra-supersymmetry currents rendering the supersymmetry of the theory $\CN=5$.\label{foot:Neq5}} always contains the terms
\bes{ \label{D8Q8char1}
D = \frac{1}{2} g (1 + \zeta+ \chi + \zeta \chi)~,
}
where $g$, $\zeta$ and $\chi$ are the fugacities for the zero-form symmetries $\BZ^{[0]}_{2, B}$, $\BZ^{[0]}_{2, \CM}$ and $\BZ^{[0]}_{2, \CC}$, respectively.

On the other hand, in the case of $(N,k)=$ (odd, even) and (odd, odd), the index of the  $[\SO(2N)_{2k} \times \USp(2N)_{-k}]/\BZ_2$ theory contains the terms
\bes{ \label{D8Q8char2}
D' = \frac{1}{2} g \zeta^{\frac{1}{2}} (1 + \zeta+ \chi + \zeta \chi)~.
}
The presence of the half-odd-integral powers of the $\BZ^{[0]}_{2, \CM}$ fugacity $\zeta$ is due to two reasons: (1) the presence of the factor $\zeta^{\sum_{i=1}^N m_i}$ in the index, and (2) the fact that we need to sum $m_i$ over $\BZ+\frac{1}{2}$ due to the $\BZ_2$ quotient, see \eqref{indABJlikemodZ2}. Such half-odd-integral powers of $\zeta$ prevents us from gauging $\BZ^{[0]}_{2, \CM}$ by summing $\zeta$ over $\pm 1$ and dividing the result by two. Therefore, the $[\Spin(2N)_{2k} \times \USp(2N)_{-k}]/\mathbb{Z}_2$ variant is anomalous for odd $N$. This argument was, in fact, extensively used in \cite{Mekareeya:2022spm, Harding:2025vov}. As pointed out in \cite[Page 19]{Harding:2025vov}, the fugacity $d \equiv g \zeta^{\frac{1}{2}}$, with $d^4=1$, corresponds to a $\BZ_4$ subgroup of the $D_8$ zero-form symmetry of the theory. Gauging this $\BZ_4$ zero-form symmetry by summing $d$ over the four fourth roots of unity and dividing by four leads to the index of the $\Spin(2N)_{2k} \times \USp(2N)_{-k}$ variant.

For reference, we report the indices for $[\SO(2N)_4 \times \USp(2N)_{-2}]/\BZ_2$ with $N=2, 3$ below:
\bes{ \label{indexexamples}
\scalebox{0.95}{
\renewcommand{\arraystretch}{1.5} 
\begin{tabular}{c|l}
\hline
$N$ & Index of $[\SO(2N)_4 \times \USp(2N)_{-2}]/\BZ_2$\\
\hline
$2$ & $1 + x + \Big[ (1 + D + \zeta + \chi) [4]_a+ (2 + D + \chi) - [2]_a \Big] x^2 +\ldots$  \\
$3$ & $1 + x + \Big[ 2+ (1 + \zeta) [4]_a -[2]_a\Big] x^2 + \Big[ (D' + \zeta + \chi) [6]_a$ \\
 & \qquad\qquad $+ (1+\zeta) [4]_a +(D'+\chi-\zeta-2)[2]_a +4) \Big] x^3 +\ldots$ \\
\hline
\end{tabular}}
}
where $[m]_a$ denotes the character of the $\SU(2)$ representation with highest weight $m$ written in terms of $a$. The Hilbert series of the Higgs (or Coulomb) branch can be obtained from the limit of the index as follows: it is equal to $\sum_{p \geq 0} C(a^{\pm 2 p} x^p) t^{2p}$, where $C(a^{\pm 2 p} x^p)$ is the coefficient of the term $a^{\pm 2 p} x^p$ in the index. Here are the explicit Hilbert series of the cases of $N=2, 3$:
\bes{ \label{HSexamples}
\scalebox{0.9}{
\begin{tabular}{c|l}
\hline
$N$ & Hilbert series of the HB or CB of $[\SO(2N)_4 \times \USp(2N)_{-2}]/\BZ_2$ \\
\hline
$2$ &  $1 + (1 + D + \zeta + \chi) t^4 + (D + \zeta + \zeta \chi) t^6 + (4 + 3 D + 2 \zeta + 2 \chi + \zeta \chi) t^8 +\ldots$  \\
$3$ & $1 + (1 + \zeta) t^4 + (D' + \zeta + \chi) t^6 + (4 + D' + 2 \zeta + \zeta \chi) t^8+\ldots$ \\
\hline
\end{tabular}}
}
Note that, upon setting $g=\chi=\zeta=1$, we obtain the Hilbert series of $\BH^N/G_N(\hat{D}_2, \BZ_2)$.

The operators associated with $D$ and $D'$ involve the monopole operators
\bes{
V^{(1)} = V_{\left(\frac{1}{2}, \frac{1}{2}, \ldots, \frac{1}{2},  -\frac{1}{2}; \, \frac{1}{2}, \frac{1}{2}, \ldots, \frac{1}{2}, \frac{1}{2} \right)}~, \qquad 
V^{(2)} = V_{\left(\frac{1}{2}, \frac{1}{2}, \ldots, \frac{1}{2},  \frac{1}{2}; \, \frac{1}{2}, \frac{1}{2}, \ldots, \frac{1}{2}, \frac{1}{2} \right)}~,
}
where we use the notation $\left( a_1, a_2, \ldots, a_N;\,  b_1, b_2, \ldots, b_N \right)$ to denote the magnetic fluxes of the gauge group $\SO(2N) \times \USp(2N)$. To form the gauge invariant quantities that contribute to the index, these bare monopole operators have to be appropriately dressed with $A^{kN}$ or $B^{kN}$, where the bifundamental half-hypermultiplets $A$ and $B$ carry fugacities $a^{\pm 1}$. As pointed out below \cite[(2.3)]{Aharony:2013kma}, the monopole operators $W^{(1)}$ and $W^{(2)}$ with fluxes $(1,1,\ldots, 1,-1)$ and  $(1,1,\ldots, 1,1)$ of the $\SO(2N)$ gauge group are exchanged by the charge conjugation symmetry $\BZ_{2,\CC}^{[0]}$. Similarly to the discussion below \cite[(4.5)]{Harding:2025vov}, neither monopole operator ($W^{(1)}$ or $W^{(2)}$)  has a definite charge conjugation parity. Instead, the linear combinations $W_\pm \equiv W^{(1)} \pm W^{(2)}$  are the monopole operators with definite (even/odd) parity under $\BZ_{2,\CC}^{[0]}$. It follows that the fugacity for $W^{(1)}$ and $W^{(2)}$ is $\frac{1}{2}(1+\chi)$. Applying similar logic to the fractional flux monopoles, we see that the fugacity associated with $V^{(1)}$ is $\frac{1}{2} g \zeta^{\frac{N-2}{2}} (1+\chi)$ and that associated with $V^{(2)}$ is $\frac{1}{2} g \zeta^{\frac{N}{2}} (1+\chi)$. Using the fact that $\zeta^2=1$, we see that if $N$ is even, the sum of these two contributions gives \eqref{D8Q8char1}, but if $N$ is odd, this gives \eqref{D8Q8char2}.

Note that $D$ is analogous to the non-invertible operator $\mathsf{D}$ defined in \cite[(8)]{Seifnashri:2024dsd} and \cite[(1.1)]{Seifnashri:2025fgd} for the $(1+1)$d lattice Hamiltonian systems with $\mathrm{Rep}(D_8)$ symmetry. In particular, our $\BZ^{[0]}_{2, \zeta}$,  $\BZ^{[0]}_{2, \chi}$ and $\BZ^{[1]}_{2}$ symmetries play the same roles as $\BZ_2^{\mathsf{e}}$, $\BZ_2^{\mathsf{o}}$, $\BZ_2^{\mathsf{V}}$ in \cite{Seifnashri:2024dsd}. In particular, the former symmetries are involved in the $(2+1)$d Type III anomaly $i \pi \int_{M_4} A_2^B \cup A_1^\CM \cup A_1^\CC$ of the $\SO(2N)_{2k} \times \USp(2N)_{-k}$ theory. An immediate consequence of this is as follows: an attempt to gauge simultaneuously the symmetry associated with $A_2^B$, $A_1^\CM$ and $A_1^\CC$ in this theory leads to $[\Pin(2N)_{2k} \times \USp(2N)_{-k}]/\BZ_2$, which is an inconsistent theory for any parity of $N$ and $k$. Moreover, similarly to the discussion below \cite[(2.10)]{Seifnashri:2025fgd}, both \eqref{D8Q8char1} and \eqref{D8Q8char2} indicate the two-dimensional irreducible representations of $D_8$ or $Q_8$, which have the same character table. Each individual term $1$, $\chi$, $\zeta$ and $\zeta \chi$ indicates the one-dimensional irreducible representations. The factor of $\frac{1}{2}$ indicates that we are not allowed to refine the fugacities $g$, $\zeta$ and $\chi$ simultaneously. In other words, the associated global symmetries do not simultaneously commute with each other (see, for example, below \cite[(B17)]{Seifnashri:2024dsd}). Note, however, that among these fugacity, if we either set $\chi= 1$ or $\zeta = 1$, or $\zeta= \chi= 1$, then the index becomes well-defined. Setting a fugacity to one amounts to turning off the background gauge field for the corresponding global
symmetry. Therefore, this discussion is consistent with the aforementioned $(2+1)$d Type III anomaly of the $\SO(2N)_{2k} \times \USp(2N)_{-k}$ theory. 

In some cases, the index and its Higgs (or Coulomb) branch or limit, namely the Hilbert series, may indicate inconsistency of the theory. Let us consider the case of $k=1$. It is forbidden by \eqref{anomABJequalranks} to obtain the $[\O(2N)_{2}^+ \times \USp(2N)_{-1}]/\BZ_2$ theory by gauging simultaneously the one-from symmetry and the charge conjugation symmetry of the $\SO(2N)_2 \times \USp(2N)_{-1}$ theory.\footnote{Note that $\O(2N)_{2}^+ \times \USp(2N)_{-1}$ is dual to the $\U(N)_{4} \times \U(N)_{-4}$ ABJM theory \cite{Aharony:2008gk}. However, we find that there is no correspondence of the $[\O(2N)_{2} \times \USp(2N)_{-1}]/\BZ_2$ variant in terms of a theory with unitary gauge groups \cite{Beratto:2021xmn}.} For the cases of $N=1$ and $2$, the index and its limit indicate the inconsistency. In the case of $[\O(2)_{2} \times \USp(2)_{-1}]/\BZ_2$, the index after factoring out the free hypermultiplet reads\footnote{Note that $[\SO(2)_{2} \times \USp(2)_{-1}]/\BZ_2$ is actually a theory of two free hypermultiplets, where the fugacities of the four chiral multiplets are given by each term in $(a+a^{-1}) D'$, with $D'$ defined by \eqref{D8Q8char2}. It is clear that gauging $\BZ_{2,\CC}^{[0]}$ by summing $\chi$ over $\pm 1$ and diving the result by two leads to an inconsistent index, with the coefficient at order $x^{\frac{1}{2}}$ being $\frac{1}{2} g \zeta^\frac{1}{2} (1+\zeta)(a+a^{-1})$. This means that each of the Higgs and Coulomb branches contains one free chiral multiplet, violating the fact that each of them is a hyperK\"ahler variety.}
\bes{
1 &+ x + \left(a^3 g+\frac{g}{a^3}+a g+\frac{g}{a}\right) x^{3/2} + \left(a^4+\frac{1}{a^4}-a^2-\frac{1}{a^2}-1\right) x^2 \\
&- \left(a^3 g+\frac{g}{a^3}+a g+\frac{g}{a}\right) x^{5/2} +\ldots~.
}
The Higgs (or Coulomb) branch limit takes the following closed form:
\bes{
\frac{1-t+t^2}{(1-t) \left(1+t^2\right)} = 1 + t^3 + t^4 + t^7  +\ldots = \PE[t^3+t^4-t^6]~.
}
Observe that the order of the pole at $t=1$ is one, indicating that the corresponding Higgs (or Coulomb) branch is one complex dimensional. However, this violates the fact that the Higgs or Coulomb branch of a 3d $\CN \geq 4$ SCFT must be a hyperK\"ahler variety, whose complex dimension must be even. In the case of $[\O(4)_{2} \times \USp(4)_{-1}]/\BZ_2$, the totally unrefined index reads $1 + 7 x + 38 x^2 + 117 x^3+\ldots$; the coefficient of $x$ is incompatible with $\CN=5$, $\CN=6$ and $\CN=8$ supersymmetry, in which case it must be $1$, $4$ and $10$, respectively \cite[Section 4.3]{Evtikhiev:2017heo}. However, the Hilbert series in this case seems to be consistent
\bes{
\frac{1 - 2 t^2 + 2 t^4 + 2 t^8 - 2 t^{10} + t^{12}}{(1 - t)^4 (1 + t)^4 (1 + t^2)^2 (1 + t^4)} = 1 + 2 t^4 + 2 t^6 + 8 t^8 + 8 t^{10} + \ldots~,
}
since it has a correct order of the pole at $t=1$ and its numerator is palindromic. For $N \geq 3$, we cannot detect the inconsistency from the series expansion or from the limit of the index.\footnote{This should be contrasted with the unitary case. For $[U(N)_{k} \times U(N)_{-k}]/\BZ_p$, where $p$ is not a divisor of $k$, which is an inconsistent theory \cite[Section 3.3]{Tachikawa:2019dvq}, the quotient $\BZ_p$ simply drops out and so the index turns out to be equal to that of $\U(N)_{k} \times \U(N)_{-k}$.  Similarly, we also found that the index for the $[\U(N+1)_k \times \U(N)_{-k}]/\BZ_k$ theory, which is also inconsistent according to \cite[(3.19)]{Tachikawa:2019dvq}, is equal to that of $\U(N+1)_k \times \U(N)_{-k}$.}

\subsection*{Anomalies and moduli spaces}

It is also intriguing that gaugings and anomalous variants can also be detected at the level of the moduli space for this class of theories. This will be discussed in further details in Sections \ref{sec:concreteexamples} and \ref{sec:anomvar}. Let us roughly describe the idea here. Suppose that, upon gauging a non-anomalous $\BZ^{[0]}_{2, S}$ where $S \in \{B, \CC, \CM, \CM\CC \}$, a non-anomalous variant $\CT$ of the $\so(2N)_{2k} \times \usp(2N)_{-k}$ theory with moduli space $\BH^{2N}/\Gamma$ becomes a non-anomalous variant $\CT'$ with moduli space $\BH^{2N}/\Gamma'$. Here $\Gamma$ and $\Gamma'$ are quaternionic reflection groups or $\BZ_2$ extensions thereof. In this case, $\Gamma'$ can be obtained from $\Gamma$ simply by adding another matrix $R_S$, listed in \eqref{listofRS_orig}, associated with $\BZ^{[0]}_{2, S}$  to the set $\mathrm{gen}(\Gamma)$ of the generators of $\Gamma$, so that $\Gamma' = \langle \mathrm{gen}(\Gamma), R_S \rangle$.  In other words, the new group $\Gamma'$ is generated by the generators of $\Gamma$ along with $R_S$. In this case, we find that $|\Gamma'| = \langle \mathrm{gen}(\Gamma), R_S \rangle = 2|\Gamma|$. On the contrary, if $\CT$ is non-anomalous and $\CT'$ is anomalous, we observe that $|\Gamma'| = \langle \mathrm{gen}(\Gamma), R_S \rangle = 4|\Gamma|$, and that the moduli space $\BH^{2N}/ \langle \mathrm{gen}(\Gamma), R_S \rangle$, instead, corresponds to a different non-anomalous variant.

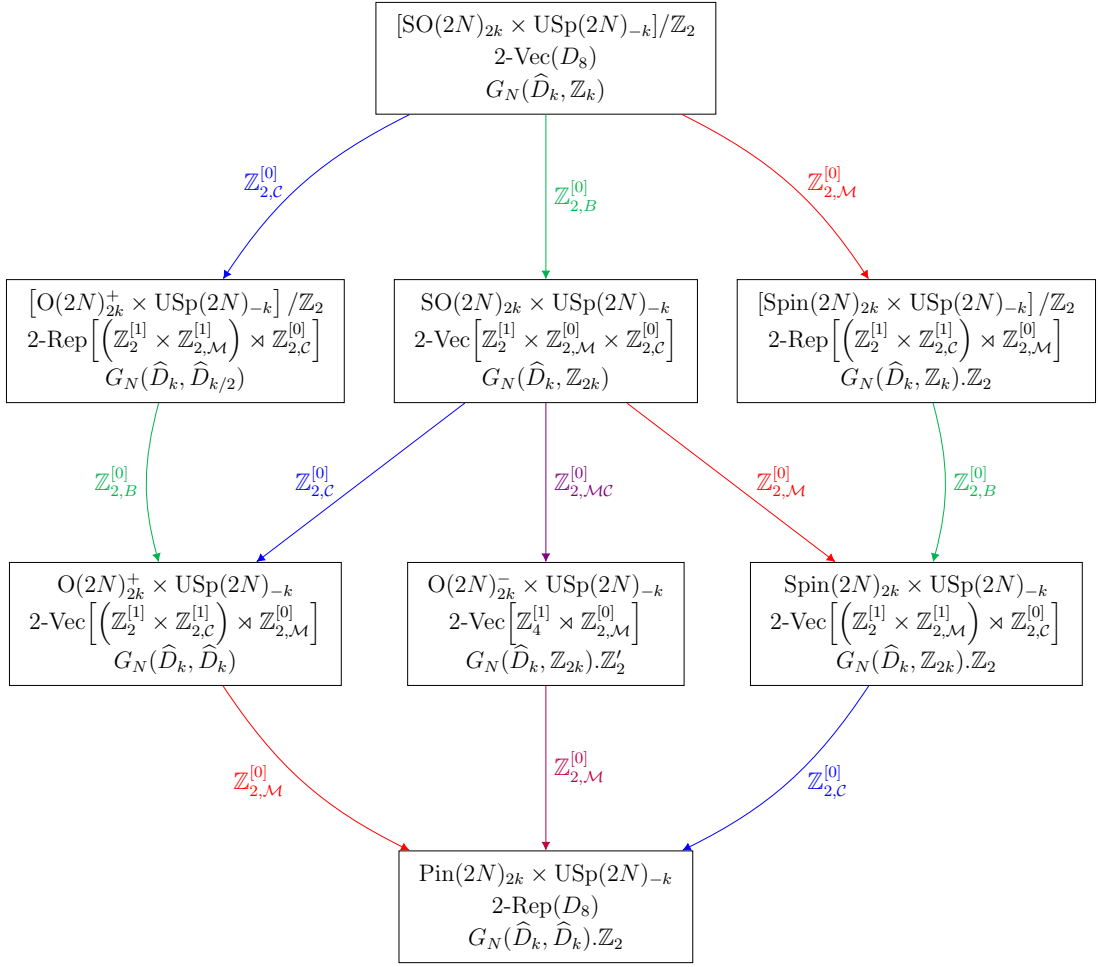
\begin{figure}
\centering
\scalebox{0.75}{
\begin{tikzpicture} 
			\node[draw] (Pin4) at (0,-7.5) {\begin{tabular}{c}
			$\Pin(2N)_{2k} \times \USp(2N)_{-k}$ \\ 2-Rep$(D_8)$ \\ $G_N(\hat{D}_k, \hat{D}_k).\BZ_2$ \end{tabular} }; 
                \node[draw] (O4m) at (0,-2.5) {\begin{tabular}{c}
			$\O(2N)^-_{2k} \times \USp(2N)_{-k}$ \\ 2-Vec$\left[\BZ_{4}^{[1]} \rtimes \BZ_{2, \CM}^{[0]}\right]$ \\ $G_N(\hat{D}_k, \BZ_{2k}) . \BZ_2'$ \end{tabular} };
                \node[draw] (O4p) at (-6.5,-2.5) {\begin{tabular}{c}
			$\O(2N)^+_{2k} \times \USp(2N)_{-k}$ \\ 2-Vec$\left[\left(\BZ_{2}^{[1]} \times \BZ_{2, \cC}^{[1]}\right) \rtimes \BZ_{2, \CM}^{[0]}\right]$ \\ $G_N(\hat{D}_k, \hat{D}_k)$ \end{tabular} };
                \node[draw] (Spin4) at (6.5,-2.5) {\begin{tabular}{c}
			$\Spin(2N)_{2k} \times \USp(2N)_{-k}$ \\ 2-Vec$\left[\left(\BZ_{2}^{[1]} \times \BZ_{2, \CM}^{[1]}\right) \rtimes \BZ_{2, \cC}^{[0]}\right]$ \\ $G_N(\hat{D}_k, \BZ_{2k}).\BZ_2$ \end{tabular} };
                \node[draw] (SO4) at (0,2.5) {\begin{tabular}{c}
			$\SO(2N)_{2k} \times \USp(2N)_{-k}$ \\ 2-Vec$\left[\BZ_{2}^{[1]} \times \BZ_{2, \CM}^{[0]} \times \BZ_{2, \cC}^{[0]}\right]$ \\ $G_N(\hat{D}_k, \BZ_{2k})$ \end{tabular} };
                \node[draw] (O4pmodZ2) at (-6.5,2.5) {\begin{tabular}{c}
			$\left[\O(2N)^+_{2k} \times \USp(2N)_{-k}\right]/\BZ_2$ \\ 2-Rep$\left[\left(\BZ_{2}^{[1]} \times \BZ_{2, \CM}^{[1]}\right) \rtimes \BZ_{2, \cC}^{[0]}\right]$ \\ $G_N(\hat{D}_k, \hat{D}_{k/2})$ \end{tabular} };
                \node[draw] (Spin4modZ2) at (6.5,2.5) {\begin{tabular}{c}
			$\left[\Spin(2N)_{2k} \times \USp(2N)_{-k} \right]/\BZ_2$ \\ 2-Rep$\left[\left(\BZ_{2}^{[1]} \times \BZ_{2, \cC}^{[1]}\right) \rtimes \BZ_{2, \CM}^{[0]}\right]$ \\ $G_N(\hat{D}_k, \BZ_k).\BZ_2$ \end{tabular} };
                \node[draw] (SO4modZ2) at (0,7.5) {\begin{tabular}{c}
			$[\SO(2N)_{2k} \times \USp(2N)_{-k}]/\BZ_2$ \\ 2-Vec$(D_8)$ \\ $G_N(\hat{D}_k, \BZ_k)$ \end{tabular} }; 
            \draw[->,blue] (SO4modZ2) to [bend right=15] node[midway, left=0.2] {\blue $\BZ_{2,\cC}^{[0]}$} (O4pmodZ2);
            \draw[->,new-green] (SO4modZ2) to node[midway,right] {\green $\BZ_{2,B}^{[0]}$} (SO4);
            \draw[->,red] (SO4modZ2) to [bend left=15] node[midway, right=0.2] {\red $\BZ_{2,\CM}^{[0]}$} (Spin4modZ2);
            \draw[->,new-green] (O4pmodZ2) to [bend right=15] node[midway, left] {\green $\BZ_{2,B}^{[0]}$} (O4p);
            \draw[->,violet] (SO4) to node[midway,right] {\violet $\BZ_{2,\CM \cC}^{[0]}$} (O4m);
            \draw[->,new-green] (Spin4modZ2) to [bend left=15] node[midway, right] {\green $\BZ_{2,B}^{[0]}$} (Spin4);
            \draw[->,blue] (SO4) to node[midway,left=0.3] {\blue $\BZ_{2,\cC}^{[0]}$} (O4p);
            \draw[->,red] (SO4) to node[midway,right=0.3] {\red $\BZ_{2,\CM}^{[0]}$} (Spin4);
            \draw[->,red] (O4p) to [bend right=15] node[midway, left=0.2] {\red $\BZ_{2,\CM}^{[0]}$} (Pin4);
            \draw[->,purple] (O4m) to node[midway,right] {\purple $\BZ_{2,\CM}^{[0]}$} (Pin4);
            \draw[->,blue] (Spin4) to [bend left=15] node[midway, right=0.2] {\blue $\BZ_{2,\cC}^{[0]}$} (Pin4);
\end{tikzpicture}
}
    \caption[D8Nevenkeven]{The $D_8$ symmetry web for variants of the $\so(2N)_{2k} \times \usp(2N)_{-k}$ ABJ theory with {\bf $N$ even and $k$ even}. Each arrow labelled by $\BZ^{[0]}_{2,x}$ connecting two boxes denotes the gauging of the zero-form symmetry $\BZ^{[0]}_{2,x}$. In each box, which is associated with a specific global form of the theory, we report the corresponding symmetry category and the quaternionic reflection group or its extension $\Gamma$ such that the moduli space is $\BH^{2N}/\Gamma$. The details of $\Gamma$ will be discussed in Section \ref{quatrefgroupmodspace}. Note that the variant $[\O(2N)^-_{2k} \times \USp(2N)_{-k}]/\BZ_2$ is anomalous and not depicted here. We also emphasise that there are two distinct variants of the $\BZ_2$ extension of the group $G_N(\hat{D}_{k}, \BZ_{2k})$ that are indicated by $\BZ_2$ and $\BZ_2'$. Moreover, in the special case of $N=2$, the group $\Gamma$ for $[\Spin(4)_{2k} \times \USp(4)_{-k}]/\BZ_2$, $\Spin(4)_{2k} \times \USp(4)_{-k}$, and $\Pin(4)_{2k} \times \USp(4)_{-k}$ turns out to be the quaternionic reflection groups $G_2(\hat{D}_{2k},\BZ_k)$, $G_2(\hat{D}_{2k},\BZ_{2k})$, and $G_2(\hat{D}_{2k}, \hat{D}_k)$ respectively; see \eqref{eq:specialcaseN2}.} \label{fig:D8Nevenkeven}
\end{figure}

\subsection*{Remarks on other gaugings}

As a final remark, one might ask what happens if we try to gauge the diagonal subgroup of the $\BZ^{[0]}_{2,B}$ and $\BZ^{[0]}_{2,\CM}$ symmetries and the diagonal subgroup of the $\BZ^{[0]}_{2,B}$ and $\BZ^{[0]}_{2,\CC}$ symmetries, which we denote by $\BZ^{[0]}_{2,B \CM}$ and $\BZ^{[0]}_{2,B \CC}$, respectively.

First, let us suppose that $\BZ^{[0]}_{2,\CM}$ is anomalous in the $\left[\SO(2N)_{2 k} \times \USp(2N)_{-k}\right]/\BZ_2$ theory, then there are some operators which are ill-quantised under $\BZ^{[0]}_{2,\CM}$, as signalled by the presence of half-odd integer powers of the fugacity $\zeta$, which has been discussed around \eref{D8Q8char2}. Then, such operators are wrongly quantised also under $\BZ^{[0]}_{2,B \CM}$, thus forbidding its gauging. This means that also $\BZ^{[0]}_{2,B \CM}$ is anomalous. Analogously, if $\BZ^{[0]}_{2,\CC}$ is anomalous in the $\left[\SO(2N)_{2 k} \times \USp(2N)_{-k}\right]/\BZ_2$ theory, then also $\BZ^{[0]}_{2,B \CC}$ is automatically anomalous.\footnote{On the other hand, recall that, when either $\BZ^{[0]}_{2,\CM}$ or $\BZ^{[0]}_{2,\CC}$ is anomalous, their diagonal subgroup $\BZ^{[0]}_{2,\CM \CC}$ is non-anomalous and can be gauged. This follows from the presence of the Type III anomaly term $\int_{M_4} \CA^B_2 \cup \CA^{\CM}_1 \cup \CA^{\CC}_1$ in the anomaly theory \eref{anomSO2LUSp2M}, as discussed in Footnote \ref{foot:A1MC}. Since there is no analogous term in the anomaly theory involving the background gauge field for $\BZ^{[0]}_{2,B}$, it follows that its diagonal subgroup with an anomalous symmetry also results in an anomalous symmetry.}

Next, let us consider the scenario in which both $\BZ^{[0]}_{2,\CM}$ and $\BZ^{[0]}_{2,\CC}$ are non-anomalous in the $\left[\SO(2N)_{2 k} \times \USp(2N)_{-k}\right]/\BZ_2$ theory. This happens when both $N$ and $k$ are even, for which sequential gauging of discrete symmetries gives rise to the $D_8$ symmetry web depicted in Figure \ref{fig:D8Nevenkeven}. In such a case, also $\BZ^{[0]}_{2,B \CM}$ and $\BZ^{[0]}_{2,B \CC}$ are non-anomalous, hence they are valid symmetries of the theory and can be gauged. Despite that, their gauging is not explicitly shown in Figure \ref{fig:D8Nevenkeven}. The point is that the theory which is reached after gauging $\BZ^{[0]}_{2,B \CM}$ ({\it resp.} $\BZ^{[0]}_{2,B \CC}$) is equivalent to the theory arising from gauging $\BZ^{[0]}_{2,\CM}$ ({\it resp.} $\BZ^{[0]}_{2,\CC}$).\footnote{This can be checked explicitly using the index, where gauging $\BZ^{[0]}_{2,B \CM}$ ({\it resp.} $\BZ^{[0]}_{2,B \CC}$) can be implemented by summing over the contributions coming from the $\left(g, \zeta = 1\right)$ and $\left(-g, \zeta = -1\right)$ sectors ({\it resp.} the $\left(g, \chi = 1\right)$ and $\left(-g, \chi = -1\right)$ sectors), and subsequently dividing by two. For instance, upon gauging $\BZ^{[0]}_{2,\CM}$ or $\BZ^{[0]}_{2,B \CM}$, the operator $D$ in \eref{D8Q8char1} becomes $\frac{1}{2} g \left(1+\chi\right)$. Upon gauging $\BZ^{[0]}_{2,\CC}$ or $\BZ^{[0]}_{2,B \CC}$, it reads $\frac{1}{2} g \left(1+\zeta\right)$ instead.} This statement admits a group theoretic explanation based on the properties of the $D_8$ group. Observe that the $D_8$ symmetry web of Figure \ref{fig:D8Nevenkeven} reproduces the lattice of subgroups of $D_8$ depicted in Figure \ref{fig:D8lattice}, where the theories arising from the $\BZ^{[0]}_{2,S}$ gauging, with $S = \{B, \CM, \CC\}$, in the former figure are associated with the boxes corresponding to the order two $D_8$ subgroups generated by $r^2$, $r s$ and $s$, respectively, in the latter figure. Under this identification, gauging the $\BZ^{[0]}_{2,B \CM}$ and $\BZ^{[0]}_{2,B \CC}$ symmetries in the $\left[\SO(2N)_{2 k} \times \USp(2N)_{-k}\right]/\BZ_2$ theory translates into reaching the order two $D_8$ subgroups generated by $r^3 s$ and $r^2 s$, respectively. The equivalence between the theories arising from the $\BZ^{[0]}_{2,\CM}$ and $\BZ^{[0]}_{2,B \CM}$ ({\it resp.} $\BZ^{[0]}_{2,\CC}$ and $\BZ^{[0]}_{2,B \CC}$) gauging follows from the fact that the $D_8$ subgroups generated by $r s$ and $r^3 s$ ({\it resp.} $s$ and $r^2 s$) belong to the same conjugacy class, as pointed out in the caption of Figure \ref{fig:D8lattice}.
\begin{figure}
\centering
\scalebox{0.7}{
\begin{tikzpicture} 
			\node[draw] (Pin4) at (0,-7.5) {\begin{tabular}{c}
			$\Pin(2N)_{2k} \times \USp(2N)_{-k}$ \\ 2-Rep$(D_8)$ \\ $G_N(\hat{D}_{k}, \hat{D}_{k}).\BZ_2$ \end{tabular} }; 
                \node[draw] (O4m) at (0,-2.5) {\begin{tabular}{c} $\O(2N)^+_{2k} \times \USp(2N)_{-k}$ \\ 2-Vec$\left[\BZ_{4}^{[1]} \rtimes \BZ_{2, \CM}^{[0]}\right]$ \\ $G_N(\hat{D}_k, \hat{D}_k)$ \end{tabular} };
                \node[draw] (O4p) at (-6.5,-2.5) {\begin{tabular}{c} $\O(2N)^-_{2k} \times \USp(2N)_{-k}$ \\ 2-Vec$\left[\left(\BZ_{2}^{[1]} \times \BZ_{2, \CM\cC}^{[1]}\right) \rtimes \BZ_{2, \CM}^{[0]}\right]$ \\ $G_N(\hat{D}_k, \BZ_{2k}) . \BZ_2'$ \end{tabular} };
                \node[draw] (Spin4) at (6.5,-2.5) {\begin{tabular}{c}
			$\Spin(2N)_{2k} \times \USp(2N)_{-k}$ \\ 2-Vec$\left[\left(\BZ_{2}^{[1]} \times \BZ_{2, \CM}^{[1]}\right) \rtimes \BZ_{2, \cC}^{[0]}\right]$ \\ $G_N(\hat{D}_{k}, \BZ_{2k}).\BZ_2$  \end{tabular} };
                \node[draw] (SO4) at (0,2.5) {\begin{tabular}{c}
			$\SO(2N)_{2k} \times \USp(2N)_{-k}$ \\ 2-Vec$\left[\BZ_{2}^{[1]} \times \BZ_{2, \CM}^{[0]} \times \BZ_{2, \cC}^{[0]}\right]$ \\ $G_N(\hat{D}_k, \BZ_{2k})$ \end{tabular} };
                \node[draw] (O4pmodZ2) at (-6.5,2.5) {\begin{tabular}{c}
			$\left[\O(2N)^-_{2k} \times \USp(2N)_{-k}\right]/\BZ_2$ \\ 2-Rep$\left[\left(\BZ_{2}^{[1]} \times \BZ_{2, \CM}^{[1]}\right) \rtimes \BZ_{2, \CM\CC}^{[0]}\right]$ \\ $G_N(\hat{D}_k, \BZ_{k}).\BZ_2'$ \end{tabular} };
                \node[draw] (Spin4modZ2) at (6.5,2.5) {\begin{tabular}{c}
			$\left[\Spin(2N)_{2k} \times \USp(2N)_{-k} \right]/\BZ_2$ \\ 2-Rep$\left[\left(\BZ_{2}^{[1]} \times \BZ_{2, \cC}^{[1]}\right) \rtimes \BZ_{2, \CM}^{[0]}\right]$ \\ $G_N(\hat{D}_{k}, \BZ_k).\BZ_2$ \end{tabular} };
                \node[draw] (SO4modZ2) at (0,7.5) {\begin{tabular}{c}
			$[\SO(2N)_{2k} \times \USp(2N)_{-k}]/\BZ_2$ \\ 2-Vec$(D_8)$ \\ $G_N(\hat{D}_k, \BZ_k)$ \end{tabular} }; 
            \draw[->,violet] (SO4modZ2) to [bend right=15] node[midway, left=0.2] {\violet $\BZ_{2,\CM \cC}^{[0]}$} (O4pmodZ2);
            \draw[->,new-green] (SO4modZ2) to node[midway,right] {\green $\BZ_{2,B}^{[0]}$} (SO4);
            \draw[->,red] (SO4modZ2) to [bend left=15] node[midway, right=0.2] {\red $\BZ_{2,\CM}^{[0]}$} (Spin4modZ2);
            \draw[->,new-green] (O4pmodZ2) to [bend right=15] node[midway, left] {\green $\BZ_{2,B}^{[0]}$} (O4p);
            \draw[->,blue] (SO4) to node[midway,right] {\blue $\BZ_{2, \cC}^{[0]}$} (O4m);
            \draw[->,new-green] (Spin4modZ2) to [bend left=15] node[midway, right] {\green $\BZ_{2,B}^{[0]}$} (Spin4);
            \draw[->,violet] (SO4) to node[midway,left=0.3] {\violet $\BZ_{2,\CM \cC}^{[0]}$} (O4p);
            \draw[->,red] (SO4) to node[midway,right=0.3] {\red $\BZ_{2,\CM}^{[0]}$} (Spin4);
            \draw[->,purple] (O4p) to [bend right=15] node[midway, left=0.2] {\purple $\BZ_{2,\CM}^{[0]}$} (Pin4);
            \draw[->,red] (O4m) to node[midway,right] {\red $\BZ_{2,\CM}^{[0]}$} (Pin4);
            \draw[->,blue] (Spin4) to [bend left=15] node[midway, right=0.2] {\blue $\BZ_{2,\cC}^{[0]}$} (Pin4);
\end{tikzpicture}
}
    \caption[D8Nevenkodd]{The $D_8$ symmetry web for variants of the $\so(2N)_{2k} \times \usp(2N)_{-k}$ ABJ theory with {\bf $N$ even and $k$ odd}. This diagram can be obtained from Figure \ref{fig:D8Nevenkeven} by exchanging $\CC$ and $\CM \CC$ in the left part of the diagram. In each box, which is associated with a specific global form of the theory, we report the corresponding symmetry category and the quaternionic relection group or its extension $\Gamma$ such that the moduli space is $\BH^{2N}/\Gamma$. The details of $\Gamma$ will be discussed in Section \ref{quatrefgroupmodspace}. Note that the variant $[\O(2N)^+_{2k} \times \USp(2N)_{-k}]/\BZ_2$ is anomalous and not depicted here. We emphasise that there are two distinct variants of the $\BZ_2$ extension of the group $G_N(\hat{D}_{k}, \BZ_{2k})$ that are indicated by $\BZ_2$ and $\BZ_2'$. Moreover, in the special case of $N=2$, the group $\Gamma$ for $[\Spin(4)_{2k} \times \USp(4)_{-k}]/\BZ_2$, $\Spin(4)_{2k} \times \USp(4)_{-k}$, and $\Pin(4)_{2k} \times \USp(4)_{-k}$ turns out to be the quaternionic reflection groups $G_2(\hat{D}_{2k},\BZ_k)$, $G_2(\hat{D}_{2k},\BZ_{2k})$, and $G_2(\hat{D}_{2k}, \hat{D}_k)$ respectively; see \eqref{eq:specialcaseN2}.} \label{fig:D8Nevenkodd}
\end{figure}
\newpage

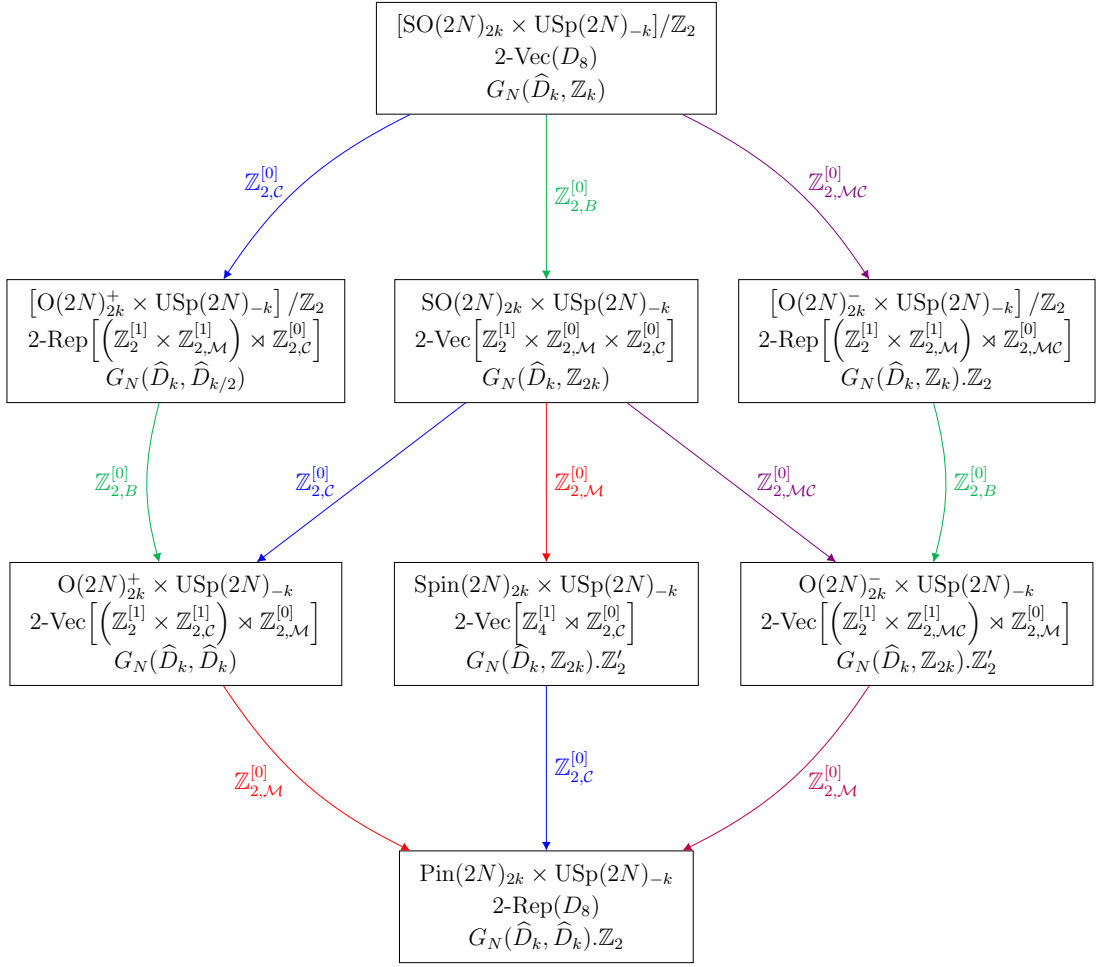
\begin{figure}
\centering
\scalebox{0.75}{
\begin{tikzpicture} 
			\node[draw] (Pin4) at (0,-7.5) {\begin{tabular}{c}
			$\Pin(2N)_{2k} \times \USp(2N)_{-k}$ \\ 2-Rep$(D_8)$ \\ $G_N(\hat{D}_{k}, \hat{D}_{k}).\BZ_2$ \end{tabular} }; 
                \node[draw] (O4m) at (0,-2.5) {\begin{tabular}{c}
			$\Spin(2N)_{2k} \times \USp(2N)_{-k}$ \\ 2-Vec$\left[\BZ_{4}^{[1]} \rtimes \BZ_{2, \CC}^{[0]}\right]$ \\ $G_N(\hat{D}_{k}, \BZ_{2k}).\BZ_2$ \end{tabular} };
                \node[draw] (O4p) at (-6.5,-2.5) {\begin{tabular}{c}
			$\O(2N)^+_{2k} \times \USp(2N)_{-k}$ \\ 2-Vec$\left[\left(\BZ_{2}^{[1]} \times \BZ_{2, \cC}^{[1]}\right) \rtimes \BZ_{2, \CM}^{[0]}\right]$ \\ $G_N(\hat{D}_k, \hat{D}_k)$ \end{tabular} };
                \node[draw] (Spin4) at (6.5,-2.5) {\begin{tabular}{c}
			$\O(2N)^-_{2k} \times \USp(2N)_{-k}$ \\ 2-Vec$\left[\left(\BZ_{2}^{[1]} \times \BZ_{2, \CM \cC}^{[1]}\right) \rtimes \BZ_{2, \CM}^{[0]}\right]$ \\ $G_N(\hat{D}_k, \BZ_{2k}) . \BZ_2'$ \end{tabular} };
                \node[draw] (SO4) at (0,2.5) {\begin{tabular}{c}
			$\SO(2N)_{2k} \times \USp(2N)_{-k}$ \\ 2-Vec$\left[\BZ_{2}^{[1]} \times \BZ_{2, \CM}^{[0]} \times \BZ_{2, \cC}^{[0]}\right]$ \\ $G_N(\hat{D}_k, \BZ_{2k})$ \end{tabular} };
                \node[draw] (O4pmodZ2) at (-6.5,2.5) {\begin{tabular}{c}
			$\left[\O(2N)^+_{2k} \times \USp(2N)_{-k}\right]/\BZ_2$ \\ 2-Rep$\left[\left(\BZ_{2}^{[1]} \times \BZ_{2, \CM}^{[1]}\right) \rtimes \BZ_{2, \cC}^{[0]}\right]$ \\ $G_N(\hat{D}_k, \hat{D}_{k/2})$ \end{tabular} };
                \node[draw] (Spin4modZ2) at (6.5,2.5) {\begin{tabular}{c}
			$\left[\O(2N)^-_{2k} \times \USp(2N)_{-k} \right]/\BZ_2$ \\ 2-Rep$\left[\left(\BZ_{2}^{[1]} \times \BZ_{2, \CM}^{[1]}\right) \rtimes \BZ_{2, \CM\CC}^{[0]}\right]$ \\ $G_N(\hat{D}_{k}, \BZ_k).\BZ_2'$ \end{tabular} };
                \node[draw] (SO4modZ2) at (0,7.5) {\begin{tabular}{c}
			$[\SO(2N)_{2k} \times \USp(2N)_{-k}]/\BZ_2$ \\ 2-Vec$(D_8)$ \\ $G_N(\hat{D}_k, \BZ_k)$ \end{tabular} }; 
            \draw[->,blue] (SO4modZ2) to [bend right=15] node[midway, left=0.2] {\blue $\BZ_{2,\cC}^{[0]}$} (O4pmodZ2);
            \draw[->,new-green] (SO4modZ2) to node[midway,right] {\green $\BZ_{2,B}^{[0]}$} (SO4);
            \draw[->,violet] (SO4modZ2) to [bend left=15] node[midway, right=0.2] {\violet $\BZ_{2,\CM\CC}^{[0]}$} (Spin4modZ2);
            \draw[->,new-green] (O4pmodZ2) to [bend right=15] node[midway, left] {\green $\BZ_{2,B}^{[0]}$} (O4p);
            \draw[->,red] (SO4) to node[midway,right] {\red $\BZ_{2,\CM}^{[0]}$} (O4m);
            \draw[->,new-green] (Spin4modZ2) to [bend left=15] node[midway, right] {\green $\BZ_{2,B}^{[0]}$} (Spin4);
            \draw[->,blue] (SO4) to node[midway,left=0.3] {\blue $\BZ_{2,\cC}^{[0]}$} (O4p);
            \draw[->,violet] (SO4) to node[midway,right=0.3] {\violet $\BZ_{2,\CM\CC}^{[0]}$} (Spin4);
            \draw[->,red] (O4p) to [bend right=15] node[midway, left=0.2] {\red $\BZ_{2,\CM}^{[0]}$} (Pin4);
            \draw[->,blue] (O4m) to node[midway,right] {\blue $\BZ_{2,\CC}^{[0]}$} (Pin4);
            \draw[->,purple] (Spin4) to [bend left=15] node[midway, right=0.2] {\purple $\BZ_{2,\CM}^{[0]}$} (Pin4);
\end{tikzpicture}
}
    \caption[D8Noddkeven]{The $D_8$ symmetry web for variants of the $\so(2N)_{2k} \times \usp(2N)_{-k}$ ABJ theory with {\bf $N$ odd and $k$ even}. This diagram can be obtained from Figure \ref{fig:D8Nevenkeven} by exchanging $\CM$ and $\CM \CC$ in the right part of the diagram. In each box, which is associated with a specific global form of the theory, we report the corresponding symmetry category and the quaternionic reflection group or its extension $\Gamma$ such that the moduli space is $\BH^{2N}/\Gamma$. The details of $\Gamma$ will be discussed in Section \ref{quatrefgroupmodspace}. Note that the variant $[\Spin(2N)_{2k} \times \USp(2N)_{-k}]/\BZ_2$ is anomalous and not depicted here. We emphasise that there are two distinct variants of the $\BZ_2$ extension of the group $G_N(\hat{D}_{k}, \BZ_{2k})$ that are indicated by $\BZ_2$ and $\BZ_2'$; for $N=2$ the latter is not a quaternionic reflection group and is explained around \eqref{gen:G2DkZ2k.Z2version2}, whereas the former, associated with $\Spin(4)_{2k} \times \USp(4)_{-k}$, is isomorphic to $G_2(\hat{D}_{2k}, \BZ_{2k})$. Moreover, for $\Pin(4)_{2k} \times \USp(4)_{-k}$, the corresponding group turns out to be the quaternionic reflection group $G_2(\hat{D}_{2k}, \hat{D}_k)$; see \eqref{eq:specialcaseN2}.} \label{fig:D8Noddkeven}
\end{figure}

\newpage
\begin{figure}
\centering
\scalebox{0.75}{
\begin{tikzpicture} 
			\node[draw] (Pin4) at (0,-7.5) {\begin{tabular}{c}
			$\Pin(2N)_{2k} \times \USp(2N)_{-k}$ \\ 2-Rep$(Q_8)$ \\ $G_N(\hat{D}_k, \hat{D}_k).\BZ_2$ \end{tabular} }; 
                \node[draw] (O4m) at (0,-2.5) {\begin{tabular}{c}
			$\O(2N)^-_{2k} \times \USp(2N)_{-k}$ \\ 2-Vec$\left[\BZ_{4}^{[1]} \rtimes \BZ_{2, \CM}^{[0]}\right]$ \\ $G_N(\hat{D}_k,\BZ_{2k}).\BZ_2'$ \end{tabular} };
                \node[draw] (O4p) at (-6.5,-2.5) {\begin{tabular}{c}
			$\O(2N)^+_{2k} \times \USp(2N)_{-k}$ \\ 2-Vec$\left[\BZ_{4}^{[1]} \rtimes \BZ_{2, \CM}^{[0]}\right]$ \\ $G_N(\hat{D}_k, \hat{D}_k)$ \end{tabular} };
                \node[draw] (Spin4) at (6.5,-2.5) {\begin{tabular}{c}
			$\Spin(2N)_{2k} \times \USp(2N)_{-k}$ \\ 2-Vec$\left[\BZ_{4}^{[1]} \rtimes \BZ_{2, \cC}^{[0]}\right]$ \\ $G_N(\hat{D}_k, \BZ_{2k}).\BZ_2$ \end{tabular} };
                \node[draw] (SO4) at (0,2.5) {\begin{tabular}{c}
			$\SO(2N)_{2k} \times \USp(2N)_{-k}$ \\ 2-Vec$\left[\BZ_{2}^{[1]} \times \BZ_{2, \CM}^{[0]} \times \BZ_{2, \cC}^{[0]}\right]$ \\ $G_N(\hat{D}_k, \BZ_{2k})$ \end{tabular} };
                \node[draw] (SO4modZ2) at (0,7.5) {\begin{tabular}{c}
			$[\SO(2N)_{2k} \times \USp(2N)_{-k}]/\BZ_2$ \\ 2-Vec$(Q_8)$ \\ $G_N(\hat{D}_k, \BZ_k)$ \end{tabular} }; 
            \draw[->,blue-green] (SO4modZ2) to [bend right=15] node[midway, left=0.2] {\bluegreen $\BZ_{4,B + \cC}^{[0]}$} (O4p);
            \draw[->,new-green] (SO4modZ2) to node[midway,right] {\green $\BZ_{2,B}^{[0]}$} (SO4);
            \draw[->,green-red] (SO4modZ2) to [bend left=15] node[midway, right=0.2] {\greenred $\BZ_{4,B + \CM}^{[0]}$} (Spin4);
            \draw[->,violet] (SO4) to node[midway,right] {\violet $\BZ_{2,\CM \cC}^{[0]}$} (O4m);
            \draw[->,blue] (SO4) to node[midway,left=0.3] {\blue $\BZ_{2,\cC}^{[0]}$} (O4p);
            \draw[->,red] (SO4) to node[midway,right=0.3] {\red $\BZ_{2,\CM}^{[0]}$} (Spin4);
            \draw[->,red] (O4p) to [bend right=15] node[midway, left=0.2] {\red $\BZ_{2,\CM}^{[0]}$} (Pin4);
            \draw[->,purple] (O4m) to node[midway,right] {\purple $\BZ_{2,\CM}^{[0]}$} (Pin4);
            \draw[->,blue] (Spin4) to [bend left=15] node[midway, right=0.2] {\blue $\BZ_{2,\cC}^{[0]}$} (Pin4);
\end{tikzpicture}
}
    \caption[Q8Noddkodd]{The $Q_8$ symmetry web for variants of the $\so(2N)_{2k} \times \usp(2N)_{-k}$ ABJ theory with {\bf $N$ odd and $k$ odd}. Note that the variants $[\O(2N)^\pm_{2k} \times \USp(2N)_{-k}]/\BZ_2$ and $[\Spin(2N)_{2k} \times \USp(2N)_{-k}]/\BZ_2$ are anomalous and not depicted here. We also emphasise that there are two distinct variants of the $\BZ_2$ extension of the group $G_N(\hat{D}_{k}, \BZ_{2k})$ that are indicated by $\BZ_2$ and $\BZ_2'$; these will be discussed in detail in Section \ref{quatrefgroupmodspace}. As before, in the special case of $N=2$, the group associated with $\Spin(4)_{2k} \times \USp(4)_{-k}$ is isomorphic to $G_2(\hat{D}_{2k}, \BZ_{2k})$, and for $\Pin(4)_{2k} \times \USp(4)_{-k}$, the group is isomorpic to $G_2(\hat{D}_{2k}, \hat{D}_k)$ respectively; see \eqref{eq:specialcaseN2}.} \label{fig:Q8Noddkodd}
\end{figure}

\section{Symplectic reflection groups and their generators}\label{quatrefgroupmodspace}
The moduli space of variants of the $\so(2N)_{2k} \times \usp(2N)_{-k}$ ABJ theory is pointed out to be $\BH^{2N}/\Gamma$, where $\Gamma$ is a quaternionic reflection group or a $\BZ_2$ extension thereof. In this section, we will explain the group $\Gamma$ in detail.

Let us start by stating four definitions of reflection groups that are closely related to each other. Here we follow the notation in \cite{etingof2001symplecticreflectionalgebrascalogeromoser}.
\begin{itemize}
\item {\bf Real reflection.} Given $V$ a vector space over $\mathbb{R}$, a real reflection is a unimodular matrix $g\in \mathrm{GL}(V)$ such that $\mathrm{rk}(\mathbf{1}-g)=1$.
\item {\bf Complex reflection.} Given $V$ a vector space over $\mathbb{C}$, a complex reflection is a unimodular matrix $g\in \mathrm{GL}(V)$ such that $\mathrm{rk}(\mathbf{1}-g)=1$.
\item {\bf Symplectic reflection.} Given $(V,\omega)$ a symplectic vector space over $\mathbb{C}$, a symplectic reflection is an element $g\in \mathrm{Sp}(V)$ such that $\mathrm{rk}(\mathbf{1}-g)=2$.
\item {\bf Quaternionic reflection.} Given $V$ a vector space over $\mathbb{H}$, a quaternionic reflection is an element $g\in \mathrm{Sp}(V)$ such that $\mathrm{rk}(\mathbf{1}-g)=1$.
\end{itemize}
We see that a symplectic reflection and a quaternionic reflection are equivalent to each other; see also \cite[Page 295]{COHEN1980293}, \cite[Page 6]{etingof2001symplecticreflectionalgebrascalogeromoser}, and \cite[Section 3]{bellamy2022}. Subsequently, we will work with symplectic reflections. A matrix group generated by symplectic reflections is called a {\it symplectic reflection group}.\footnote{A symplectic reflection group (or a quaterternionic reflection group) is called {\it proper} if it cannot be uplifted from a complex reflection group.} Moreover, we treat the symplectic reflection group $G$ as a matrix group with its module $V$ and representation $R$ specified.

Note that the moduli space of a 3d $\CN=8$ ({\it resp.} $\CN=6$) SCFT, with no non-anomalous one-form symmetry, is known to be $\BH^{2N}/\Gamma$, where $\Gamma$ is a real ({\it resp.} complex) reflection group \cite{Tachikawa:2019dvq}. Similarly, it was shown in \cite{Deb:2024zay} that, for $\CN=5$ SCFTs with no non-anomalous one-form symmetry, the moduli space takes the same form as above, but with $\Gamma$ a quaternionic reflection group. This statement also holds for some other variants with a one-form symmetry such as $\SO(2N)_{2k} \times \USp(2N)_{-k}$ and $\O(2N)_{2k}^+ \times \USp(2N)_{-k}$. However, we show that there are variants of $\CN=5$ ABJ theories whose $\Gamma$ is not a quaternionic reflection group, but a $\BZ_2$ extension thereof. We will briefly explain what we mean by a $\BZ_2$ extension.

Symplectic reflection groups are classified in \cite{COHEN1980293} (see also \cite{Deb:2024zay}).
On $\mathbb{H}\cong\mathbb{C}^2$, the symplectic reflection groups are the finite subgroups $\Gamma_{ADE}$ of $\mathrm{SU}(2)$. The $A$-type subgroup $\Gamma_{A_{n-1}}\cong\mathbb{Z}_n$ is an uplift of a complex reflection group. The $D$-type $\Gamma_{D_{n+2}}\cong\hat{D}_n$,\footnote{In this notation, $\hat{D}_n$ denotes a dicyclic group of order $4n$. Specifically, $\hat{D}_1 \cong \BZ_4$, and $\hat{D}_2 \cong Q_8$, the quaternion group of order eight.} and $E$ type $\Gamma_{E_6}\cong\hat{T}$, $\Gamma_{E_7}\cong\hat{O}$, $\Gamma_{E_8}\cong\hat{I}$ subgroups are intrinsic symplectic reflection groups.
On general $\mathbb{H}^N$, the action of a quaternionic reflection group is
\begin{equation}  \label{defGNKH}
    G_N(K,H) \cong (K^{N-1} \times H) \rtimes S_N~,
\end{equation}
where $K$ is a finite $ADE$ subgroup of $\mathrm{SU}(2)$, and $H$ is a normal subgroup of $K$ with the additional restriction that $K/H$ has to be Abelian if $N\geq 3$. In the special case of $N=1$, we simply have $G_1(K,H) = H$, which is a finite subgroup of $\SU(2)$, as mentioned above. If we choose $K$ to be $\mathbb{Z}_{pk}$ and $H=\mathbb{Z}_{k}$ where $p,k\in\mathbb{Z}_{>0}$, the symplectic reflection group $G_N(\mathbb{Z}_{pk},\mathbb{Z}_{k})$ is an uplift of the complex reflection group $G(pk,k,N)$\footnote{The full classification of complex reflection groups is given in \cite{ShephardTodd}.
The full classification of real reflection groups is given in \cite{realreflection}.} on $\mathbb{C}^N$ to $\mathbb{C}^{2N}$. Note that, from \eqref{defGNKH}, the order of the group $G_N(K, H)$ is
\begin{equation}
|G_N(K, H)| = N! |K|^{N-1} |H|~.
\end{equation}

Given a representation $R$ of a symplectic reflection group (or its extension), its action on the vector space $V$ is uniquely specified. One can naturally ``double'' the group action to $V \oplus V$ by considering the representation $R\oplus R$. This concept is particularly useful for us in the following way. While the full moduli space of the $\CN=5$ SCFT is $\BH^{2N}/\Gamma$, the space $\BH^N/\Gamma$ can be regarded as the Higgs or Coulomb branch of a 3d $\CN=5$ SCFT being viewed as an $\CN=4$ theory. We emphasise again that the limit of the index computes the Hilbert series of $\BH^N/\Gamma$, and not of the full moduli space $\BH^{2N}/\Gamma$. It is the limit of the index that allows us to verify in a field theoretic way that we have the correct $\Gamma$ for each variant of the $\CN=5$ SCFT.

\subsection{Generators of symplectic reflection groups}
Let us define the following $2 \times 2$ matrices:
\begin{equation}
I = \diag(1,1)~, \quad J=\begin{pmatrix} 0 & 1 \\ -1 & 0 \end{pmatrix}~, \quad E_n = \diag(\omega_n, \omega_n^{-1})~,
\end{equation}
with $\omega_n = \exp(2\pi i /n)$.

Let $P_j$ (with $j=1,\ldots, N-1$) be a matrix representation of the transposition $(j, j+1)$. In particular, the $2\times 2$ block-matrix in the block-positions $(j,j+1)$, $(j+1,j)$, and $(m, m)$ for $m\neq \{j, j+1\}$, are the identity matrix $I$, and the other entries are zero. The set $\{ P_j\}$ corresponds to the collection of transpositions $\{ (1,2), (2,3), \ldots, (N-1,N) \}$ that generates $S_N$. Furthermore, we define
\begin{equation}
R_1=\mathrm{diag}(J,J^{-1},I,\dots,I)~, \qquad 
R_2=\mathrm{diag}(E_{2k},E_{2k}^{-1},I,\dots,I)~,
\end{equation}
where these are generators of $\hat{D}_k$.\footnote{In the special case of $N=1$, we simply have $R_1 = J$ and $R_2 = E_{2k}$ for the generators of $\hat{D}_k$.} For the $N=2$ case, we also define a symplectic reflection
\begin{equation}
R_3 = \diag(E_k, I)~.
\end{equation}

As an example, the generators of $G_2(\hat{D}_k, \BZ_k)$ are $P_1, R_1, R_2, R_3$. The generators of $G_N(\hat{D}_k, \BZ_k)$ for $N \geq 3$ are $P_1, \ldots, P_{N-1}, R_1, R_2$. Note that, for $N\geq 3$, there is no analog of $R_3$. The reason is as follows. Consider an operation $\beta$ which takes one of the $E_{2k}$ blocks in $R_2$ to its inverse, then one can construct $R_3=\beta(R_2)R_2$. Here, $\beta$ can be regarded as an inner automorphism, composed of actions of $S_N$ and $R_1$.

Suppose that we gauge a non-anomalous $\BZ^{[0]}_{2,S}$ symmetry, where $S$ takes values in $\{B, \CM, \CC, \CM\CC\}$, in a theory $\CT$ whose moduli space is $\BH^{2N}/\Gamma$, and obtain a new non-anomalous theory $\CT'$ whose moduli space is $\BH^{2N}/\Gamma'$. Then, $\Gamma'$ can be constructed by inserting the extra generator $R_S$ into the set of generators of the group $\Gamma$, where we define
\begin{equation} \label{listofRS_orig}
\begin{aligned}
R_B &=\mathrm{diag}(E_{2k},I,\dots,I)~, \\
R_\CM &=\mathrm{diag}(E_{4k},E_{4k},\dots,E_{4k})~, \\
R_\CC &=\mathrm{diag}(J,I,\dots,I)~, \\ 
R_{\CM\CC} & = R_\CM R_\CC~.
\end{aligned}
\end{equation}

To illustrate this point, let us consider $\SO(2N)_{2k} \times \USp(2N)_{-k}$, obtained from $[\SO(2N)_{2k} \times \USp(2N)_{-k}]/\BZ_2$ by gauging $\BZ_{2,B}^{[0]}$. The group $\Gamma$ associated with the latter is $\Gamma = G_N(\hat{D}_k, \BZ_k)$, whose set of generators is $\{P_1, \ldots, P_{N-1}, R_1, R_2\}$. To obtain the group $\Gamma'$ associated with the former theory, we simply add $R_B$ into the set of generators; thus, we have $\Gamma' = \langle P_1, \ldots, P_{N-1}, R_1, R_2, R_B \rangle$. However, some of these generators are redundant, for example $R_2 = P_1 R_B^{-1} P_1 R_B$, and so we can rewrite $\Gamma'$ as $\Gamma' = \langle P_1, \ldots, P_{N-1}, R_1, R_B \rangle$. We call the set of generators after removing the redundant ones the set of {\it reduced generators}.

\subsection{Symplectic reflections and their \texorpdfstring{$\BZ_2$}{Z2} extensions}
The transposition $P_j$ is a symplectic reflection, since it satisfies $\mathrm{rk}(\mathbf{1}-P_j)=2$. The generators $R_B$ and $R_\CC$, however, are not symplectic reflections, but can be combined with permutations to form them; for example, $P_1R_B$ and $P_1R_\CC$ are symplectic reflections. Similarly, the group $G_2(\hat{D}_k,\mathbb{I})$, generated by $\{P_1, R_1, R_2\}$, is a symplectic reflection group because an equivalent set of generators, $\{P_1, P_1R_1, P_1R_2\}$, consists entirely of symplectic reflections.

For $R_\CM$ and $R_{\CM\CC}$, the situation differs for $N=2$ and $N\geq3$. For $N=2$, we can use the alternative generators
\begin{equation} \label{gen:G2DkZ2k.Z2version2}
\tilde{R}_\CM=\diag(E_{4k},  E_{4k}^{-1})~,\qquad \tilde{R}_{\CM\CC}=\tilde{R}_\CM R_\CC~.
\end{equation}
These two choices are equivalent, as their actions on $\BH^2 \cong \BC^4$ are related by a change of complex structure. An important observation is that $\tilde{R}_\CM$ can be combined with $P_1$ to form the symplectic reflection $P_1\tilde{R}_{\CM}$. However, $\tilde{R}_{\CM\CC}$ is not, in general, a symplectic reflection, nor can it be converted into one by multiplication with other generators.\footnote{The situation here is similar to that of the $\left[\mathrm{SU}(N)_k\times \mathrm{SU}(N)_k\right]/\BZ_m$ theory discussed in \cite[Section 3.2]{Tachikawa:2019dvq}, whose moduli space is guaranteed to be an orbifold of a complex reflection group only when $N=2$.} Consequently, for any variant involving the gauge group $\O(4)^-$, the associated discrete group $\Gamma$ is not a symplectic reflection group.

For $N\geq3$, it is not possible to transform $R_\CM$ or $R_{\CM\CC}$ into symplectic reflections, either by a change of complex structure of $\BC^{4N}$ or by multiplication with other generators. Therefore, any group containing $R_\CM$ or $R_{\CM \CC}$ as a generator is not a symplectic reflection group for $N \geq 3$.

The generator $R_\CM$ arises from a specific construction. It is obtained by replacing the diagonal blocks associated with the $\BZ_{2k}$ subgroup in $G_N(\hat{D}_k, \BZ_k)$ with blocks corresponding to a $\BZ_{4k}$ subgroup. Due to this enlargement from $\BZ_{2k}$ to $\BZ_{4k}$, the inclusion of $R_\CM$ (or $R_{\CM\CC}$) into the generating set is called a \textit{$\BZ_2$ extension}.\footnote{Note that $G_N(\hat{D}_k, \BZ_k).\BZ_2$ is indeed a central extension of $G_N(\hat{D}_k, \BZ_k)$ by $\BZ_2$ characterised by the short exact sequence $1  \rightarrow  \BZ_2  \rightarrow  G_N(\hat{D}_k, \BZ_k).\BZ_2   \rightarrow G_N(\hat{D}_k, \BZ_k)  \rightarrow   1$, where $\BZ_2$ is a centre of $G_N(\hat{D}_k, \BZ_k).\BZ_2$. This statement can also be generalised to other extensions discussed in this paper.} For example, the group associated with $[\Spin(2N)_{2k} \times \USp(2N)_{-k}]/\BZ_2$, generated by $\{P_1, \ldots, P_{N-1}, R_1, R_\CM \}$, is denoted by $G_N(\hat{D}_k,\BZ_k).\BZ_2$. We also denote by $G_N(\hat{D}_k,\BZ_k).\BZ_2'$ the group associated with $[\O(2N)^-_{2k} \times \USp(2N)_{-k}]/\BZ_2$, which is generated by $\{P_1, \ldots, P_{N-1}, R_1, R_{\CM \CC} \}$. In the special case of $N=2$, some of these $\BZ_2$ extensions are themselves symplectic reflection groups:
\begin{equation} \label{eq:specialcaseN2}
\begin{aligned}
{}[\Spin(4)_{2k} \times \USp(4)_{-k}]/\BZ_2 &\quad \longleftrightarrow \quad G_2(\hat{D}_{k}, \BZ_k).\BZ_2 \cong G_2(\hat{D}_{2k}, \BZ_k)~, \\
\Spin(4)_{2k} \times \USp(4)_{-k} &\quad \longleftrightarrow \quad G_2(\hat{D}_{k}, \BZ_{2k}).\BZ_2 \cong G_2(\hat{D}_{2k}, \BZ_{2k})~,\\
\Pin(4)_{2k} \times \USp(4)_{-k} &\quad \longleftrightarrow \quad G_2(\hat{D}_{k}, \hat{D}_{k}).\BZ_2 \cong G_2(\hat{D}_{2k}, \hat{D}_{k})~.
\end{aligned}
\end{equation}

\subsection{Concrete examples and summary} \label{sec:concreteexamples}
We present the results for $N=2$ in Table \eqref{tab:resultsN2}.
\bes{ \label{tab:resultsN2}
\hspace{-0.4cm}
\scalebox{0.77}{
\renewcommand{\arraystretch}{1.3} 
\begin{tabular}{c|c|c|c|c}
\hline
Group   &  Theory & Generators & Reduced generators & Order\\
\hline \hline
$G_2(\hat{D}_k,\mathbb{I})$ &  &  $P_1, R_1, R_2$ & $P_1, R_1, R_2$ & $8k$ \\
\hline \hline
$G_2(\hat{D}_k,\mathbb{Z}_k)$ & $[\mathrm{SO}(4)_{2k}\times \mathrm{USp}(4)_{-k}]/\mathbb{Z}_2$ &  $P_1, R_1, R_2, R_3$ & $P_1, R_1, R_2, R_3$ & $8k^2$ \\
\hline \hline
$G_2(\hat{D}_k,\mathbb{Z}_{2k})$ & $\mathrm{SO}(4)_{2k}\times \mathrm{USp}(4)_{-k}$ &  $P_1, R_1, R_2, R_3, R_B$ & $P_1, R_1,R_B$  & $16k^2$  \\
\hline
$G_2(\hat{D}_k,\hat{D}_{k/2})$ &  $[\mathrm{O}(4)^+_{4}\times \mathrm{USp}(4)_{-k}]/\BZ_2$ &  $P_1, R_1, R_2, R_3, R_\CC$ & $P_1, R_2,R_\CC$  & $16k^{2*}$    \\ 
\hline
$G_2(\hat{D}_{2k},\mathbb{Z}_k)$ & $[\mathrm{Spin}(4)_{2k}\times \mathrm{USp}(4)_{-k}]/\mathbb{Z}_2$ & $P_1, R_1, R_2, R_3, \tilde{R}_{\CM}$ & $P_1, R_1, R_3 ,\tilde{R}_{\CM}$  & $16k^2$  \\
\hline 
$G_2(\hat{D}_k,\mathbb{Z}_k).\mathbb{Z}_2'$ & $[\mathrm{O}(4)^-_{2k}\times \mathrm{USp}(4)_{-k}]/\BZ_2$ & $P_1, R_1, R_2, R_3, \tilde{R}_{\CM\CC}$ & $P_1, R_1, \tilde{R}_{\CM\CC}$ & $16k^{2**}$ \\
\hline \hline
$G_2(\hat{D}_k,\hat{D}_k)$ & $\mathrm{O}(4)^{+}_{2k}\times \mathrm{USp}(4)_{-k}$ & $P_1, R_1, R_2, R_3, R_B, R_\CC$ & $P_1, R_B, R_\CC$ & $32k^2$  \\ 
\hline
$G_2(\hat{D}_{2k},\mathbb{Z}_{2k})$ & $\mathrm{Spin}(4)_{2k}\times \mathrm{USp}(4)_{-k}$ & $P_1, R_1, R_2, R_3, R_B, \tilde{R}_{\CM}$ & $P_1, R_1, R_B, \tilde{R}_{\CM}$ & $32k^2$  \\ 
\hline
$G_2(\hat{D}_{k},\mathbb{Z}_{2k}).\mathbb{Z}_2'$ & $\mathrm{O}(4)^{-}_{2k}\times \mathrm{USp}(4)_{-k}$ &  $P_1,
R_1, R_2, R_3, R_B, \tilde{R}_{\CM\CC}$ & $P_1,
R_1, R_B, \tilde{R}_{\CM\CC}$ & $32k^2$   \\
\hline \hline
$G_2(\hat{D}_{2k},\hat{D}_k)$ & $\mathrm{Pin}(4)_{2k}\times \mathrm{USp}(4)_{-k}$ & $P_1, R_1,
R_2,
R_3,
R_B,
R_\CC,
\tilde{R}_{\CM}$ & $P_1,
R_B,
R_\CC,
\tilde{R}_{\CM}$ & $64k^2$  \\
\hline
\end{tabular}}
}
Remarks on the $N=2$ cases:
\begin{itemize}
    \item[${}^*$] This case applies only when $k$ is even. For odd $k$, the presence of $R_\CC$ (along with $P_1, R_1, R_2, R_3$) implies the presence of $R_B$, since $R_B=R_\CC^2 R_3$; thus, the group becomes $G_2(\hat{D}_k,\hat{D}_{k})$ with order $32k^2$. This implies that $[\O(4)^+_{2k} \times \USp(4)_{-k}]/\BZ_2$ is anomalous for odd $k$.
    \item[${}^{**}$] This case applies only when $k$ is odd. For even $k$, the presence of $\tilde{R}_{\CM \CC}$ (along with $P_1, R_1, R_2, R_3$) implies the presence of $R_B$, since we have $R_B=P_1(\tilde{R}_{\CM \CC})^{-2k}P_1^{-1}(\tilde{R}_{\CM \CC})^{2k}$; thus the group becomes $G_2(\hat{D}_k,\BZ_{2k}).\BZ_2'$ with order $32k^2$. This implies that $[\O(4)^-_{2k} \times \USp(4)_{-k}]/\BZ_2$ is anomalous for even $k$.
\end{itemize}

The results for $N \geq 3$ are summarised in Table \eqref{tab:resultsNgeq3}.
\bes{ \label{tab:resultsNgeq3}
\hspace{-0.9cm}
\scalebox{0.77}{
\renewcommand{\arraystretch}{1.3} 
\begin{tabular}{c|c|c|c|c}
\hline
Group   &  Theory & Generators & Reduced generators & Order\\
\hline \hline
$G_N(\hat{D}_k,\mathbb{Z}_k)$ & $[\mathrm{SO}(2N)_{2k}\times \mathrm{USp}(2N)_{-k}]/\mathbb{Z}_2$ &  $\{ P_j \}, R_1, R_2$ & $\{ P_j \}, R_1, R_2$ & $k \times (4k)^{N-1} \times N!$ \\
\hline \hline
$G_N(\hat{D}_k,\mathbb{Z}_{2k})$ & $\mathrm{SO}(2N)_{2k}\times \mathrm{USp}(2N)_{-k}$ &  $\{ P_j \}, R_1,
R_2, R_B$ & $\{ P_j \}, R_1, R_B$  & $2k \times (4k)^{N-1} \times N!$  \\
\hline
$G_N(\hat{D}_k,\hat{D}_{k/2})$ &  $[\mathrm{O}(2N)^+_{2k}\times \mathrm{USp}(2N)_{-k}]/\BZ_2$ &  $\{ P_j \}, R_1,
R_2, R_\CC$ & $\{ P_j \}, R_2, R_\CC$ & $2k \times (4k)^{N-1} \times N!^{*}$    \\ 
\hline
$G_N(\hat{D}_k,\mathbb{Z}_k).\mathbb{Z}_2$ & $[\mathrm{Spin}(2N)_{2k}\times \mathrm{USp}(2N)_{-k}]/\mathbb{Z}_2$ & $\{ P_j \}, R_1,
R_2, R_\CM$ & $\{ P_j \}, R_1, R_\CM$ & $2k \times (4k)^{N-1} \times N!^{\dagger}$  \\
\hline 
$G_N(\hat{D}_k,\mathbb{Z}_k).\mathbb{Z}'_2$ & $[\mathrm{O}(2N)^-_{2k}\times \mathrm{USp}(2N)_{-k}]/\BZ_2$ & $\{ P_j\},
R_1, R_2, R_{\CM\CC}$ & $\{ P_j\},
R_1, R_{\CM\CC}$ & $2k \times (4k)^{N-1} \times N!^{**}$ \\
\hline \hline
$G_N(\hat{D}_k,\hat{D}_k)$ & $\mathrm{O}(2N)^{+}_{2k}\times \mathrm{USp}(2N)_{-k}$ &  $\{ P_j\},
R_1,R_2,R_B,R_\CC$ & $\{ P_j\},R_B,R_\CC$ & $(4k)^N \times N!$  \\ 
\hline
$G_N(\hat{D}_k,\mathbb{Z}_{2k}).\mathbb{Z}_2$ & $\mathrm{Spin}(2N)_{2k}\times \mathrm{USp}(2N)_{-k}$ & $\{ P_j\},
R_1, R_2, R_B, R_\CM$ & $\{ P_j\},
R_1, R_B, R_\CM$ & $(4k)^N \times N!$  \\ 
\hline
$G_N(\hat{D}_k,\mathbb{Z}_{2k}).\mathbb{Z}_2'$ & $\mathrm{O}(2N)^{-}_{2k}\times \mathrm{USp}(2N)_{-k}$ &  $\{ P_j\},
R_1, R_2, R_B, R_{\CM\CC}$ & $\{ P_j\},
R_1, R_B, R_{\CM\CC}$ & $(4k)^N \times N!$   \\
\hline \hline
$G_N(\hat{D}_k,\hat{D}_k).\mathbb{Z}_2$ & $\mathrm{Pin}(2N)_{2k}\times \mathrm{USp}(2N)_{-k}$ & $\{ P_j \}, R_1,
R_2,
R_B,
R_\CC,
R_\CM$ & $\{ P_j \},
R_B,
R_\CC,
R_\CM$ & $2\times(4k)^N \times N!$  \\
\hline
\end{tabular}}
}
Here, $\{P_j\}$ denotes the set $\{P_1, \ldots, P_{N-1}\}$. Remarks on these cases:
\begin{itemize}
    \item[${}^*$] This case is valid only when $k$ is even. For odd $k$, the presence of $R_\CC$ (along with $\{ P_j \}, R_1, R_2$) implies the presence of $R_B$, since $R_B$ can be constructed from $R_2$, $R_\CC$ and certain $P_j$; the group becomes $G_N(\hat{D}_k,\hat{D}_{k})$ with order $N!(4k)^N$. This implies that $[\O(2N)^+_{2k} \times \USp(2N)_{-k}]/\BZ_2$ is anomalous for odd $k$.
    \item[${}^{\dagger}$] This case is valid only when $N$ is even. For odd $N$, the presence of $R_\CM$ (along with $\{ P_j \}, R_1, R_2$) implies the presence of $R_B$, since $R_B$ can be constructed from $R_1$, $R_\CM$ and certain $P_j$; the group becomes $G_N(\hat{D}_k,\BZ_{2k}).\BZ_2$ with order $N!(4k)^N$. This implies that $[\Spin(2N)_{2k} \times \USp(2N)_{-k}]/\BZ_2$ is anomalous for odd $N$.
    \item[${}^{**}$] This case is valid only when $(N,k)$ is (even, odd) or (odd, even). For other parities, the presence of $R_{\CM \CC}$ (along with $\{ P_j \}, R_1, R_2$) implies the presence of $R_B$, since $R_B$ can be constructed from $R_1$, $R_\CM R_\CC$ and certain $P_j$, and the group becomes $G_N(\hat{D}_k,\BZ_{2k}).\BZ_2'$ with order $N!(4k)^N$. This implies that $[\O(2N)^-_{2k} \times \USp(2N)_{-k}]/\BZ_2$ is anomalous when $N$ and $k$ are both even or both odd.
\end{itemize}
We have explicitly checked that the groups obtained from the generators listed above are in agreement with those discussed in \cite[Appendix A]{Deb:2024zay}.

Finally, we point out that the moduli space of the variant $\mathrm{O}(2N)_{2k}^+\times \mathrm{USp}(2N)_{-k}$ is $\mathbb{C}^{4N}/G_N(\hat{D}_k,\hat{D}_k)$. This is, in fact, isomorphic to $\Sym^N(\BC^4/\hat{D}_k)$, which is in agreement with the notable result of \cite{Aharony:2008gk}, where it was proposed that this variant is the worldvolume theory of $N$ M$2$-branes probing a $\mathbb{C}^4/\hat{D}_k$ singularity.

\subsection{Anomalous variants} \label{sec:anomvar}
As discussed previously, if gauging a non-anomalous $\BZ^{[0]}_{2, S}$ symmetry of a given non-anomalous theory $\CT$ (with moduli space $\BH^{2N}/\Gamma$) leads to another non-anomalous theory $\CT'$ (with moduli space $\BH^{2N}/\Gamma'$), then the new group is $\Gamma' = \langle \mathrm{gen}(\Gamma), R_S \rangle$, where $\mathrm{gen}(\Gamma)$ is a set of generators of $\Gamma$. We find that in such cases,
\begin{equation} \label{twicevol}
|\Gamma'| = 2|\Gamma|~.
\end{equation}
This is consistent with the principle that the relative volume of the base of the associated Calabi-Yau cone must increase by a factor of two when a non-anomalous $\BZ_2$ zero-form symmetry is gauged (see, for example, \cite{Hanany:2018dvd}).\footnote{If $H$ is the base of the Calabi-Yau hyperK\"ahler cone associated with theory $\CT$, then $\text{vol}(H/\Gamma) = 2\text{vol}(H/\Gamma')$; see \eg~ \cite{Bergman:2001qi, Martelli:2005tp, Martelli:2006yb}.} This statement holds for the entries in Tables \eqref{tab:resultsN2} and \eqref{tab:resultsNgeq3}.

On the other hand, if $\CT$ is non-anomalous, but the theory $\CT'$ obtained by gauging $\BZ^{[0]}_{2, S}$ is anomalous, we observe that the order of the new group $\langle \mathrm{gen}(\Gamma), R_S \rangle$ becomes four times the order of $\Gamma$, not twice. Furthermore, the resulting quotient space does not correspond to the moduli space of $\CT'$; instead, it describes the moduli space of a different non-anomalous theory $\CT''$. This observation aligns with the remarks accompanying Tables \eqref{tab:resultsN2} and \eqref{tab:resultsNgeq3}. We summarise the anomalous variants for each parity of $N$ and $k$ in \eqref{tab:summaryNk}.

This pattern can be generalised further. Consider a non-anomalous theory $\mathcal{T}$ with an {\it anomalous} finite discrete Abelian zero-form symmetry $G$. Let $\mathcal{T}'$ be an anomalous theory obtained by gauging $G$ from $\mathcal{T}$, and let  $\mathcal{T}''$ be a non-anomalous theory obtained by gauging $G'$ from $\mathcal{T}$, where $G'$ is the minimal non-anomalous extension of $G$.\footnote{By the term ``minimal extension'', we mean as follows. Suppose that $G'$ is a non-anomalous extension of $G$ by $A$ described by $1 \rightarrow A \rightarrow G' \rightarrow G \rightarrow 1$. If, for any other non-anomalous extension $G''$ of $G$ by $A'$ described by $1 \rightarrow A' \rightarrow G'' \rightarrow G \rightarrow 1$, $A$ is a normal subgroup of $A'$, then $G'$ is a minimal extension. The construction in \cite{Tachikawa:2017gyf} guarantees an $A'$ can always be found, and so $A$ can be acquired by examining the subgroups of $A'$. Also, in \cite{Robbins:2021lry}, a method to directly reduce $A'$ to $A$ is provided.}
If $\mathcal{M}_\CT$ is the moduli space of $\mathcal{T}$, then the action of the anomalous symmetry $G$ on $\mathcal{M}_\CT$ does not form a closed orbit.\footnote{By the term ``closed orbit", we mean as follows. For two generic points $x,y\in \mathcal{M}_\CT$, if $y\in G(x)$ implies $G(x) = G(y)$, then $G(x)$ is a closed orbit.} Instead, it generates the $G'$ action on $\mathcal{M}_\CT$, which forms a closed orbit. On a vector space $V$, this means that the action of $G$ is not a linear representation $G\to \mathrm{GL}(V)$, but an $A$-projective representation $G\to\mathrm{GL}(V)/A$, which determines a linear representation $G'\to GL(V)$ of its covering group $G'$.\footnote{In general, the $A$-projective representation is classified by the group cohomology $H^2(G,A)$.} By quotienting the action of $G$ through a closed orbit, we obtain the moduli space $\mathcal{M}_\CT/G'$, which is the moduli space of $\mathcal{T}''$.

\subsection{Hilbert series}
In this subsection, we report the Hilbert series of $\BH^N/\Gamma$, which is the Higgs or Coulomb branch of the $\CN=5$ SCFT in question (viewed as an $\CN=4$ theory). Using the generators described previously, we can construct the group elements of $\Gamma$. The Hilbert series can be computed using the Molien discrete formula:
\bes{
\mathrm{HS}[\BH^N/\Gamma](t) = \frac{1}{|\Gamma|} \sum_{M \in \Gamma} \frac{1}{\det(\mathbf{1}- t M)}~.
}
For each case listed in \eqref{tab:resultsN2} and \eqref{tab:resultsNgeq3}, we have verified that the Hilbert series is in agreement with the Higgs/Coulomb branch limit of the superconformal index, up to a sufficiently high order in the series expansion. Note that the Hilbert series of certain variants of the ABJ theory whose $\Gamma$ is a quaternionic reflection group were reported in \cite{Deb:2024zay}. Let us present certain cases whose $\Gamma$ is not a quaternionic reflection group as follows:
 \bes{
\scalebox{0.8}{ 
\renewcommand{\arraystretch}{1.3} 
\begin{tabular}{c|l}
\hline
$\O(4)^-_{2k} \times \USp(4)_{-k}$ & Hilbert series of $\BH^2/G_2(\hat{D}_k,\BZ_{2k}).\BZ_2'$ \\
\hline
     & $\frac{1 - 2 t^2 + t^4 + t^6 + t^8 - 2 t^{10} + t^{12}}{(1 - t)^4 (1 + t)^4 (1 + t^2)^2 (1 + t^4)}$ \\
$k=2$     & $= 1 + t^4 + t^6 + 5 t^8 + 4 t^{10}+ 9 t^{12} +10 t^{14} + 19 t^{16}+\ldots$ \\
& $ = \PE[t^4 + t^6 + 4 t^8 + 3 t^{10} + 3 t^{12} + 2 t^{14} - 3 t^{16}+\ldots]$ \\
\hline 
& $\frac{1 - t^2 + t^4 - 2 t^6 + 3 t^8 - 2 t^{10} + 4 t^{12} - 2 t^{14} + 3 t^{16} -2 t^{18} + t^{20} - t^{22} + t^{24}}{(1 - t)^4 (1 + t)^4 (1 + t^2)^2 (1 - t + t^2)^2 (1 + t + t^2)^2 (1 + t^4) (1 - t^2 + t^4)}$\\
$k=3$ & $=1 + t^4 + 3 t^8 + t^{10} + 6 t^{12} + 4 t^{14} + 10 t^{16}+ \ldots$ \\
&$=\PE[t^4 + 2 t^8 + t^{10} + 3 t^{12} + 3 t^{14} + t^{16} - 6 t^{18} - 17 t^{20}+\ldots ]$\\
\hline
\end{tabular}}
}
\bes{
\scalebox{0.75}{ 
\renewcommand{\arraystretch}{1.3} 
\begin{tabular}{c|l}
\hline
$[\O(4)^-_{2k} \times \USp(4)_{-k}]/\BZ_2$ & Hilbert series of $\BH^2/G_2(\hat{D}_k,\BZ_{k}).\BZ_2'$ (with $k$ odd) \\
\hline
& $\frac{1-t^2+t^4-t^6+3 t^8-t^{10}+4 t^{12}-t^{14}+3 t^{16}-t^{18}+t^{20}-t^{22}+t^{24}}{(1-t)^4 (1+t)^4 \left(1+t^2\right)^2 \left(1-t+t^2\right)^2 \left(1+t+t^2\right)^2 \left(1+t^4\right) \left(1-t^2+t^4\right)}$ \\
$k=3$ & $= 1 + t^4 + t^6 + 4 t^8 + 3 t^{10} + 9 t^{12} + 9 t^{14}+ 16 t^{16} + \ldots$ \\
& $= \PE[t^4 + t^6 + 3 t^8 + 2 t^{10} + 4 t^{12} + 3 t^{14} - t^{16} +\ldots]$ \\
\hline
& $\frac{1 - t^2 + t^4 - t^6 + t^8 - t^{10} + 3 t^{12} - t^{14} + 3 t^{16} - 2 t^{18} + 3 t^{20} - t^{22} + 3 t^{24} - t^{26} + t^{28} - t^{30} + t^{32} - t^{34} + t^{36}}{(1 - t)^4 (1 + t)^4 \left(1 + t^2\right)^2 \left(1 + t^4\right) \left(1 - t + t^2 - t^3 + t^4\right)^2 \left(1 + t + t^2 + t^3 + t^4\right)^2 \left(1 - t^2 + t^4 - t^6 + t^8\right)}$ \\
$k=5$ & $=1 + t^4 + 2 t^8 + t^{10} + 4 t^{12} + 3 t^{14} + 7 t^{16} + 5 t^{18} + 12 t^{20} +\ldots $ \\
& $=\PE[t^4 + t^8 + t^{10} + 2 t^{12} + 2 t^{14} + 2 t^{16} + t^{18} + 2 t^{20}+\ldots ]$ \\
\hline
\end{tabular}}
}
\bes{
\scalebox{0.75}{ 
\renewcommand{\arraystretch}{1.3} 
\begin{tabular}{c|l}
\hline
$\Pin(6)_{2k} \times \USp(6)_{-k}$ & Hilbert series of $\BH^3/G_3(\hat{D}_k,\hat{D}_k).\BZ_2$ \\
\hline
& $\frac{1 - 2t^{2} + 2t^{4} - 3t^{6} + 5t^{8} - 4t^{10} + 6t^{12} - 8t^{14} + 9t^{16} - 9t^{18} + 11t^{20} - 10t^{22} + \text{palindrome} + t^{44}}{(1-t)^{6} (1+t)^{6} \left(1+t^{2}\right)^{3} \left(1 - t + t^{2}\right)^{2} \left(1+ t + t^{2}\right)^{2} \left(1+t^{4}  \right)^{3} \left(1-t^2+t^4\right) \left(1-t^4+t^8\right)}$ \\
$k=2$ & $= 1 + t^{4} + 4t^{8} + 2t^{10} + 8t^{12} + 5t^{14} + 18t^{16} + 16t^{18} + 34t^{20} + \ldots$ \\
& $= \PE[t^{4} + 3t^{8} + 2t^{10} + 4t^{12} + 3t^{14} + 4t^{16} + 5t^{18} + t^{20} +\ldots]$ \\
\hline
& $\frac{1}{(1 - t)^4 (1 + t)^4 \left(1 + t^2\right)^2 \left(1 + t^4\right) \left(1 - t + t^2 - t^3 + t^4\right)^2 \left(1 + t + t^2 + t^3 + t^4\right)^2 \left(1 - t^2 + t^4 - t^6 + t^8\right)} \times$ \\
& \quad {\footnotesize $(1 - 3t^{2} + 5t^{4} - 7t^{6} + 9t^{8} - 11t^{10} + 14t^{12} - 16t^{14} + 19t^{16} - 22t^{18} + 25t^{20} $} \\
$k=3$ & \quad {\footnotesize $ - 28t^{22} + 31t^{24} - 32t^{26} + 34t^{28} - 34t^{30} + 34t^{32} +\text{palindrome} + t^{64})$} \\
 & $=1 + t^{4} + 2t^{8} + 5t^{12} + 2t^{14} + 9t^{16} + 5t^{18} + 15t^{20} +\ldots$ \\
& $=\PE[t^{4} + t^{8} + 3t^{12} + 2t^{14} + 3t^{16} + 3t^{18} + 3t^{20}+\ldots ]$ \\
\hline
\end{tabular}}
 }

\section{Orthosymplectic ABJ theories with unequal ranks} \label{sec:unequalranks}
We now consider the case with unequal ranks, which deserves a separate discussion from the equal-rank case due to differing 't Hooft anomalies and moduli space structures. There are two possibilities to consider, namely theories with the gauge algebras
\bes{\label{eq21101}
\so(2N+2x)_{2k} \times \usp(2N)_{-k}~, \qquad \so(2N)_{2k} \times \usp(2N+2x)_{-k}~.
}
Note that the case of $x=0$, namely that with equal ranks, was discussed in Section \ref{sec:ABJMequalranks}.
The Chern-Simons levels are such that the theories have at least $\CN=5$ supersymmetry. 
As pointed out in \cite{Aharony:2008gk} (see also \cite[(4.9), (4.10)]{Honda:2017nku}), the theories with the $\O^+$-type gauge group for particular values of $k$ and $x$ enjoy the following dualities:
\bes{\label{ABJMdualitiesunequalrank}
\scalebox{1}{$
\begin{split}
&\O(2N+2x)^+_{2k} \times \Usp(2N)_{-k}\,\, \leftrightarrow \,\, \O(2N+2(k-x+1))^+_{-2k}\times \Usp(2N)_{k}~,  \\
&\O(2N)^+_{2k} \times \Usp(2N+2x)_{-k}\,\, \leftrightarrow \,\, \O(2N)^+_{-2k}\times \Usp(2N+2(k-x-1))_{k}~,
\end{split}$}
}
where $x$ is restricted to $0 \leq x \leq k+1$ in the first duality and $0 \leq x \leq k-1$ in the second one. If $x >k+1$ in the former or $x > k-1$ in the latter, then supersymmetry is broken. Moreover, when the equalities hold, the theories with unequal ranks turn out to be dual to theories with equal ranks:
\bes{\label{ABJMdualitiesunequalranktoequalrank}
\O(2N+2(k+1))^+_{2k}\times \Usp(2N)_{-k}\,\, &\leftrightarrow \,\, \O(2N)^+_{-2k}\times \Usp(2N)_{k}\\
\O(2N)^+_{2k}\times \Usp(2N+2(k-1))_{-k} \,\, &\leftrightarrow \,\, \O(2N)^+_{-2k}\times \Usp(2N)_{k}~.
}
We shall henceforth take $0 < x < k \pm 1$ for each case in the following analysis.

The anomaly theory for the theory $\SO(2N+2x)_{2k} \times \USp(2N)_{-k}$ is
\bes{ \label{anomuneqrank1}
i \pi \int_{M_4} \mathcal{A}^B_2 \cup \Big[(N+x) \mathcal{A}^{\mathcal{M}}_1 \cup \mathcal{A}^{\mathcal{M}}_1 &+ k \mathcal{A}^{\mathcal{C}}_1 \cup \mathcal{A}^{\mathcal{C}}_1 + \mathcal{A}^{\mathcal{M}}_1 \cup \mathcal{A}^{\mathcal{C}}_1 + \left(kx\right) \mathcal{A}^B_2 \Big]~,}
whereas that for the theory
$\SO(2N)_{2k} \times \USp(2N+2x)_{-k}$ is
\bes{i \pi \int_{M_4} \mathcal{A}^B_2 \cup \Big[N \mathcal{A}^{\mathcal{M}}_1 \cup \mathcal{A}^{\mathcal{M}}_1 &+ k \mathcal{A}^{\mathcal{C}}_1 \cup \mathcal{A}^{\mathcal{C}}_1 + \mathcal{A}^{\mathcal{M}}_1 \cup \mathcal{A}^{\mathcal{C}}_1 + \left(kx\right) \mathcal{A}^B_2 \Big]~.}
We focus on the case where the $\BZ_{2}^{[1]}$ one-form symmetry can be gauged, namely when $k x$ is even, and will discuss the case in which $kx$ is odd in Section \ref{sec:commentsanomoneform}. In the first ({\it resp.} second) case, if both $N+x$ and $k$ ({\it resp.} both $N$ and $k$) are odd, the corresponding symmetry category is $Q_8$; otherwise, it is $D_8$. The symmetry webs in these cases are, therefore, similar to those of the equal-rank cases depicted in Figures \ref{fig:D8Nevenkeven}--\ref{fig:Q8Noddkodd}. Note that the $\BZ^{[0]}_{2,S}$ symmetry (with $S \in  \{ B, \CM, \CC, \CM\CC \}$) acts non-trivially on the $\left[\SO(2N+2x)_{2k} \times \USp(2N)_{-k}\right]/\BZ_2$ and $\left[\SO(2N)_{2k} \times \USp(2N+2x)_{-k}\right]/\BZ_2$ theories, but we will see below that some of them act trivially on the moduli space.

We can now turn to the study of the moduli space. For the case of the theory $\left[\SO(2N+2x)_{2k} \times \USp(2N)_{-k}\right]/\BZ_2$, we find that the moduli space is always $\BH^{2N}/G_N(\hat{D}_k , \hat{D}_k)$.  On the other hand, for the $\left[\SO(2N)_{2k} \times \USp(2N+2x)_{-k}\right]/\BZ_2$ theory, the moduli space is $\BH^{2N}/ G_N(\hat{D}_k, \BZ_{2k})$.\footnote{We verified that the Higgs or Coulomb branch limit of the index for each theory matches the Hilbert series of the corresponding moduli space $\BH^{N}/\Gamma$. However, this method is unable to detect the presence of a radical ideal, should one exist, as is the case for the unitary ABJ theory \cite{Giacomelli:2024sex}.} We will explain these results towards the end of this Section.  Starting from these variants, one can gauge the $\BZ_{2, S}^{[0]}$ symmetry to obtain the other variants. The moduli space of each non-anomalous variant is $\BH^{2N}/\Gamma$, where $\Gamma$ can be obtained by adding an appropriate generator $R_S$ to the set of generators of $\Gamma$ associated with the variant prior to gauging, precisely as described in Section \ref{quatrefgroupmodspace}. The group $\Gamma$ for each variant of the $\so(2N+2x)_{2k} \times \usp(2N)_{-k}$ theory is reported below, categorised by all parity combinations of $N+x$ and $k$.
\bes{\label{tabSON+x}
\hspace{-0.9cm}
\scalebox{0.77}{
\renewcommand{\arraystretch}{1.3} 
\begin{tabular}{c|c|c|c|c}
\hline
 Theory  & $N+x$ even, $k$ even & $N+x$ odd, $k$ even & $N+x$ even, $k$ odd & $N+x$ odd, $k$ odd\\
\hline \hline
$\left[\mathrm{SO}(2N+2x)_{2k}\times \mathrm{USp}(2N)_{-k}\right]/\mathbb{Z}_2$& $G_N(\hat{D}_k,\hat{D}_k)$ & $G_N(\hat{D}_k,\hat{D}_k)$& $G_N(\hat{D}_k,\hat{D}_k)$& $G_N(\hat{D}_k,\hat{D}_k)$\\
\hline 
$\mathrm{SO}(2N+2x)_{2k}\times \mathrm{USp}(2N)_{-k}$ &  $G_N(\hat{D}_k,\hat{D}_k)$ &  $G_N(\hat{D}_k,\hat{D}_k)$ &  $G_N(\hat{D}_k,\hat{D}_k)$& $G_N(\hat{D}_k,\hat{D}_k)$\\
\hline
 $\left[\mathrm{O}(2N+2x)^+_{2k}\times \mathrm{USp}(2N)_{-k}\right]/\BZ_2$&  $G_N(\hat{D}_k,\hat{D}_k)$&  $G_N(\hat{D}_k,\hat{D}_k)$&  Anomalous&  Anomalous\\ 
\hline
 $\left[\mathrm{Spin}(2N+2x)_{2k}\times \mathrm{USp}(2N)_{-k}\right]/\mathbb{Z}_2$& $G_N(\hat{D}_k,\hat{D}_k).\mathbb{Z}_2$& Anomalous& $G_N(\hat{D}_k,\hat{D}_k).\mathbb{Z}_2$ &  Anomalous\\
\hline 
$\left[\mathrm{O}(2N+2x)^-_{2k}\times \mathrm{USp}(2N)_{-k}\right]/\BZ_2$ & Anomalous &  $G_N(\hat{D}_k,\hat{D}_k).\mathbb{Z}_2$&  $G_N(\hat{D}_k,\hat{D}_k).\mathbb{Z}_2$&  Anomalous\\
\hline 
 $\mathrm{O}(2N+2x)^{+}_{2k}\times \mathrm{USp}(2N)_{-k}$& $G_N(\hat{D}_k,\hat{D}_k)$& $G_N(\hat{D}_k,\hat{D}_k)$ & $G_N(\hat{D}_k,\hat{D}_k)$& $G_N(\hat{D}_k,\hat{D}_k)$\\ 
\hline
 $\mathrm{Spin}(2N+2x)_{2k}\times \mathrm{USp}(2N)_{-k}$& $G_N(\hat{D}_k,\hat{D}_k).\mathbb{Z}_2$ & $G_N(\hat{D}_k,\hat{D}_k).\mathbb{Z}_2$& $G_N(\hat{D}_k,\hat{D}_k).\mathbb{Z}_2$ & $G_N(\hat{D}_k,\hat{D}_k).\mathbb{Z}_2$\\ 
\hline
 $\mathrm{O}(2N+2x)^{-}_{2k}\times \mathrm{USp}(2N)_{-k}$&$G_N(\hat{D}_k,\hat{D}_k).\mathbb{Z}_2$ &$G_N(\hat{D}_k,\hat{D}_k).\mathbb{Z}_2$ &$G_N(\hat{D}_k,\hat{D}_k).\mathbb{Z}_2$& $G_N(\hat{D}_k,\hat{D}_k).\mathbb{Z}_2$\\
\hline 
 $\mathrm{Pin}(2N+2x)_{2k}\times \mathrm{USp}(2N)_{-k}$& $G_N(\hat{D}_k,\hat{D}_k).\mathbb{Z}_2$ &$G_N(\hat{D}_k,\hat{D}_k).\mathbb{Z}_2$&$G_N(\hat{D}_k,\hat{D}_k).\mathbb{Z}_2$& $G_N(\hat{D}_k,\hat{D}_k).\mathbb{Z}_2$\\
\hline
\end{tabular}}
}
We emphasise that, in contrast to the equal-rank case, $\Gamma$ may be the same for different variants of the gauge group, so such theories have the same moduli space. In particular, it is clear from \eqref{tabSON+x} that $\BZ^{[0]}_{2, B}$ acts trivially on the moduli space of $\left[\SO(2N+2x)_{2k} \times \USp(2N)_{-k}\right]/\BZ_2$ for any parity of $N+x$ and $k$, whereas $\BZ^{[0]}_{2, \CC}$ acts trivially on the moduli space of the theories when $k$ is even, but is anomalous for $k$ odd.\footnote{However, gauging $\BZ^{[0]}_{2, \CC}$ of $\SO(2N+2x)_{2k} \times \USp(2N)_{-k}$ is allowed for any parity of $N+x$ and $k$, and it acts trivially on the moduli space of such theory.} Gauging any of these non-anomalous symmetries will not affect the moduli space, even if the index changes upon such gauging.

Let us illustrate this point using the following example. The index of the theory $\left[\SO(6)_{4} \times \USp(4)_{-2}\right]/\BZ_2$, where $N=2, \, x=1, \, k=2$, can be computed using \eqref{indABJlikemodZ2} and is given by 
\bes{
 1 + x &+ \Big[ 2 + (1 + \zeta)[4]_a - [2]_a \Big] x^2 \\
& + \Big[ 3 + (D' - \zeta - 1)[2]_a + (1 + \zeta)[4]_a + \zeta [6]_a \Big] x^3 + \ldots~,
}
where $D'$ is defined in \eqref{D8Q8char2}. Recall that the Higgs or Coulomb branch limit can be obtained as $\sum_{p \geq 0} C(a^{\pm 2p} x^p) t^{2p}$, where $C(a^{\pm 2p} x^p)$ is the coefficient of the term $a^{\pm 2p} x^p$ in the index. This yields the Hilbert series:
\bes{
1 &+ (1 + \zeta) t^4 + \zeta t^6 + (4 + 2\zeta) t^8 + (2 + 2\zeta) t^{10} + (6 + 5\zeta) t^{12} + (4 + 6\zeta) t^{14} \\
& + (12 + 9\zeta) t^{16} + (10 + 10\zeta) t^{18} + (18 + 16\zeta) t^{20}+ \ldots~,
}
which is independent of $g$ and $\chi$. This means that $\BZ_{2, B}^{[0]}$ and $\BZ_{2, \CC}^{[0]}$ act trivially on the Higgs or Coulomb branch operators of $\left[\SO(6)_{4} \times \USp(4)_{-2}\right]/\BZ_2$. Any non-anomalous variant obtained by gauging either of these two symmetries (but not $\BZ_{2, \CM}^{[0]}$) thus has the same moduli space as this theory, namely $\BH^{4}/G_2(\hat{D}_2, \hat{D}_2)$. It is also interesting to point out that, although $D'$ appears in the index, the Hilbert series depends only on $\zeta$. This means that the $\left[\SO(6)_{4} \times \USp(4)_{-2}\right]/\BZ_2$ theory possesses the $D_8$ zero-form symmetry, but only $\BZ_{2, \CM}^{[0]}$ acts non-trivially on the moduli space. Since half-odd-integral powers of $\zeta$ appear in the index via $D'$, it follows that $\BZ_{2,\CM}^{[0]}$ cannot be gauged,\footnote{On the other hand, $\BZ_{2,\CC}^{[0]}$ is gaugable, as can be seen from \eqref{anomuneqrank1}.} as previously mentioned. As a consequence, the variant $\left[\Spin(6)_{4} \times \USp(4)_{-2}\right]/\BZ_2$ is anomalous.

For definiteness, let us focus on the case of $\left[\SO(2N+2x)_{2k} \times \USp(2N)_{-k}\right]/\BZ_2$, with both $N+x$ and $k$ even. We now explain why the $\BZ_{2, B}^{[0]}$ and $\BZ_{2,\CC}^{[0]}$ symmetries act trivially on the moduli space. Note that the other variants can be obtained by sequentially gauging $\BZ_{2,B}^{[0]}$, $\BZ_{2,\CM}^{[0]}$ or $\BZ_{2,\CC}^{[0]}$ of this variant.
As discussed extensively in Section \ref{quatrefgroupmodspace}, the generators of the group $G_N(\hat{D}_{k}, \hat{D}_k)$ that gives the moduli space $\BH^{2N}/G_N(\hat{D}_{k}, \hat{D}_k)$ are as follows:
\bes{\label{eqrefunequalranksgenerators}
 \{P_j\}~,
 R_1~,
R_2~,
R_B~,
R_\CC~.}
Gauging the $\BZ_{2,B}^{[0]}$ or $\BZ_{2,\CC}^{[0]}$ symmetries amounts to adding $R_B$ or $R_\CC$ to the above set of generators. However, since $R_B$ and $R_\CC$ are already present in \eqref{eqrefunequalranksgenerators}, the group $G_N(\hat{D}_{k}, \hat{D}_k)$ does not change upon such gauging. Thus, the moduli space of the resulting theory remains $\BH^{2N}/G_N(\hat{D}_{k}, \hat{D}_k)$. On the other hand, the generator $R_\CM$ associated with $\BZ_{2,\CM}^{[0]}$ is not present in \eqref{eqrefunequalranksgenerators}. Adding it to \eqref{eqrefunequalranksgenerators} results in a $\BZ_2$ extension of $G_N(\hat{D}_{k}, \hat{D}_k)$, namely $G_N(\hat{D}_{k}, \hat{D}_k).\BZ_2$. Note that this is no longer a quaternionic reflection group for $N>2$; however, for $N=2$, it is isomorphic to the quaternionic reflection group $G_2(\hat{D}_{2k}, \hat{D}_k)$. Finally, let us consider gauging the symmetry $\BZ_{2,\CM\CC}^{[0]}$, which amounts to adding $R_{\CM \CC} = R_\CM R_\CC$ to \eqref{eqrefunequalranksgenerators}, which leads to the anomalous variant $\left[\O(2N+2x)^-_{2k} \times \USp(2N)_{-k}\right]/\BZ_2$. We see that the order of the resulting group $G_N(\hat{D}_{k}, \hat{D}_k).\BZ_2$ increases only by a factor of two (not by a factor of four as in Section \ref{sec:ABJMequalranks}) with respect to the group $G_N(\hat{D}_{k}, \hat{D}_k)$. Nevertheless, in the same spirit as Section \ref{sec:ABJMequalranks}, it can be clearly seen from \eqref{tabSON+x} that the latter group, resulting from a gauging that is forbidden by the 't Hooft anomalies, corresponds to the other non-anomalous variants.  Finally, we remark that this argument applies to other parities of $N+x$ and $k$. 

We now turn to the case of $\left[\SO(2N)_{2k} \times \USp(2(N+x))_{-k}\right]/\BZ_2$. We report the group $\Gamma$ corresponding to the moduli space $\BH^{2N}/\Gamma$ as follows:
\bes{
\hspace{-0.8cm}
\scalebox{0.77}{
\renewcommand{\arraystretch}{1.3} 
\begin{tabular}{c|c|c|c|c}
\hline
 Theory  & $N$ even, $k$ even & $N$ odd, $k$ even & $N$ even, $k$ odd & $N$ odd, $k$ odd\\
\hline \hline
$\left[\mathrm{SO}(2N)_{2k}\times \mathrm{USp}(2N+2x)_{-k}\right]/\mathbb{Z}_2$& $G_N(\hat{D}_k, \BZ_{2k})$ & $G_N(\hat{D}_k, \BZ_{2k})$& $G_N(\hat{D}_k, \BZ_{2k})$& $G_N(\hat{D}_k, \BZ_{2k})$\\
\hline 
$\mathrm{SO}(2N)_{2k}\times \mathrm{USp}(2N+2x)_{-k}$ & $G_N(\hat{D}_k, \BZ_{2k})$ & $G_N(\hat{D}_k, \BZ_{2k})$& $G_N(\hat{D}_k, \BZ_{2k})$& $G_N(\hat{D}_k, \BZ_{2k})$\\
\hline
 $\left[\mathrm{O}(2N)^+_{2k}\times \mathrm{USp}(2N+2x)_{-k}\right]/\BZ_2$&  $G_N(\hat{D}_k,\hat{D}_k)$&  $G_N(\hat{D}_k,\hat{D}_k)$&  Anomalous&  Anomalous\\ 
\hline
 $\left[\mathrm{Spin}(2N)_{2k}\times \mathrm{USp}(2N+2x)_{-k}\right]/\mathbb{Z}_2$& $G_N(\hat{D}_k, \BZ_{2k}).\mathbb{Z}_2$& Anomalous& $G_N(\hat{D}_k, \BZ_{2k}).\mathbb{Z}_2$ &  Anomalous\\
\hline 
$\left[\mathrm{O}(2N)^-_{2k}\times \mathrm{USp}(2N+2x)_{-k}\right]/\BZ_2$ & Anomalous &  $G_N(\hat{D}_k, \BZ_{2k}).\mathbb{Z}_2'$&  $G_N(\hat{D}_k, \BZ_{2k}).\mathbb{Z}_2'$&  Anomalous\\
\hline 
 $\mathrm{O}(2N)^{+}_{2k}\times \mathrm{USp}(2N+2x)_{-k}$& $G_N(\hat{D}_k,\hat{D}_k)$& $G_N(\hat{D}_k,\hat{D}_k)$ & $G_N(\hat{D}_k,\hat{D}_k)$& $G_N(\hat{D}_k,\hat{D}_k)$\\ 
\hline
 $\mathrm{Spin}(2N)_{2k}\times \mathrm{USp}(2N+2x)_{-k}$& $G_N(\hat{D}_k, \BZ_{2k}).\mathbb{Z}_2$ & $G_N(\hat{D}_k, \BZ_{2k}).\mathbb{Z}_2$& $G_N(\hat{D}_k, \BZ_{2k}).\mathbb{Z}_2$ & $G_N(\hat{D}_k, \BZ_{2k}).\mathbb{Z}_2$\\ 
\hline
 $\mathrm{O}(2N)^{-}_{2k}\times \mathrm{USp}(2N+2x)_{-k}$&$G_N(\hat{D}_k, \BZ_{2k}).\mathbb{Z}_2'$ &$G_N(\hat{D}_k, \BZ_{2k}).\mathbb{Z}_2'$ &$G_N(\hat{D}_k, \BZ_{2k}).\mathbb{Z}_2'$& $G_N(\hat{D}_k, \BZ_{2k}).\mathbb{Z}_2'$\\
\hline 
 $\mathrm{Pin}(2N)_{2k}\times \mathrm{USp}(2N+2x)_{-k}$& $G_N(\hat{D}_k,\hat{D}_k).\mathbb{Z}_2$ &$G_N(\hat{D}_k,\hat{D}_k).\mathbb{Z}_2$&$G_N(\hat{D}_k,\hat{D}_k).\mathbb{Z}_2$& $G_N(\hat{D}_k,\hat{D}_k).\mathbb{Z}_2$\\
\hline
\end{tabular}}
}
For this class of theories, the symmetry $\mathbb{Z}_{2,B}^{[0]}$ acts trivially on the moduli space, due to the fact that the matrix $R_B$ is already present in the quaternionic reflection group $G_N(\hat{D}_k, \BZ_{2k})$ associated with the ``mother" $\left[\mathrm{SO}(2N)_{2k}\times \mathrm{USp}(2N+2x)_{-k}\right]/\mathbb{Z}_2$ theory, from which all other variants can be obtained by gauging discrete zero-form symmetries. 

We see that $\BZ_{2, \CC}^{[0]}$ acts non-trivially on the moduli space of $\mathrm{SO}(2N)_{2k}\times \mathrm{USp}(2N+2x)_{-k}$, but acts trivially on that of $\mathrm{SO}(2N+2x)_{2k}\times \mathrm{USp}(2N)_{-k}$. This phenomenon can be explained as follows. Recall that, for the $\SO(N_c)$ gauge theory with $N_f$ hypermultiplets in the vector representation, a baryon, which is constructed using the epsilon tensor of the $\SO(N_c)$ gauge group and is odd under the charge conjugation symmetry, can acquire a non-zero vacuum expectation value if $N_f \geq N_c$ \cite{Argyres:1996eh} (see also \cite[Appendix B.3]{Cremonesi:2014uva}). In the former class of theories, the effective number of flavours of the $\SO(2N)$ gauge group is $N_f= 2N+2x$ and there is a non-zero vacuum expectation value of the baryon in this case. On the contrary, for the latter class, the effective number of flavours for $\SO(2N+2x)$ is $N_f= 2N$, so there is no baryon in this case. As a result, gauging $\BZ_{2, \CC}^{[0]}$ in the former theory turns $G_N(\hat{D}_k, \BZ_{2k})$ into $G_N(\hat{D}_k, \hat{D}_{k})$, and doing so in the latter theory leaves $G_N(\hat{D}_k, \hat{D}_{k})$ unchanged.

Let us now argue why the moduli space of $\left[\mathrm{SO}(2N+2x)_{2k}\times \mathrm{USp}(2N)_{-k}\right]/\mathbb{Z}_2$ is $\BH^{2N}/G_N(\hat{D}_k , \hat{D}_k)$, whereas that of $\left[\mathrm{SO}(2N)_{2k}\times \mathrm{USp}(2N+2x)_{-k}\right]/\mathbb{Z}_2$ turns out to be $\BH^{2N}/G_N(\hat{D}_k , \BZ_{2k})$. It was pointed out in \cite{Aharony:2008gk} (see also \cite[Appendix D.2]{Deb:2024zay}) that the moduli spaces of $\mathrm{O}(2N+2x)^+_{2k}\times \mathrm{USp}(2N)_{-k}$ and that of $\mathrm{O}(2N)^+_{2k}\times \mathrm{USp}(2N+2x)_{-k}$ are $\Sym^N(\BC^{4}/\hat{D}_k) = \BH^{2N}/G_N(\hat{D}_k , \hat{D}_k)$. Due to the argument in the preceding paragraph, we see that the moduli space of $\mathrm{SO}(2N+2x)_{2k}\times \mathrm{USp}(2N)_{-k}$ remains $\BH^{2N}/G_N(\hat{D}_k , \hat{D}_k)$, whereas that of $\mathrm{SO}(2N)_{2k}\times \mathrm{USp}(2N+2x)_{-k}$ becomes $G_N(\hat{D}_k,\BZ_{2k})$. Due to the trivial action of $\mathbb{Z}_{2,B}^{[0]}$ on the moduli spaces of $\left[\mathrm{SO}(2N)_{2k}\times \mathrm{USp}(2N+2x)_{-k}\right]/\mathbb{Z}_2$ and $\left[\mathrm{SO}(2N+2x)_{2k}\times \mathrm{USp}(2N)_{-k}\right]/\mathbb{Z}_2$, the result follows.

\subsection{Comments on the cases with anomalous one-form symmetry} \label{sec:commentsanomoneform}
Finally, let us also comment on the cases in which $kx$ is odd. Since condition \eref{Z21formcond} is not satisfied, the $\BZ^{[1]}_2$ one-form symmetry associated with the diagonal subgroup of the centres of the gauge groups is anomalous, hence cannot be gauged. The attempt to gauge this one-form symmetry turns out to be a trivial operation at the level of the index.
Indeed, we will see that any monopole operators containing half-odd-integral gauge fluxes do not contribute to the index.\footnote{Note that a similar phenomenon also appears in the $\left[\U(N+x)_{k} \times \U(N)_{-k}\right]/\BZ_p$ variants of the unitary ABJ theories, where the $\BZ_p$ quotient is consistent if $\frac{k x}{p^2} \in \BZ$ \cite{Tachikawa:2019dvq}. If one tries to quotient the original ABJ theory by $\BZ_I$, \ie considers the $\left[\U(N+x)_{k} \times \U(N)_{-k}\right]/\BZ_I$ theory, where $\BZ_I$ does not satisfy the consistency condition on the quotient, then there are ill-quantised monopole operators under the gauge group which drop out of the integral upon computing the index. For instance, the index of the anomalous $\left[\U(3)_4 \times \U(1)_{-4}\right]/\BZ_4$ theory equals the one of the non-anomalous $\left[\U(3)_4 \times \U(1)_{-4}\right]/\BZ_2$ theory. We thank Gabi Zafrir for pointing this out to us.}

For simplicity, let us focus on the $\SO(2 N+2 x)_{2 k} \times \USp(2N)_{-k}$ theory, and investigate the behaviour of the monopole operator with magnetic fluxes $\left(\frac{1}{2},\ldots,\frac{1}{2};\frac{1}{2},\ldots,\frac{1}{2}\right)$, which would appear in the anomalous $\left[\SO(2 N+ 2 x)_{2 k} \times \USp(2N)_{-k}\right]/\BZ_2$ variant. Such a bare monopole operator is not gauge invariance, indeed it would contribute to the index as $\left(\prod_{i=1}^{N+x}z_i^k\right) \left(\prod_{j=1}^N u_j^{-k}\right)$. In order to make it gauge invariant, it must be dressed by appropriate products of chiral fields components contributing to the index as $\left(\prod_{i=1}^{N+x}z_i^{-k}\right) \left(\prod_{j=1}^N u_j^{k}\right)$. Since the matter content of the theory is in the bifundamental representation of $\SO(2 N+2 x) \times \USp(2N)$, the chiral fields can be parametrised with the gauge fugacities as $\left[\sum_{i=1}^{N+x} \left(z_i + z_i^{-1}\right)\right] \left[\sum_{j=1}^{N} \left(u_j + u_j^{-1}\right)\right]$. In other words, each chiral field component carries an $\SO(2N+2x)$ fugacity $z_i^{s_1}$ and a $\USp(2N)$ fugacity $u_j^{s_2}$, with $s_1, s_2 = \pm 1$. In particular, observe that each fugacity $z_i^{\pm 1}$ and $u_j^{\pm 1}$ coming from products of the chiral fields components has to appear with the same multiplicity at the end in order to restore gauge invariance. However, this cannot happen since there is an odd disparity $x$ between the number of $\SO(2N+2x)$ and $\USp(2N)$ fugacities. The natural interpretation is then that the anomalous sector of the index coming from this particular monopole, which would appear in the theory in which the anomalous $\BZ_2^{[1]}$ one-form symmetry is gauged, is identically equal to zero, since there are no gauge invariant contributions which would survive the computation of residues in the index, meaning that such anomalous contribution gets killed by the integration process.

To clarify better this point, let us consider the case $N=2$, $x = 1$ and $k=1$, \ie we take the theory to be $\SO(6)_2 \times \USp(4)_{-1}$. In such a case, the anomaly of the $\BZ_2^{[1]}$ one-form symmetry can be investigated by looking at the bare monopole operator with fluxes $\left(\frac{1}{2},\frac{1}{2},\frac{1}{2};\frac{1}{2},\frac{1}{2}\right)$, which contributes $z_1 z_2 z_3 u_1^{-1} u_2^{-1}$ to the index, and has dimension zero. In order to make it gauge invariant, it must be dressed with chiral fields which compensate such contribution. This can be done for the $\SO(6)$ part by taking, for instance, the product of three components in the chiral fields, for example, $z_1^{-1} u_1$, $z_2^{-1} u_2$ and $z_3^{-1} u_1^{-1}$. However, this particular combination fails to preserve gauge invariance related to the $\USp(4)$ fugacity $u_1$, since the total combination reads $u_1^{-1} u_1 u_1^{-1} = u_1^{-1}$.\footnote{For odd $kx$, one can show that any attempt to dress such a monopole operator fails, due to the fact that there is always a disparity in fugacity $u_i$, for some $i$.} Upon looking at the integrand in the index, the monopole operator dressed with the product of such components appears at order $x^{\frac{3}{2}}$, but then drops out when the integral is performed.\footnote{On the other hand, for the case of $k=2$, the monopole operator with fluxes $\left(\frac{1}{2},\frac{1}{2},\frac{1}{2};\frac{1}{2},\frac{1}{2}\right)$ carries gauge fugacities $(z_1 z_2 z_3 u_1^{-1} u_2^{-1})^2$; however, this can be dressed using the components of the chiral fields in the following set: $\{ z_i^{-1} u_j | i,j =1,2 \} \cup \{z_3^{-1} u_1, z_3^{-1} u_1^{-1} \} $, and this gauge invariant quantity appears at order $x^3$ in the index.}

\section{Comments on the case with $\so(2N+1)$ gauge algebra}\label{Sec:Soodd}
We now discuss theories with the gauge algebra $\so(2N+1)_{2k} \times \usp(2M)_{-k}$. Only two distinct variants exist, namely
\bes{
\SO(2N+1)_{2k} \times \USp(2M)_{-k}~, \qquad
\Spin(2N+1)_{2k} \times \USp(2M)_{-k}~.
}
This follows from the trivial centre of the $\SO(2N+1)$ group. Furthermore, since the $\BZ_2$ centre of the $\USp(2M)$ gauge group is screened by the bifundamental half-hypermultiplets, the theory possesses no one-form symmetry. It is therefore meaningless to discuss the $\left[\SO(2N+1)_{2k} \times \USp(2M)_{-k}\right]/\mathbb{Z}_2$ variant. Moreover, as pointed out in \cite[(3.107)]{Beratto:2021xmn} using the index, the charge conjugation symmetry  $\mathbb{Z}_{2,\mathcal{C}}^{[0]}$ acts trivially on the $\SO(2N+1)_{2k} \times \USp(2M)_{-k}$ theory, since the corresponding fugacity can be reabsorbed with a gauge transformation and thus disappears from the index. The only remaining zero-form symmetry to discuss for the $ \SO(2N+1)_{2k} \times \USp(2M)_{-k}$ theory is the magnetic symmetry $\mathbb{Z}_{2,\mathcal{M}}^{[0]}$. This symmetry is non-anomalous, which can be seen as follows: the fugacity $\zeta$ associated with $\mathbb{Z}_{2,\mathcal{M}}^{[0]}$ appears in the index as $\zeta^{\sum_{i=1}^{N} m_i}$, where $m_i \in \BZ$ are the magnetic fluxes of $\SO(2N+1)$. Since the $m_i$ take only integral values, $\zeta$ necessarily appears with an integral power. An explicit computation shows that $\mathbb{Z}_{2,\mathcal{M}}^{[0]}$ acts non-trivially, and gauging this symmetry leads to the $\Spin(2N+1)_{2k} \times \USp(2M)_{-k}$ theory.

Analogous to \eqref{ABJMdualitiesunequalrank}, the following dualities also hold \cite{Aharony:2008gk}:
\bes{\label{ABJMdualitiesunequalrank1}
\scalebox{0.95}{$
\begin{split}
\O(2N+2x+1)^+_{2k} \times \Usp(2N)_{-k}\,\, &\leftrightarrow \,\, \O(2N+2(k-x)+1)^+_{-2k}\times \Usp(2N)_{k}~,  \\
\O(2N+1)^+_{2k} \times \Usp(2N+2x)_{-k}\,\, &\leftrightarrow \,\, \O(2N+1)^+_{-2k}\times \Usp(2N+2(k-x))_{k}~.
\end{split}$}
}
In both cases, $x$ is restricted to $0 \leq x \leq k$. Supersymmetry is broken if $x > k$. When $x=k$, these unequal-rank theories become dual to equal-rank theories.

The moduli space for these theories is $\BH^{2N}/\Gamma$, where $\Gamma$ is given below:
\bes{\label{tabSOodd}
\scalebox{0.9}{
\renewcommand{\arraystretch}{1.2}
\begin{tabular}{c|c}
\hline
 Theory  &  $\Gamma$\\
\hline 
$\mathrm{SO}(2N+1)_{2k}\times \mathrm{USp}(2N)_{-k}$& $G_N(\hat{D}_k,\hat{D}_k)$\\ 
\hline
 $\mathrm{Spin}(2N+1)_{2k}\times \mathrm{USp}(2N)_{-k}$& $G_N(\hat{D}_k,\hat{D}_k).\mathbb{Z}_2$\\
\hline \hline
$\mathrm{SO}(2N+2x+1)_{2k}\times \mathrm{USp}(2N)_{-k}$& $G_N(\hat{D}_k,\hat{D}_k)$\\ 
\hline
 $\mathrm{Spin}(2N+2x+1)_{2k}\times \mathrm{USp}(2N)_{-k}$& $G_N(\hat{D}_k,\hat{D}_k).\mathbb{Z}_2$\\
\hline \hline
$\mathrm{SO}(2N+1)_{2k}\times \mathrm{USp}(2N+2x)_{-k}$& $G_N(\hat{D}_k,\hat{D}_k)$\\ 
\hline
 $\mathrm{Spin}(2N+1)_{2k}\times \mathrm{USp}(2N+2x)_{-k}$& $G_N(\hat{D}_k,\hat{D}_k).\mathbb{Z}_2$\\
\hline 
\end{tabular}}
}
We have checked that the Hilbert series of the Higgs (or Coulomb) branch, $\BH^{N}/\Gamma$, computed using the prescription in Section \ref{quatrefgroupmodspace}, agrees with the corresponding limit of the index. We also note that the moduli space of $\SO(5)_{2k} \times \USp(4)_{-k}$ was discussed in \cite{Deb:2024zay}.

\section{SCFTs based on the $F(4)$ superalgebra} \label{sec:F4}
Let us consider the 3d $\CN = 3$ $\Spin(7)_{k_1} \times \SU(2)_{k_2}$ gauge theory, with two half-hypermultiplets $Q_1$, $Q_2$ in the $\left(\vec{8},\vec{2}\right)$ representation of the gauge group. As pointed out in \cite{Deb:2024zay}, following the analysis of \cite{Schnabl:2008wj}, when $2 k_1 + 3 k_2 = 0$, supersymmetry gets enhanced to $\CN = 5$. In particular, the authors of the latter reference studied the moduli space of the $\Spin(7)_{-3 k} \times \SU(2)_{2 k}$ theory, with $k$ integer, and of its $\BZ_2$ quotient, which is identified with the $\BZ_2$ one-form symmetry associated with the diagonal subgroup of the $\BZ_2 \times \BZ_2$ centre symmetry of $\Spin(7) \times \SU(2)$. Given that the instanton number of $\Spin(7)/\BZ_2$, denoted by $l_{\Spin(7)/\BZ_2}$, is integer \cite[(3.56)]{Witten:2000nv}, and that of $\SU(2)/\BZ_2$, denoted by $l_{\SU(2)/\BZ_2}$, is half-integer \cite[(3.38)]{Witten:2000nv}, the variation of the Chern-Simons action under a gauge transformation associated with the $\BZ_2$ quotient in question yields a trivial phase $\exp\left[2 \pi i \left(-3 k l_{\Spin(7)/\BZ_2} + 2 k l_{\SU(2)/\BZ_2}\right)\right] = 1$, meaning that the $\BZ_2$ one-form symmetry is non-anomalous for any integer value of $k$ and can thus be gauged \cite{Deb:2024zay}. Such $\BZ_2$ one-form symmetry is responsible for turning the $\Spin(7)$ gauge group into an $\SO(7)$ gauge group, which is centreless and does not admit any further $\BZ_2$ quotient.

In order to analyse the other possible global variants of the theory, namely the ones with $\O(7)^\pm$ and $\Pin(7)$ gauge factors, we have to take into account the action of the charge conjugation symmetry associated with the $\Spin(7)$ gauge group in the original theory. To this purpose, we recall that, in the $\so(2L+1)$ case, the orthogonal variant is simply $\O(2 L +1) \cong \SO(2L+1) \times \BZ_2$ and a generic $\O(2L+1)$ holonomy of determinant $\chi$ can be put in the form $\left(z_1, z^{-1}_1, \ldots,z_L, z^{-1}_L, \chi\right)$. In the following, we claim that charge conjugation acts trivially on the matter content in the $(\vec{8}, \vec{2})$ representation of the $\Spin(7)_{-3 k} \times \SU(2)_{2 k}$ gauge group. This can be deduced by employing the branching rules from $\so(7)$ to $\su(2)^3$. In particular, the vector representation of $\so(7)$ branches into the $\left(\vec{2},\vec{2},\vec{1}\right) \oplus \left(\vec{1},\vec{1},\vec{3}\right)$ representation of $\su(2)^3$, where the corresponding character can be refined with the charge conjugation fugacity $\chi$ as \footnote{Observe that the character of the vector representation of $\so(7)$ written in this way this is related to the convention adopted in \eref{indexSpin7SU2modZ2}, namely
\bes{ \label{7sottosu2cube}
\vec{7}_{\so(7)} = \left(\sum_{i=1}^3 z_i + z^{-1}_i\right) + \chi~,
}
via the fugacity map $z_1 = x_1 x_2$, $z_2 = \frac{x_1}{x_2}$, $z_3 = x_3^2$. \label{foot:7so7}}
\bes{
\vec{7}_{\so(7)} = \left(x_1 + x^{-1}_1\right) \left(x_2 + x^{-1}_2\right) + \left(x_3^2 + x^{-2}_3 + \chi \right)~.
}
It follows that charge conjugation acts non-trivially on the character of the antisymmetric representation of $\so(7)$, whose definition
\bes{ \label{21so7}
\vec{21}_{\so(7)} = \frac{1}{2} \Big\{&\left[\left(x_1 + x^{-1}_1\right) \left(x_2 + x^{-1}_2\right) + \left(x_3^2 + x^{-2}_3 + \chi \right)\right]^2 \\ 
- &\left[\left(x_1 + x^{-1}_1\right) \left(x_2 + x^{-1}_2\right) + \left(x_3^2 + x^{-2}_3 + \chi \right)\right]_{x_i \rightarrow x^2_i, \,\ \chi \rightarrow \chi^2 = 1}\Big\}
}
is in agreement with the expression presented in \cite[(2.6)]{Harding:2025vov}, upon exploiting the fugacity map detailed in Footnote \ref{foot:7so7}. On the other hand, the spinor representation of $\so(7)$ branches to $\su(2)^3$ as $\left(\vec{2},\vec{1},\vec{2}\right) \oplus \left(\vec{1},\vec{2},\vec{2}\right)$, where the $\su(2)$ triplet, whose character can be refined with $\chi$ as in \eref{7sottosu2cube}, is now absent. We then conclude that charge conjugation acts trivially on the spinor representation of $\so(7)$, hence the theory in question only admits the $\Spin(7) \times \SU(2)$ variant and its $\BZ_2$ quotient, namely $\left[\Spin(7)_{-3 k} \times \SU(2)_{2 k}\right]/\BZ_2$.

An interesting observation is that, in the special case $k = 1$, the $[\Spin(7)_{-3} \times \SU(2)_{2}]/\BZ_2$ gauge theory actually possesses $\CN=6$ supersymmetry. The crucial point is that the further enhancement of supersymmetry from $\CN=5$ to $\CN=6$ does not originate from the mechanism described in \cite{Schnabl:2008wj} in this case, for which the enhancement to $\CN=6$ supersymmetry is due to monopoles instead. This statement can be demonstrated explicitly by means of the index, whose expression up to order $x^2$, which can be computed as detailed in \eref{indexSpin7SU2modZ2}, reads
\bes{ \label{indSpintSU2modZ2k1}
1+ \left({\cerulean 1 + \zeta [2]_a} \right) x + \Big[\left(2+\zeta\right) [4]_a + [2]_a + 3 - \left({\cerulean 1+[2]_a}+{\claret [2]_a}\right)\Big] x^2 + \ldots~,
}
where we denote with $[m]_a$ the character corresponding to the $\su(2)$ representation with highest weight $m$ written in terms of the variable $a$, which is the fugacity associated with the ``axial symmetry''.\footnote{The action of the ``axial symmetry'' in question can be understood by adopting the 3d $\CN=2$ formalism, with manifest $\SO(2)$ R-symmetry, and looking at the effective superpotential of the $\Spin(7)_{-3 k} \times \SU(2)_{2 k}$ theory, obtained after integrating out the massive adjoint chiral fields associated with the $\Spin(7)$ and $\SU(2)$ gauge groups. If we label by $A, B, C, D = 1, \ldots, 16$ the indices of the $(\vec{8},\vec{2})$ representation and by $p,q =1, \ldots, 24$ the indices of the adjoint representation of the gauge group, with generators $T^p_{AB}$, the effective superpotential is given by \cite[(1.5)]{Schnabl:2008wj} $2 f_{ABCD} \epsilon^{\alpha \gamma} \epsilon^{\beta \delta} Q^A_\alpha Q^B_\beta Q^C_\gamma Q^D_\delta$, where $\alpha, \beta, \gamma, \delta = 1,2$ are $\su(2)$ indices and $f_{ABCD}=K_{pq}T^p_{AB}T^q_{CD}$, with $K_{pq}$ being the inverse Chern-Simons coefficient. This superpotential manifests an $\su(2)$ flavour symmetry, which we refer to as ``axial symmetry'', that transforms the two half-hypermultiplets as a doublet. This $\su(2)$ flavour symmetry is responsible for the enhancement of supersymmetry from $\CN=3$ to $\CN=5$, since it combines with the $\SO(3)$ R-symmetry, which is not manifest in $\CN=2$ notation, to form the $\SO(5)$ R-symmetry.} The contributions due to the $\CN=3$ flavour currents are highlighted in {\cerulean cerulean}, whereas the term coloured in {\claret claret} indicates the presence of three $\CN=3$ extra-supersymmetry currents, which explain the enhancement from $\CN=3$ to $\CN=6$ supersymmetry. A crucial role is played by the monopole operator with magnetic fluxes $\left(1,0,0;\frac{1}{2}\right)$, where the first three entries are associated with the $\so(7)$ gauge factor, and the last entry stands for the $\su(2)$ gauge flux. Such a monopole operator has dimension $-2$ and carries gauge charges $-3$ and $2$ under the $\so(7)$ Cartan element corresponding to $z_1$ and the $\su(2)$ gauge factor, respectively. In order to make this bare monopole operator gauge invariant, it needs to be dressed with six chiral fields chosen from copies of $Q_{1,2}$, hence it contributes to the index as $\zeta [2]_a$ at order $x$. Let us explain this point more in detail. The matter content of the theory transforms in the $(\vec{8},\vec{2})$ representation of the gauge group, as well as in the $\vec{2}$ representation of the ``axial symmetry''. The two half-hypermultiplets $Q_{1,2}$ carry fugacities associated with the said symmetries which are parametrised according to the character of this combined gauge and ``axial'' representation, where this can be expressed as
\bes{ \label{chso7su2su2}
\left(\sum_{s_1, \ldots, s_3=\pm 1} z_1^{\frac{s_1}{2}} z_2^{\frac{s_2}{2}} z_3^{\frac{s_3}{2}}\right) \left(u + \frac{1}{u}\right) \left(a+\frac{1}{a}\right)~.
}
Gauge invariant quantities can be built by taking appropriate products of chiral fields components parametrised by \eref{chso7su2su2}, whose collective contribution cancels the factor $z_1^{-3} u^{2}$, which obstructs gauge invariance of the bare monopole operator with fluxes $\left(1,0,0;\frac{1}{2}\right)$. In other words, we are looking for a set of components of chiral fields, whose product contribute to the index as $z_1^3 u^{-2} a^p$, where $p$ can be any integer. Such gauge invariant quantities can be built by considering the following parametrisations of the chiral fields components:
\bes{ 
\begin{array}{lll}
x_1 = z_1^{\frac{1}{2}} z_2^{\frac{1}{2}}z_3^{-\frac{1}{2}} u a &,~ x_2 = z_1^{\frac{1}{2}} z_2^{-\frac{1}{2}}z_3^{\frac{1}{2}} u a &,~ x_3 = z_1^{\frac{1}{2}} z_2^{\frac{1}{2}}z_3^{-\frac{1}{2}} u^{-1} a ~, \\ x_4 = z_1^{\frac{1}{2}} z_2^{-\frac{1}{2}}z_3^{\frac{1}{2}} u^{-1} a &,~ x_5 = z_1^{\frac{1}{2}} z_2^{\frac{1}{2}}z_3^{-\frac{1}{2}} u a^{-1} &,~ x_6 = z_1^{\frac{1}{2}} z_2^{-\frac{1}{2}}z_3^{\frac{1}{2}} u a^{-1} ~, \\ x_7 = z_1^{\frac{1}{2}} z_2^{\frac{1}{2}}z_3^{-\frac{1}{2}} u^{-1} a^{-1} &,~ x_8 = z_1^{\frac{1}{2}} z_2^{-\frac{1}{2}}z_3^{\frac{1}{2}} u^{-1} a^{-1} &,~ x_9 = z_1^{\frac{1}{2}} z_2^{-\frac{1}{2}}z_3^{-\frac{1}{2}} u^{-1} a ~, \\ x_{10} = z_1^{\frac{1}{2}} z_2^{\frac{1}{2}}z_3^{\frac{1}{2}} u^{-1} a^{-1}~.& &
\end{array}
}
We can then construct dimension three operators by taking products of six such combinations, which, combined with negative dimension of the bare monopole, yield dimension one dressed monopole operators appearing at order $x$ in the index. Such gauge invariant operators can be built, for instance, by considering the products $x_7 x_8 \prod_{i=1}^4 x_i = z_1^3 u^{-2} a^2$, $\prod_{i=3}^8 x_i = z_1^3 u^{-2} a^{-2}$ and $x_1 x_2 \prod_{i=7}^{10} x_i = z_1^3 u^{-2}$, where the corresponding dressed monopole operators contribute to the index as $\zeta a^2 x$, $\zeta a^{-2} x$ and $\zeta x$, respectively. These three terms form the expected contribution $\zeta [2]_a$ appearing at order $x$ in \eref{indSpintSU2modZ2k1}. Taking into account also the relevant operator $Q_1 Q_2$, it follows that there are four terms appearing at order $x$ in the index, which is a necessary condition for supersymmetry to enhance to $\CN=6$ \cite{Evtikhiev:2017heo}. 

On the other hand, for $k > 1$, the bare monopole operator associated with the magnetic fluxes $\left(1,0,0;\frac{1}{2}\right)$ carries gauge charges $-3 k$ and $2 k$ under the $\so(7)$ Cartan element corresponding to $z_1$ and the $\su(2)$ gauge factor. Hence, in order to give rise to gauge invariant operators, this bare monopole needs to be dressed with at least $6 k$ chiral fields. Such gauge invariant dressed monopole operators have dimension $3 k -2$, hence they appear at order $x^{3 k -2}$ in the index of the $\left[\Spin(7)_{-3 k} \times \SU(2)_{2 k}\right]/\BZ_2$ theory. For $k > 1$, they are not present at order $x$, where only the relevant operator $Q_1 Q_2$ appears, in agreement with the necessary condition for $\CN=5$ supersymmetry enhancement of \cite{Evtikhiev:2017heo}. It follows that supersymmetry gets enhanced to $\CN=6$ just for $k=1$, whereas, for $k > 1$, the $\left[\Spin(7)_{-3 k} \times \SU(2)_{2 k}\right]/\BZ_2$ theory possesses enhanced $\CN=5$ supersymmetry.

We can also consider the $\Spin(7)_{-3 k} \times \SU(2)_{2 k}$ theory, whose index can be derived, as explained in \eref{indexSpin7SU2}, from the one of the $\left[\Spin(7)_{-3 k} \times \SU(2)_{2 k}\right]/\BZ_2$ theory by summing over $\zeta=\pm1$ in the latter, and dividing by two. In the case $k=1$, the expression \eref {indSpintSU2modZ2k1} becomes
\bes{ \label{indSpintSU2modZ2k1modZ2}
1+ x + \Big[2 \left([4]_a + 1 \right) - [2]_a\Big] x^2 + \ldots~,
}
where there is just a single relevant operator $Q_1 Q_2$, appearing at order $x$. The negative terms at order $x^2$, namely $a^2 +1 + a^{-2}$, reveal that there are two extra-supersymmetry currents, which are compatible with the enhanced $\CN=5$ supersymmetry, see Footnote \ref{foot:Neq5}. This means that, upon gauging the discrete zero-form symmetry associated with $\zeta$, supersymmetry gets broken from $\CN=6$ down to $\CN=5$ in the case $k=1$.

As a final remark, we observe that, upon taking either the Higgs or the Coulomb branch limit of the indices \eref{indSpintSU2modZ2k1} and \eref{indSpintSU2modZ2k1modZ2} as explained in \eref{limitsindex}, these yield the Hilbert series of $\BC^2/\BZ_4$ and $\BC^2/\hat{D}_2$, respectively, in agreement with \cite[(3.37) and (3.38)]{Deb:2024zay}. Note that the series expansion of the former Hilbert series contains a term at order $t^2$, corresponding to the contribution coming from the dressed monopole with fluxes $\left(1,0,0;\frac{1}{2}\right)$, which is instead absent in the latter Hilbert series. For $k > 1$, the moduli spaces of the $\left[\Spin(7)_{-3 k} \times \SU(2)_{2 k}\right]/\BZ_2$ and $\Spin(7)_{-3 k} \times \SU(2)_{2 k}$ theories are instead of the form $\BH^2/\hat{D}_{3k}$ and $\BH^2/\hat{D}_{6k}$, respectively \cite{Deb:2024zay}.

\acknowledgments
We would like to thank Riccardo Comi, Amihay Hanany and Gabi Zafrir for interesting discussions. W.H. also acknowledges the DESY Theory Group, Hamburg, in particular Craig Lawrie, and the Abdus Salam Centre for Theoretical Physics, Imperial College London, especially Amihay Hanany, for hospitality during the realisation of this project. N.M.~gratefully acknowledges support from the Simons Center for Geometry and Physics, Stony Brook University, at which some of the research for this paper was performed during the 22nd Simons Physics Summer Workshop 2025. This work is supported in part by the INFN. W.H. and N.M.'s research is partially supported by the MUR-PRIN grant No. 2022NY2MXY (Finanziato dall'Unione europea -- Next Generation EU, Missione 4 Componente 1 CUP H53D23001080006, I53D23001330006). D.L.'s research is partially supported by STFC Consolidated Grants ST/T000791/1 and ST/X000575/1.

\appendix

\section{The superconformal index} \label{app:index}
In this Appendix, we collect the expressions for the superconformal index \cite{Bhattacharya:2008zy,Bhattacharya:2008bja,Kim:2009wb,Imamura:2011su,Kapustin:2011jm,Dimofte:2011py,Aharony:2013dha,Aharony:2013kma} of the theories considered in this paper. We adopt of the convention of \cite{Aharony:2013kma}. For the $\left[\SO(2L)_{2k_1}\times \USp(2M)_{k_2}\right]/\BZ_2$ theory, we denote by $g$, $\zeta$ and $\chi$ the fugacities for the zero-form symmetries $\BZ^{[0]}_{2, B}$, $\BZ^{[0]}_{2, \CM}$ and $\BZ^{[0]}_{2, \CC}$, respectively.  We also turn on the fugacity $a$ associated with the ``axial symmetry'', where each of the $\SO(2L) \times \USp(2M)$ bifundamental half-hypermultiplets carries charges $1$ and $-1$. The index for the $\left[\SO(2L)_{2k_1}\times \USp(2M)_{k_2}\right]/\BZ_2$ theory with $\chi=+1$ is given by
\bes{ \label{indABJlikemodZ2}
\scalebox{0.87}{$
\renewcommand{\arraystretch}{1.3} 
\begin{split}
&\CI \left\{\left[\SO(2L)_{2k_1}\times \USp(2M)_{k_2}\right]/\BZ_2\right\}(x;a;g;  \zeta; \chi=+1) \\
& = \frac{1}{L! 2^{L-1}} \times \frac{1}{M! 2^M} \,\, \sum_{e=0}^1 g^e \,\, \sum_{(m_1, \ldots, m_L) \in \left(\BZ+\frac{e}{2} \right)^L} \sum_{(n_1, \ldots, n_M) \in \left(\BZ+\frac{e}{2} \right)^M} \zeta^{\sum_i m_i} \\
& \quad \times \oint \left( \prod_{\alpha=1}^L \frac{d z_\alpha}{2\pi i z_\alpha} \prod_{\beta =1}^M \frac{d u_\beta}{2\pi i u_\beta} \right)\prod_{\alpha=1}^L z_\alpha^{2k_1 m_\alpha}\prod_{\beta=1}^M u_\beta^{2k_2 n_\beta} \\
& \quad \times \CZ_{\text{vec}}^{\SO(2L)}(x; \vec z; \vec m; \chi = +1) \CZ_{\text{vec}}^{\USp(2M)}(x; \vec u; \vec n)  \\
& \quad \times \prod_{\alpha=1}^L\prod_{\beta=1}^M \prod_{s_1, s_2=\pm 1} \CZ_{\text{chir}}^{1/2} (x; a z_\alpha^{s_1} u_\beta^{s_2};s_1 m_\alpha +s_2 n_\beta)~\CZ_{\text{chir}}^{1/2} (x; a^{-1} z_\alpha^{s_1} u_\beta^{s_2};s_1 m_\alpha +s_2 n_\beta)~,
\end{split}$}
}
where the contribution of a chiral multiplet with $R$-charge $R$ is
\bes{
\CZ_{\text{chir}}^{R} (x; z; m ) = (x^{1-R} z^{-1})^{|m|/2} \prod_{j=0}^\infty \frac{1-(-1)^m z^{-1} x^{|m|+2-R+2j}}{1-(-1)^m z x^{|m|+R+2j}}~,
}
and the vector multiplet contributions are
\bes{
\scalebox{0.9}{$
\begin{split}
&\CZ_{\text{vec}}^{\SO(2L)}(x; \vec z; \vec m; \chi = +1) \\
&=  \prod_{1 \leq a< b\leq L} \prod_{s_1, s_2 = \pm 1} x^{-|s_1 m_a +s_2 m_b|/2} \Big(1-(-1)^{|s_1 m_a +s_2 m_b|} z_a^{s_1} z_b^{s_2} x^{|s_1 m_a +s_2 m_b|} \Big)~,
\end{split}
$}
}
\bes{ \label{vecadjUSp2N}
\scalebox{0.98}{$
\begin{split}
&\CZ_{\text{vec}}^{\USp(2M)}(x; \vec u; \vec n) =  \prod_{\ell=1}^M x^{-|2n_\ell|} \prod_{s=\pm 1}  (1-(-1)^{2 s n_\ell} u_\ell^{2s} x^{|2n_\ell|} ) \\
&\qquad \times \prod_{1 \leq a< b\leq M} \,\, \prod_{s_1, s_2 = \pm 1} x^{-|s_1 n_a +s_2 n_b|/2}  \Big(1-(-1)^{|s_1 n_a +s_2 n_b|} u_a^{s_1} u_b^{s_2} x^{|s_1 n_a +s_2 n_b|} \Big)~.
\end{split}
$}
}
On the other hand, the index for $\chi=-1$ is
\bes{
\scalebox{1}{$
\begin{split}
&\CI\left\{\left[\SO(2L)_{2k_1} \times \USp(2M)_{-k_2}\right]/\BZ_2\right\}(x;a; \zeta; \chi=-1) \\
&= \CI\left[\SO(2L)_{2k_1}\times \USp(2M)_{-k_2}\right](x;a;  \zeta; \chi=-1) \\
& = \frac{1}{(L-1)! 2^{L-1}} \times \frac{1}{M! 2^M} \,\, \sum_{(m_1, \ldots, m_L) \in \BZ^L} \sum_{(n_1, \ldots, n_L) \in \BZ^M} \zeta^{\sum_i m_i} \\
& \qquad \times \oint \left( \prod_{\alpha, \beta =1}^L \frac{d z_\alpha}{2\pi i z_\alpha}  \frac{d u_\beta}{2\pi i u_\beta} \right) \prod_{\alpha=1}^L z_\alpha^{2k_1 m_\alpha}\prod_{\beta=1}^M u_\beta^{2k_2 n_\beta} \\
& \qquad \times \CZ_{\text{vec}}^{\SO(2L)}(x; \vec z; \vec m; \chi = -1) \CZ_{\text{vec}}^{\USp(2M)}(x; \vec u; \vec n)  \\
& \qquad \times \prod_{i=1}^{n} \prod_{\alpha, \beta=1}^N \prod_{s, s_1, s_2=\pm 1} \CZ_{\text{chir}}^{1/2} (x; a^s f_i z_\alpha^{s_1} u_\beta^{s_2};s_1 m_\alpha +s_2 n_\beta) \Bigg|_{z_L=1, \, z_L^{-1}=-1, \, m_L=0}~,
\end{split}$}
}
where the $\SO(2L)$ vector multiplet in this case is
\bes{ \label{vecchi-1}
&\CZ_{\text{vec}}^{\SO(2L)}(x; \vec z; \vec m; \chi = -1) = \Big[ \CZ_{\text{vec}}^{\SO(2L)}(x; \vec z; \vec m; \chi = +1) \Big]_{z_L = 1,\, z_L^{-1}=-1, \, m_L =0} \\
& = \prod_{1 \leq a< b\leq L-1} \,\, \prod_{s_1, s_2 = \pm 1} x^{-|s_1 m_a +s_2 m_b|/2}  \Big(1-(-1)^{|s_1 m_a +s_2 m_b|} z_a^{s_1} z_b^{s_2} x^{|s_1 m_a +s_2 m_b|} \Big) \\ & \quad \,\,\, \times \prod_{\ell=1}^{L-1} x^{-|2 m_\ell|} (1- (-1)^{2 m_\ell} z_\ell^{2} x^{|2 m_\ell|}) (1- (-1)^{2 m_\ell} z_l^{-2} x^{|2 m_\ell|})~.
}
The index of the $\SO(2L)_{2k_1}\times \USp(2M)_{k_2}$ theory can then by obtained as
\bes{
\frac{1}{2} \sum_{g = \pm 1} \CI \left\{\left[\SO(2L)_{2k_1}\times \USp(2M)_{k_2}\right]/\BZ_2\right\}(x;a;g;  \zeta; \chi)~.
}
We also report the index for the $\SO(2N+1)_{2k_1} \times \USp(2M)_{-k_2}$ theory:
\bes{ \label{indABJlikeSOodd}
\scalebox{0.9}{$
\begin{split}
&\CI\left\{\left[\SO(2N+1)_{2k_1}\times \USp(2M)_{k_2}\right]\right\}(x;a;g;  \zeta; \chi) \\
& = \frac{1}{N! 2^{N}} \times \frac{1}{M! 2^M} \,\, \sum_{(m_1, \ldots, m_N) \in \left(\BZ\right)^N} \sum_{(n_1, \ldots, n_M) \in \left(\BZ\right)^M} \zeta^{\sum_i m_i} \\
& \quad \times \oint \left( \prod_{\alpha=1}^N \frac{d z_\alpha}{2\pi i z_\alpha} \prod_{\beta =1}^M \frac{d u_\beta}{2\pi i u_\beta} \right)\prod_{\alpha=1}^N z_\alpha^{2k_1 m_\alpha}\prod_{\beta=1}^M u_\beta^{2k_2 n_\beta}   \\
& \quad \times \CZ_{\text{vec}}^{\SO(2(N+1))}(x; \vec z; \vec m; \chi) \CZ_{\text{vec}}^{\USp(2M)}(x; \vec u; \vec n)  \\
& \quad \times \prod_{\alpha=1}^N\prod_{\beta=1}^M \prod_{s_1, s_2=\pm 1} \CZ_{\text{chir}}^{1/2} (x; a z_\alpha^{s_1} u_\beta^{s_2};s_1 m_\alpha +s_2 n_\beta)~\CZ_{\text{chir}}^{1/2} (x; a^{-1} z_\alpha^{s_1} u_\beta^{s_2};s_1 m_\alpha +s_2 n_\beta)\\&
\quad \times \prod_{\beta=1}^M \prod_{s_1=\pm 1} \CZ_{\text{chir}}^{1/2} (x; a  u_\beta^{s_1}\chi;s_1 n_\beta)\CZ_{\text{chir}}^{1/2} (x; a^{-1}  u_\beta^{s_1}\chi;s_1 n_\beta)~,
\end{split}$}
}
where the $\SO(2N+1)$ vector contribution is given by:
\bes{
\scalebox{0.9}{$
\begin{split}
&\CZ_{\text{vec}}^{\SO(2N+1)}(x; \vec z; \vec m; \chi) =  \prod_{\ell=1}^N \prod_{s=\pm 1} x^{-| m_\ell|/2} (1-(-1)^{m_\ell} \chi z^s_\ell x^{|m_\ell|} )  \\ & \qquad \times \prod_{1 \leq a< b\leq N} \,\, \prod_{s_1, s_2 = \pm 1} x^{-|s_1 m_a +s_2 m_b|/2} \Big(1-(-1)^{|s_1 m_a +s_2 m_b|} z_a^{s_1} z_b^{s_2} x^{|s_1 m_a +s_2 m_b|} \Big)~.
\end{split}
$}
}
Note that, as pointed out in \cite[(3.107)]{Beratto:2021xmn}, in the index \eqref{indABJlikeSOodd} the fugacity $\chi$ for the $\mathbb{Z}_{2,C}^{[0]}$ symmetry can be reabsorbed by a gauge transformation.

Finally, let us also report the index of the $\left[\Spin(7)_{-3 k} \times \SU(2)_{2 k}\right]/\BZ_2$ theory, with two half-hypermultiplets in the $(\vec{8},\vec{2})$ representation of the gauge group, discussed in Section \ref{sec:F4}. This is given by the following expression:
\bes{ \label{indexSpin7SU2modZ2}
\scalebox{0.96}{$
\begin{split}
&\CI\left\{\left[\Spin(7)_{-3 k} \times \SU(2)_{2 k}\right]/\BZ_2\right\}(x;a; \zeta) \\
& = \frac{1}{96} \sum_{\epsilon =0}^1 \,\, \sum_{(m_1, m_2, m_3) \in \BZ^3} \,\, \sum_{n \in \BZ + \frac{\epsilon}{2}} \zeta^{m_1 + m_2 + m_3} \\
& \qquad \times \oint \left( \prod_{j =1}^3 \frac{d z_j}{2\pi i z_j} z_j^{-3 k m_j} \right) \frac{d u}{2\pi i u} u^{4 k n}  \CZ_{\text{vec}}^{\SO(7)}(x; \vec z; \vec m; \chi = 1) \CZ_{\text{vec}}^{\SU(2)}(x; u; n)  \\
& \qquad \times \prod_{s = \pm 1} \,\, \prod_{s_1, \ldots, s_4=\pm 1} \CZ_{\text{chir}}^{1/2} \left(x; a^s z_1^{\frac{s_1}{2}} z_2^{\frac{s_2}{2}} z_3^{\frac{s_3}{2}} u^{s_4};\frac{s_1}{2} m_1 + \frac{s_2}{2} m_2 + \frac{s_3}{2} m_3 + s_4 n\right)~,
\end{split}
$}
}
where, as usual, we denote by $a$ is the fugacity for the``axial symmetry'', under which the two half-hypermultiplets carry charges $\pm 1$. The magnetic fluxes $(m_1, m_2, m_3)$ and $n$ associated with the $\so(7)$ and $\su(2)$ gauge factors, respectively, have to satisfy the Dirac quantisation condition, namely 
\bes{ \label{DiracqcondSpin7SU2}
\frac{s_1}{2} m_1 + \frac{s_2}{2} m_2 + \frac{s_3}{2} m_3 + s_4 n \in \BZ~, \quad \text{with} \quad s_1, \ldots, s_4 = \pm 1~.
}
In particular, when the global form of the gauge group is $\left[\Spin(7)_{-3 k} \times \SU(2)_{2 k}\right]/\BZ_2$, the parameter $\epsilon$ involved in the summation in \eref{indexSpin7SU2modZ2} takes values in $\{0, 1\}$, meaning that the flux $n$ can be either integer or half-integer, whereas the fluxes $(m_1, m_2, m_3)$ are integers, with no further parity restriction. It follows that the index depends on the $\BZ_2$ fugacity $\zeta$ associated with the zero-form magnetic symmetry of $\SO(7)$, which would be absent if the global form of the gauge group were $\Spin(7)$. Hence, the $\BZ_2$ quotient in the $\left[\Spin(7)_{-3 k} \times \SU(2)_{2 k}\right]/\BZ_2$ theory has the effect of turning the $\Spin(7)$ gauge group into $\SO(7)$, as well as acting non-trivially on the $\su(2)$ gauge factor. On the other hand, the index of the $\Spin(7)_{-3 k} \times \SU(2)_{2 k}$ variant can be implemented by setting $\epsilon = 0$ in \eref{indexSpin7SU2modZ2}, which is equivalent to summing over integer values of $n$. In order for the Dirac quantisation condition \eref{DiracqcondSpin7SU2} to be satisfied, the $\so(7)$ magnetic fluxes are constrained by the requirement that $m_1 + m_2 + m_3$ has to be even. As a consequence, the $\BZ_2$ fugacity $\zeta$ appears only with even powers and the index does not depend on it anymore, which is precisely the expected behaviour for the index involving strictly the $\Spin(7)$ gauge group. Indeed, the same index expression can be obtained by gauging the $\BZ_2$ magnetic symmetry of $\SO(7)$ in the $\left[\Spin(7)_{-3 k} \times \SU(2)_{2 k}\right]/\BZ_2$ theory by summing over $\zeta = \pm 1$ and dividing by two, which corresponds to turning $\SO(7)$ into $\Spin(7)$. Summarising, we have that
\bes{ \label{indexSpin7SU2}
\scalebox{0.94}{$
\CI\left[\Spin(7)_{-3 k} \times \SU(2)_{2 k}\right](x;a) = \frac{1}{2} \sum\limits_{\zeta = \pm 1} \CI\left\{\left[\Spin(7)_{-3 k} \times \SU(2)_{2 k}\right]/\BZ_2\right\}(x;a; \zeta)~.
$}
}

Moreover, the indices for various variants of the orthogonal gauge group can be computed as follows \cite[(6.13)]{Aharony:2013kma}:
\bes{ \label{indvariants}
\scalebox{1}{$
\begin{split}
\CI_{\O(N)^+}(\zeta) &=\frac{1}{2} \left[ \CI_{\SO(N)}(\zeta; \chi=+1)+\CI_{\SO(N)}(\zeta; \chi=-1) \right]~, \\
\CI_{\Spin(N)}(\chi) &=\frac{1}{2} \left[ \CI_{\SO(N)}(\zeta=+1; \chi)+\CI_{\SO(N)}(\zeta=-1; \chi) \right]~, \\
\CI_{\O(N)^-}(\zeta) &=\frac{1}{2} \left[ \CI_{\SO(N)}(\zeta; \chi=1)+\CI_{\SO(N)}(-\zeta; \chi=-1) \right]~, \\
\CI_{\Pin(N)} &=\frac{1}{2} \left[ \CI_{\Spin(N)}(\chi=+1)+\CI_{\Spin(N)}(\chi=-1) \right]~,
\end{split}$}}
where, for conciseness, we omit various parameters that are not relevant and display explicitly only the variants of the orthogonal group. The index for the $\SO(N)$ variant refined with respect to both $\zeta$ and $\chi$ is then
\bes{
\CI_{\SO(N)}(\zeta; \chi) = &\frac{1}{2} \left[ \CI_{\SO(N)} (\zeta, \chi=+1) + \CI_{\SO(N)}(\zeta,  \chi=-1) \right] \\
+ &\frac{1}{2} \left[ \CI_{\SO(N)}(\zeta, \chi=+1) -\CI_{\SO(N)}(\zeta, \chi=-1) \right] \chi~.
}

As pointed out in \cite{Razamat:2014pta}, the Coulomb branch limit of the index of a 3d $\CN=4$ theory can be obtained as $\sum_{p=0}^\infty C(a^{-2p} x^p) t^{2p}$, whereas the Higgs branch limit of the index is given by $\sum_{p=0}^\infty C(a^{2p} x^p) t^{2p}$, where $C(a^{\pm2p} x^p)$ denote the coefficients of the terms $a^{\pm 2p} x^p$ in the series expansion of the index. 
Note that taking these coefficients in the series expansion of the index is equivalent to set \bes{ \label{limitsindex}
x=hc~, \quad a=(hc^{-1})^{1/2}~,}
and then send $h\rightarrow 0$ (with $c=t$) in order to have the Coulomb branch Hilbert series, and $c\rightarrow 0$ (with $h=t$) for the Higgs branch Hilbert series. 
Since the theories discussed in this paper have at least $\CN=5$ supersymmetry, it is expected that the two branches of the moduli space are the same, and that the two limits of the index are equal.

\bibliographystyle{JHEP}
\bibliography{bibli.bib}

\end{document}